\documentclass[onecolumn,pre,floats,aps,amsmath,amssymb,nofootinbib]{revtex4-2}
\usepackage{graphicx}
\usepackage{caption}
\usepackage{subfig}
\usepackage{bm}
\usepackage{verbatim}
\usepackage{microtype}
\usepackage{amsmath}
\usepackage{siunitx}
\usepackage{physics}
\usepackage{amssymb}
\usepackage{slashed}
\usepackage{hyperref}
\newcommand{\bea}{\begin{eqnarray}}
\newcommand{\eea}{\end{eqnarray}}
\newcommand{\bs}{\boldsymbol}
\newcommand{\Romatre}{Dipartimento di Fisica, Universit\`a  Roma Tre and INFN, Sezione di Roma Tre,\\ Via della Vasca Navale 84, I-00146 Rome, Italy}
\newcommand{\RomatreINFN}{Istituto Nazionale di Fisica Nucleare, Sezione di Roma Tre,\\ Via della Vasca Navale 84, I-00146 Rome, Italy}
\newcommand{\Romadue}{Dipartimento di Fisica and INFN, Universit\`a di Roma ``Tor Vergata",\\ Via della Ricerca Scientifica 1, I-00133 Roma, Italy}
\newcommand{\LaSapienza}{Physics Department and INFN Sezione di Roma La Sapienza,\\ Piazzale Aldo Moro 5, 00185 Roma, Italy}
\newcommand{\soton}{Department of Physics and Astronomy, University of Southampton,\\ Southampton SO17 1BJ, UK}

\begin{document}
\title{Virtual Photon Emission in Leptonic Decays of Charged Pseudoscalar Mesons}
\author{G.\,Gagliardi}\affiliation{\RomatreINFN}
\author{V.\,Lubicz}\affiliation{\Romatre} 
\author{G.\,Martinelli}\affiliation{\LaSapienza}
\author{F.\,Mazzetti}\affiliation{\Romatre}
\author{C.T.\,Sachrajda}\affiliation{\soton}
\author{F.\,Sanfilippo}\affiliation{\RomatreINFN}
\author{S.\,Simula}\affiliation{\RomatreINFN}
\author{N.\,Tantalo}\affiliation{\Romadue}

\date{\today}
\begin{abstract}
We study the radiative leptonic decays $P\to\ell\nu_\ell\,\ell^{\prime\,+}\ell^{\prime\,-}$, where $P$ is a pseudoscalar meson and $\ell$ and $\ell^\prime$ are charged leptons. In such decays the emitted photon is off-shell and, in addition to the ``point-like" contribution in which the virtual photon is emitted either from the lepton or the meson treated as a point-like particle, four structure-dependent (SD) form factors contribute to the amplitude. 
We present a strategy for the extraction of the SD form factors and implement it in an exploratory lattice computation of the decay rates for the four channels of kaon decays ($\ell,\ell^\prime=e,\mu$). It is the SD form factors which describe the interaction between the virtual photon and the internal hadronic structure of the decaying meson, and in our procedure we separate the SD and point-like contributions to the amplitudes. We demonstrate that the form factors can be extracted with good precision and, in spite of the unphysical quark masses used in our simulation ($m_\pi\simeq 320\,$MeV and $m_K\simeq 530$\,MeV), the results for the decay rates are in reasonable semiquantitative agreement with experimental data (for the channels where these exist). Following this preparatory work, the emphasis of our future work will be on obtaining results at physical quark masses and on the control of the systematic uncertainties associated with discretisation and finite-volume errors.
\end{abstract}

\maketitle
 
 \section{Introduction}

The comparison of precise theoretical predictions for flavour-changing processes, in particular those which are suppressed in the Standard Model (SM), with experimental measurements is a fruitful approach to searches for new physics. For example, there have been experimental results suggesting the violation of Lepton Flavour Universality which is an important feature of the SM\,(see e.g. Refs.\cite{LHCb:2021trn,Crivellin:2021sff} and references therein). In order to confirm the presence of new physics and to elucidate its underlying structure it is important to study as many such processes as possible. In this paper we consider weak decays of the form $P\to\ell\nu_\ell\,\ell^{\prime +}\ell^{\prime\,-}$, where $\ell$ and $\ell^\prime$ are charged leptons, for which the decay rates start at $O(\alpha_\mathrm{em}^2)$.

For each decay $P\to\ell\nu_\ell\,\ell^{\prime +}\ell^{\prime\,-}$, the computation of the decay rate  requires the knowledge of four Structure Dependent (SD) hadronic form factors, that depend on the invariant masses of the two leptonic pairs $\ell\,\nu_\ell$ and $\ell^{\prime\,+}\ell^{\prime\,-}$ as well as of the leptonic decay constant $f_P$ (see Eqs.\,(\ref{hadronic})\,-\,(\ref{Hmunu_sd}) below). 
The ``point-like" (or inner-bremssstrahlung) contribution to the decay rate, in which the virtual photon is emitted either from the lepton $\ell$ or from the meson $P$ treated as a point-like particle, is readily calculable in perturbation theory, requiring only the well-known value of $f_P$ as the non-perturbative input. 
The SD form factors describe the interaction between the virtual photon and the internal hadronic structure of the decaying meson and their computation in Lattice Quantum Chromodyanmics (LQCD) is the subject of this paper. 
This work is a natural extension of our detailed studies and computations of isospin breaking corrections to leptonic decays
\,\cite{Carrasco2014,Lubicz:2016xro,Giusti:2017dwk,DiCarlo:2019thl} and to the calculation of leptonic radiative decays of the type $P\to\ell\nu_\ell\gamma$ where $\gamma$ is a real photon\,\cite{Desiderio2020,Frezzotti:2020bfa}.

Experimentally, only a few measurements exist. For the pion, the only measured decay rate is for the process $\pi^+\to e^+\,\nu_e\,e^+\,e^-$, for which the Particle Data Group (PDG) reports a branching ratio of $(3.2 \pm 0.5)\times 10^{-9}$ \cite{pdg}.  For kaon decays measurements of the (partial) branching ratios have been performed by the E865 experiment at the Brookhaven National Laboratory AGS for the decays $K^+\to e^+\,\nu_e\,e^+\,e^-$, $K^+\to \mu^+\,\nu_\mu\,e^+\,e^-$ and $K^+\to e^+\,\nu_e\,\mu^+\,\mu^-$\,\cite{Poblaguev_2002,Ma_2006}.
The branching ratios are found to be of $O(10^{-8})$. For decays with an $e^+e^-$ pair in the final state a lower limit of about $150\,\textrm{MeV}$ is imposed on the invariant mass of the lepton pair.
Without such a cut the branching ratio would be dominated by the point-like contribution in the low $e^+e^-$ invariant mass region which is of $O(10^{-5})$, so that the relevant SD contribution would not be detectable.
For D mesons there are no data, while for B mesons there is an upper bound on $\mathrm{BR}(B^+ \to \mu^+\, \nu_\mu\,\mu^+\, \mu^-$) of $1.6 \times 10^{-8}$ \cite{Aaij_2019}. 
 
In this paper we present the general strategy for the computation of the SD form-factors and then implement the procedure in an exploratory lattice simulation for kaon decays, i.e. for $P=K$. The computation is performed using a single gauge ensemble of $N_f=2+1+1$ flavours of twisted mass fermions generated by the European Twisted Mass Collaboration (ETMC) on a $32^3\times 64$ lattice with lattice spacing $a=0.0885$\,fm and with unphysical light-quark masses such that the pion and kaon masses are $m_\pi\simeq 320$\,MeV and $m_K\simeq 530$\,MeV. Further details of the ensemble are given at the beginning of Sec.\,\ref{ffnum}. Our method enables us to determine each of the four SD form factors contributing to the amplitude with good precision, and to study their dependence on the kinematic variables. Using these form factors one can reconstruct separately all the contributions to the branching ratios; the point-like contribution, the SD one and that coming from the interference between the two.
There has been one previous lattice study of these decays, in which a method was presented and implemented to compute the branching ratio for the decays $K\to\ell\nu_\ell\,\ell^{\prime +}\ell^{\prime\,-}$ without separating the point-like contribution and determining the SD form factors themselves\,\cite{xu}.  The exploratory computations in Ref.\,\cite{xu} were performed on a single gauge ensemble on a $24^3\times48$ lattice with $a\simeq0.093$\,fm and with quark masses corresponding to $m_\pi\simeq 352$\,MeV and $m_K\simeq 506$\,MeV.  

In our computation and also that in Ref.\,\cite{xu}, the kaon mass is smaller that twice the pion mass, $m_K<2m_\pi$, so that there are no contributions of the form $K\to\pi\pi\,\ell\nu_\ell\to\ell\nu_\ell\gamma$, with an on-shell $\pi\pi\,\ell\nu_\ell$ intermediate state. With physical quark masses, contributions with such an intermediate state are present in the region of phase space in which $k^2>4m_\pi^2$, where $k$ is the four-momentum of the virtual photon. This leads to finite-volume effects which decrease only as inverse powers of the volume and not exponentially\,\cite{Lellouch:2000pv,Kim:2005gf,Briceno:2019opb}. This effect is particularly important for the decays of heavy mesons, where there are many more possible on-shell intermediate states.  This issue, together with a complete study of all the systematic effects (due to discretization, finite volume and unphysical quark masses) will be object of our future studies.

For kaon decays, in addition to the lattice results from the computations reported here and in Ref.\,\cite{xu}, theoretical information about the form factors comes from 
Chiral Perturbation Theory (ChPT), which has been used at next-to-leading order (NLO) to estimate their values and their contribution to the branching ratios\,\cite{Bijnens:1994me}. It is worth noting that at NLO order in ChPT the form factors are constants, i.e. independent of the kinematical variables. 
In spite of the unphysical quark masses used in our simulation it has been interesting and instructive to compare our results with those from experiment (where available) and from NLO ChPT, as well as with those from Ref.\,\cite{xu}.  
Perhaps surprisingly, as can be seen from Tabs.\,\ref{tab_Br1}\,-\,\ref{tab_Br4} below,
the results are generally in reasonable semi-quantitative agreement but with some differences. In particular we speculate that the form factor $H_1$, defined in Eq.\,(\ref{Hmunu_sd}) may have to increase by $O(20\%)$ in order to get precise agreement with the experimental data (although there are also discrepancies in the experimental determination of $H_1$ from different decay channels). It will be important therefore, after this successful exploratory computation, to focus our future work on controlling and reducing the systematic uncertainties in order to obtain robust results at physical quark masses and in the continuum and infinite-volume limits. It will then be interesting to see whether the form factor $H_1$ will indeed change or whether there will be a different explanation for the differences between the experimentally observed rates and our current results. 

For heavy mesons ChPT does not apply, and the one theoretical predictions is presented in \cite{Danilina_2020} for $B$ decays, where a Vector Meson Dominance model has been used. The prediction for the $B^+\to\mu^+\,\nu_\mu\,\mu^+\,\mu^-$ branching ratio of \cite{Danilina_2020}, however, is almost four times larger than the experimental upper limit obtained in Ref.\,\cite{Aaij_2019}.
It is therefore clear that a non-perturbative, model independent lattice evaluation of the SD form factors is required.

The plan for the remainder of this paper is as follows. In Sec.\,\ref{hadten} we define the hadronic tensor and the form factors into which it is decomposed. These are the main target of our lattice calculation. This is followed in Sec.\,\ref{lattcorr} by an explanation of how the hadronic tensor can be determined from lattice computations and in Sec.\,\ref{sdff} by the presentation of our strategy for extracting the four SD form factors from the hadronic tensor. In Sec.\,\ref{ffnum} we present the details of the numerical computation of the SD form factors and in Sec.\,\ref{brnum} we use these form factors to compute (partially) integrated branching ratios for the four channels of kaon decays, $K^+\to e^+\nu_e\,\mu^+\mu^-$, $K^+\to \mu^+\nu_\mu\, e^+e^-$, $K^+\to e^+\nu_e\,e^+e^-$ and $K^+\to \mu^+\nu_\mu\,\mu^+\mu^-$. We also compare our results to the experimental measurements (where these exist) and to the predictions of ChPT. We present a summary and our conclusions in Sec.\,\ref{sec:concs}. There are two appendices. In Appendix \ref{kernels} we collect the formulae used to obtain the branching ratios from the form factors for the two channels in which $\ell\neq\ell^\prime$. The corresponding formulae for the other two channels, i.e. when $\ell=\ell^\prime$ are too lengthy to present here, but are available from the authors upon request. In Appendix\,\ref{kto0} 
we discuss the non-trivial limit of the relevant lattice correlation function as the four momentum of the photon, $k$, goes to zero, $k\to 0$. This is a key element in the subtraction of the point-like term from the hadronic matrix element, which itself is a necessary step to extract each of the SD form factors.

\section{The Hadronic Tensor in Minkowskian and Euclidean Space-Time}\label{hadten}
\begin{figure}
	\subfloat{%
		\includegraphics[scale=0.20]{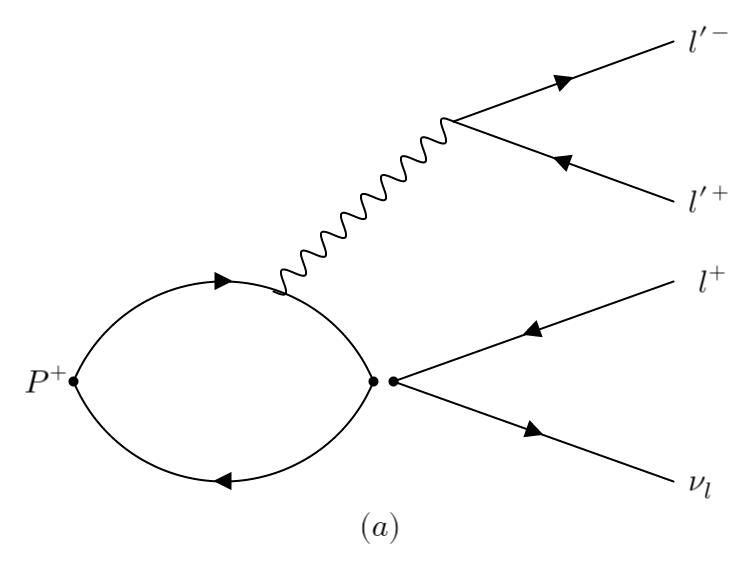}}
\hspace{0.5cm}
	\subfloat{%
		\includegraphics[scale=0.20]{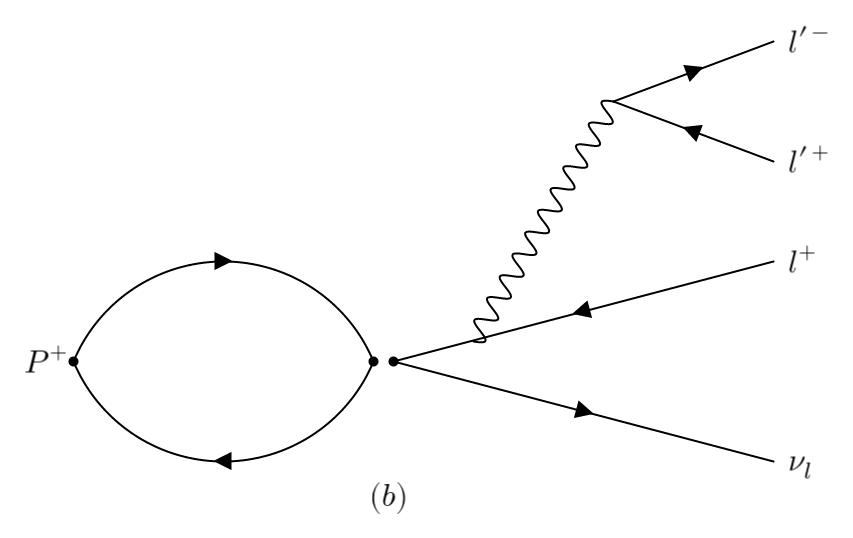}} 
	\caption{\it Diagrams contributing to the process $P^+\to l^+\,\nu_l\,l'^+\,l'^-$. We work in the \textit{electroquenched} approximation in which the sea quarks are electrically neutral so contributions from disconnected diagrams are neglected (see Fig.\,\ref{fig:disctheta} and the corresponding discussion).}
	\label{feynman}
\end{figure}

At lowest order in the electroweak interaction, $P^+\to l^+\,\nu_l\,l'^+\,l'^-$ decays are obtained from the diagrams depicted in Fig.\,\ref{feynman}. If $l=l'$, we also need to consider the diagrams obtained by interchanging the two identical charged leptons.
The diagram \ref{feynman}(b) can readily be computed in perturbation theory, with the meson decay constant as the only required non-perturbative input. In diagram \ref{feynman}(a) the non-perturbative hadronic contribution to the matrix element factorizes, and is encoded in the following tensor:
\bea\label{tensor}
H_W^{\mu\nu}(k,p)=\int d^4x\, e^{ik\cdot x}\mel{0}{T[J_{\mathrm{em}}^\mu(x) J_W^\nu(0)]}{P(p)}\label{hadronic}\,,
\eea
where $k=(E_\gamma, \boldsymbol{k})$ is the four-momentum of the virtual photon and $p=(E, \boldsymbol{p})$ is that of the incoming pseudoscalar meson $P$. The meson and photon energies satisfy $E=\sqrt{m_P^2+\boldsymbol{p}^2}$ and $E_\gamma=\sqrt{k^2+\boldsymbol{k}^2}$.
The two operators
\bea\label{eq:currents}
J^\mu_{\mathrm{em}}(x)=\sum_f q_f \bar{\psi}_f(x)\gamma^\mu\psi_f(x)\,\quad J_W^\nu(x)=J_V^\nu(x)-J_A^\nu(x)=\bar{\psi}_D(x)\left(\gamma^\nu-\gamma^\nu\gamma_5\right)\psi_U(x)\,,
\eea
are respectively the electromagnetic hadronic current and the hadronic weak current expressed in
terms of the quark fields $\psi_f$ having electric charge $q_f$ in units of the charge of the positron; $\psi_U$ and $\psi_D$ indicate the fields of an up-type or a down-type quark. In Eq.\,(\ref{eq:currents}) we have written the weak current, $J_W^\nu$, corresponding to a positively charged meson $P^+$; for a negatively charged meson we make the replacement $D\leftrightarrow U$.

The hadronic tensor can be decomposed into form factors which are scalar functions encoding the non-perturbative strong dynamics. Following Ref.\,\cite{Carrasco_2015}, we write:
\bea
H_W^{\mu\nu}&=&H^{\mu\nu}_{\mathrm{pt}}+H^{\mu\nu}_{\mathrm{SD}}\,,\\
H^{\mu\nu}_{\mathrm{pt}}&=&f_P\left[g^{\mu\nu}-\frac{(2p-k)^\mu(p-k)^\nu}{(p-k)^2-m_P^2}\right]\,,\\
H^{\mu\nu}_{\mathrm{SD}}&=&\frac{H_1}{m_P}\left(k^2g^{\mu\nu}-k^\mu k^\nu\right)+\frac{H_2}{m_P}\frac{\left[(k\cdot p-k^2)k^\mu-k^2\left(p-k\right)^\mu\right]}{(p-k)^2-m_P^2}\left(p-k\right)^\nu+\frac{F_A}{m_P}\left[(k\cdot p-k^2)g^{\mu\nu}-(p-k)^\mu k^\nu\right]\nonumber\\
& &\hspace{0.5in}-i\frac{F_V}{m_P}\epsilon^{\mu\nu\alpha\beta}k_\alpha p_\beta\,.\label{Hmunu_sd}
\eea
With this decomposition we have separated the point-like contribution to the hadronic tensor from the structure-dependent one. The former depends only on the meson decay constant and is obtained by assuming a point-like meson. The SD contribution describes the interaction between the virtual photon and the hadronic structure of the pseudoscalar meson. The SD form factors, $H_1$, $H_2$, $F_A$ and $F_V$, are scalar functions of $k^2$ and $(p-k)^2$. 
Note that, compared to our earlier work, see for example
Eq.\,(B4) of Ref.\cite{Carrasco_2015}  or Eq.\,(3) of Ref.\,\cite{Carrasco2014}, we have modified the definitions of $H_{1,2}$ by a factor of $m_P$ and introduced the denominator $(p-k)^2-m_P^2$ in the factor multiplying $H_2$. 
In these earlier papers we were studying radiative corrections to leptonic decays with a real photon in the final state for which the form factors $H_{1,2}$ do not contribute.
With the definitions in Eq.\,(\ref{Hmunu_sd}) 
all four form factors are now dimensionless and finite in the infrared limit. 
The main goal of this lattice study is to compute the SD form factors in order to reconstruct the full matrix element and subsequently the branching ratio for the decay. We do this in a way which separates the point-like contribution from that which depends on the hadronic structure. 

In order to show how the hadronic tensor can be extracted from Euclidean correlation functions it is useful to express $H_W^{\mu\nu}(k,p)$ in terms of the contributions coming from the two different time–orderings. By inserting a complete set of intermediate states we obtain the contributions from the two separate time-orderings ($t_x<0$ and $t_x>0$, where $x=(t_x,\bs{x})$) as 
\bea
& &H_W^{\mu\nu}(k,p)=H^{\mu\nu}_{W,1}(k,p)+H^{\mu\nu}_{W,2}(k,p)\,,
\eea
where
\bea
H^{\mu\nu}_{W,1}(k,p)&=&-i\sum_{n_f:{\boldsymbol{p}_{n_f}
		=\boldsymbol{p}-\boldsymbol{k}}}\frac{\mel{0}{J^\nu_W(0)}{n_f}\mel{n_f}{J^\mu_{\mathrm{em}}(0)}{P(p)}}{(E_\gamma + E_{n_f}-E)}\,,\label{eq:HmunuM1}\\
H^{\mu\nu}_{W,2}(k,p)&=&-i\sum_{n:\boldsymbol{p}_n=\boldsymbol{k}}\frac{\mel{0}{J^\mu_{\mathrm{em}}(0)}{n}\mel{n}{J^\nu_W(0)}{P(p)}}{(E_n-E_\gamma-i\epsilon)}\,,\label{eq:HmunuM2}
\eea
and the sums over the intermediate states implicitly include the phase-space integration. The states 
$|n_f\rangle$ have the same flavour quantum numbers as the initial meson $P$, while the states $|n\rangle$ have zero additive flavour quantum numbers. For example, if we consider the decay of a $K^+$ and $J_W^\nu=\bar{s}\gamma^\nu(1-\gamma^5)u$, then the $|n_f\rangle$ have strangeness $S=-1$ and the $|n\rangle$ have $S=0$.

Lattice correlation functions can only be computed in Euclidean space-time, thus we have to translate the Minkowski Green function to the corresponding Euclidean one. By making the naive Wick rotation $ t \to -it$ we obtain the Euclidean expression
\bea\label{euccor}
H^{\mu\nu}_E(k,p)=-i\int d^4x\, e^{tE_\gamma-i\boldsymbol{k}\cdot\boldsymbol{x}}\mel{0}{T[J_{\mathrm{em}}^\mu(x) J_W^\nu(0)]}{P(p)}\,.
\eea
As before, we insert a complete set of intermediate states and obtain contributions from each of the two time-orderings:
\bea
H^{\mu\nu}_E(k,p)&=&H^{\mu\nu}_{E,1}(k,p)+H^{\mu\nu}_{E,2}(k,p)\,,\\
H^{\mu\nu}_{E,1}(k,p)&=&-i\sum_{n_f:\boldsymbol{p}_{n_f}=\boldsymbol{p}-\boldsymbol{k}}\mel{0}{J^\nu_W(0)}{n_f}\mel{n_f}{J^\mu_{\mathrm{em}}(0)}{P(p)} \int_{-\infty}^0 dt_x\,e^{t_x(E_\gamma + E_{n_f}-E)}\label{he1}\,,\\
H^{\mu\nu}_{E,2}(k,p)&=&-i\sum_{n:\boldsymbol{p}_n=\boldsymbol{k}}\mel{0}{J^\mu_{\mathrm{em}}(0)}{n}\mel{n}{J^\nu_W(0)}{P(p)}\int^{+\infty}_0 dt_x\,e^{-t_x(E_n-E_\gamma)}\label{he2}\,.
\eea
If the conditions 
\bea\label{conditions}
E_\gamma+E_{n_f}-E>0\label{condition1}\,,\\
E_n-E_\gamma>0\label{condition2}\,,
\eea
are satisfied, the time integrals converge and we have
\bea
H^{\mu\nu}_{E}=-i\sum_{n_f:\boldsymbol{p}_{n_f}=\boldsymbol{p}-\boldsymbol{k}}\frac{\mel{0}{J^\nu_W(0)}{n_f}\mel{n_f}{J^\mu_{\mathrm{em}}(0)}{P(p)}}{E_\gamma + E_{n_f}-E}-i\sum_{n:\boldsymbol{p}_n=\boldsymbol{k}}\frac{\mel{0}{J^\mu_{\mathrm{em}}(0)}{n}\mel{n}{J^\nu_W(0)}{P(p)}}{E_n-E_\gamma}\,.
\eea
If the inequalities (\ref{condition1}) and (\ref{condition2}) are satisfied then the Wick rotation leaves the hadronic
tensor (\ref{tensor}) unchanged, and thus the lattice calculation with Euclidean time can be done without particular difficulties. In such situations, the $i\epsilon$ in the second line of Eq.\,(\ref{eq:HmunuM2}) is also unnecessary. 
On the other hand for external momenta such that the inequalities (\ref{condition1})-(\ref{condition2}) are not satisfied then the time
integrals in Euclidean space-time diverge at large $t_x$. 
The above is a consequence of the analytic structure of the $T$-product in Eq.\,(\ref{tensor}):
the presence of singularities (poles or cuts) in Minkowski space can prevent the possibility of making a naive Wick rotation. The presence of such singularities implies the existence of intermediate states with energies which are smaller than the external ones resulting in integrals over $t_x$ which grow exponentially with the upper limit of integration. The conditions (\ref{condition1}) and (\ref{condition2}) correspond to the requirement
that the internal states contributing to the correlation function all have energies larger than that of the external states (see \cite{maiani} for more details).

In the above discussion we have not specified what the pseudoscalar meson $P$ is but in this paper we apply the formalism to the decays of a kaon. For $t_x<0$, i.e. when the electromagnetic current is inserted before the weak operator,  the internal lightest state is given by a kaon with spatial momentum $\boldsymbol{p}-\boldsymbol{k}$, and it can be readily shown that the condition (\ref{condition1}) is satisfied for every choice of the external momenta $\boldsymbol{p}$ and $k$.  
On the other hand, for $t_x>0$, i.e. when the weak current is inserted before the electromagnetic one, the lowest-energy internal state is given by two pions with the same spatial momentum $\boldsymbol{k}$ as the virtual photon. Thus the condition (\ref{condition2}) is satisfied only for $k^2<4m_\pi^2$ and for larger photon virtualities the correlator in Euclidean time is divergent.  On a finite spatial lattice the spectrum of states $|n\rangle$ is discrete and so there is only a finite number of states with $k^2>4m_\pi^2$ and in practice the number of such states is small and the terms with the exponentially growing exponentials can be explicitly subtracted, thus extending the validity of the method beyond the region $k^2<4m_\pi^2$. The remaining issue is then the correction for the non-exponential finite-volume effects (analogous to those corrected by the Lellouch-L\"uscher factor in $K\to\pi\pi$ decays\,\cite{Lellouch:2000pv}). We postpone a discussion of this issue to a future publication and for 
now we restrict our analysis, presented in Sec.\,\ref{ffnum}, to kaon decays with an unphysical pion mass such that $m_K<2m_\pi$. Thus, two-pion internal states are always heavier than the external states and so conditions (\ref{condition1}) and (\ref{condition2}) are both satisfied.

Now that we have discussed the the analytic continuation to Euclidean space-time, we proceed to the presentation of our strategy for extracting the SD form factors from suitable three-point lattice correlation functions. 

\section{The hadronic tensor from lattice correlation functions}\label{lattcorr}

The principal ingredient in evaluating the decay amplitude on a Euclidean lattice, with finite space-time volume $V=L^3\times T$, is the correlation function 
\bea\label{eq:MW}
M_W^{\mu\nu}(t_x,t;\boldsymbol{k},\boldsymbol{p})=T\langle J^\nu_W(t)\hat{J}^\mu_{\mathrm{em}}(t_x,\boldsymbol{k})\hat{P}(0,\boldsymbol{p})\rangle_{LT}\,,
\eea
where $\langle ... \rangle_{LT}$ denotes the average over the gauge field configurations at finite L and T. Note that in Eq.\,(\ref{eq:MW}) we have placed the interpolating operator $\hat{P}(0,\bs{p})$ at time 0 and the weak current $J_W(t)$ at time $t$. The three operators in Eq.(\ref{eq:MW}) are as follows:\\[0.1cm]
\mbox{}\hspace{0.2in}$\bullet$ $\hat{P}(0,\bs{p})$ is the spatial Fourier transform of the interpolating operator for the decaying pseudoscalar meson at time $t=0$:
\begin{equation}
\hat{P}(0,\boldsymbol{p})=\sum_{\boldsymbol{z}}e^{i\boldsymbol{p}\cdot \boldsymbol{z}} P(0,\boldsymbol{z})\,,
\end{equation}
where $P(0,z)=i\overline{\psi}_U(0,\bs{z})\gamma_5\psi_D(0,\bs{z})$ for a positively charged meson or $P(0,z)=i\overline{\psi}_D(0,\bs{z})\gamma_5\psi_U(0,\bs{z})$ for a negatively charged one 
and $\psi_{U,D}$ indicate the fields of up-type and down-type quarks respectively. In this paper we study kaon decays so $U=u$ and $D=s$.\\[0.1cm]
\mbox{}\hspace{0.2in}$\bullet$ The renormalised  hadronic weak current, $J_W^\nu(t)= J_V^\nu(t)-J_A^\nu(t)$ is placed at a generic time $t$ and at the origin in space. The vector and axial currents, $J_V^\nu(t)$ and $J_A^\nu(t)$ respectively, 
satisfy the continuum Ward identities (up to discretisation effects). In the Twisted-Mass discretisation of the fermionic action\,\cite{Frezzotti:2000nk}, the vector and axial vector  currents we use are  given by  
    \bea
    J_V^\nu(t)=Z_A\, \bar \psi_{D}(t)\gamma^\nu\psi_{U}(t)\;,
    \quad
    J_A^\nu(t)=Z_V\, \bar \psi_{D}(t)\gamma^\nu\gamma_5\psi_{U}(x) \label{eq:jWVA},
    \eea
 for a positively charged meson or their Hermitian conjugates for a negatively charged one, where  $Z_{A,V}$  are  the renormalisation factors ensuring that the Ward identities are satisfied\,\footnote{Note that the renormalisation factors to be used in Twisted-Mass at maximal twist are chirally-rotated with respect to the ones of standard Wilson fermions\,\cite{Frezzotti_2004}. This is a consequence of the fact that the up-type and down-type quark fields in the action are discretised with opposite values of the Wilson parameter.}.\\[0.1cm]
\mbox{}\hspace{0.2in}$\bullet$ The electromagnetic current, $J_{\mathrm{em}}^\mu(t_x,\boldsymbol x)$, is defined by 
    \bea J_{\mathrm{em}}^\mu(t_x,\boldsymbol x) =  \sum_f \, q_f\, J_f^\mu(t_x,\boldsymbol x) \, , 
    \eea
    where $f$ is the flavour index and the charge $q_f$ is equal to $2/3$ for up-type quarks and to $-1/3$ for down-type quarks.  
    A possible choice for the lattice electromagnetic current is the local operator $ J_f^\mu(t_x,\boldsymbol x) = Z^{\rm loc}_V \bar q_f(t_x,\boldsymbol x)  \gamma^\mu q_f(t_x,\boldsymbol x)$, where $Z^{\rm loc}_V$ is the finite renormalisation constant of the vector current ($Z^{\rm loc}_V=Z_A$ with Twisted-Mass at maximal twist). 
We choose instead to use the exactly conserved lattice vector current which with  Twisted-Mass Fermions at maximal twist is given by\,\footnote{With twisted boundary conditions we use the corresponding conserved current given by Eq.\,(B10) of Ref.\,\cite{Desiderio2020}.}
    \begin{flalign}
    J_f^\mu(x)=-\left\{
    \bar \psi_f(x)\frac{ i\,r_f\gamma_5-\gamma^\mu}{2}\, U_\mu(x)\psi_f(x+\hat \mu)
    -
    \bar \psi_f(x+\hat \mu)\frac{ i\,r_f\gamma_5+\gamma^\mu}{2}U_\mu(x)^\dagger \psi_f(x)
    \right\}\,  .\label{eq:jfdef}
    \end{flalign}
    In Eq.\,(\ref{eq:jfdef}), $U_\mu(x)$ are the QCD link variables and $r_f=\pm 1$ is the Wilson parameter of the flavour $f$ \cite{de_Divitiis_2013}. 
    The spatial momentum $\boldsymbol k$ of the current is assigned by defining
    \begin{flalign}
    \hat{J}^\mu_{\mathrm{em}}(t_x, \boldsymbol k) =  \sum_{\boldsymbol x} e^{-i\boldsymbol k\cdot (\boldsymbol x+\boldsymbol{\hat \imath}/2)}\, J_{\mathrm{em}}^\mu(t_x,\boldsymbol x) \,  .
    \end{flalign}
    
\begin{figure}[!t]
\begin{center}
\includegraphics[width=0.425\textwidth]{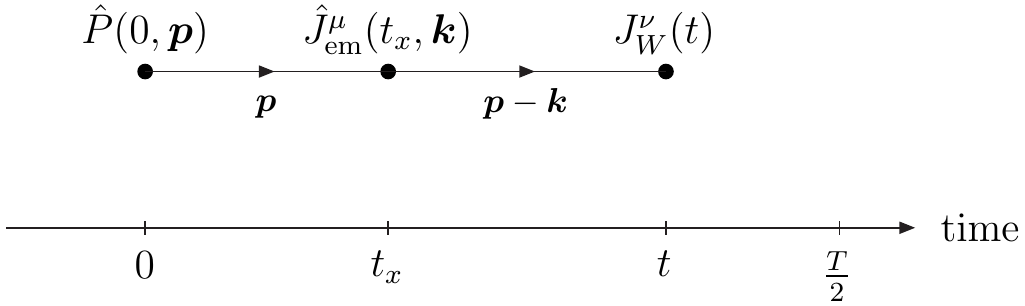}\qquad\qquad
\includegraphics[width=0.425\textwidth]{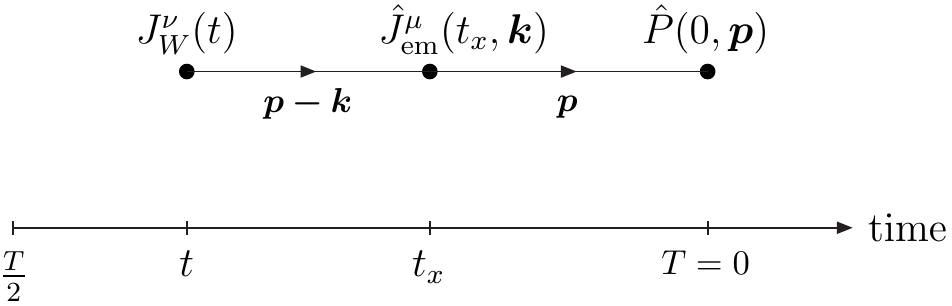}
\end{center}
\caption{\it{Schematic diagrams representing the correlation function $C^{\mu\nu}_W(t,  E_\gamma,\boldsymbol k,\boldsymbol p)$ used to extract the form factors, see Eqs.\,(\ref{eq:MW}) and (\ref{eq:Cmunudef}).  The interpolating operator for the meson $\hat{P}$ and the weak current $J_W$  are placed at fixed times $0$ and $t$ and the electromagnetic current $\hat{J}_{\mathrm{em}}$ is inserted at $t_x$ which is integrated over $0 \le t_x\le T$, where $T$ is the temporal extent of the lattice. The left and right panels correspond to the leading contributions to the correlation functions for $t<T/2$ and $t>T/2$ respectively, with mesons propagating with momenta $\bs{p}$ or $\bs{p-k}$. } \hspace*{\fill}
\label{fig:correlator}}
\end{figure}

In order to obtain the decay amplitude, we need to integrate $M_W^{\mu\nu}(t_x,t;\boldsymbol{k},\boldsymbol{p})$ over $t_x$, as seen for example in Eq.\,(\ref{hadronic}). To this end we construct the function:
\bea\label{eq:correlator22}
C_W^{\mu\nu}(t,E_\gamma, \boldsymbol{k}, \boldsymbol{p})&=&-i\theta\left(T/2-t\right)\sum^T_{t_x=0}\left(\theta\left(T/2-t_x\right)e^{E_\gamma\,t_x}+\theta\left(t_x-T/2\right)e^{-E_\gamma(T-t_x)} \right)M_W^{\mu\nu}(t_x,t;\boldsymbol{k},\boldsymbol{p})\nonumber\\
& &-i\theta\left(t-T/2\right)\sum^T_{t_x=0}\left(\theta\left(T/2-t_x\right)e^{-E_\gamma\,t_x}+\theta\left(t_x-T/2\right)e^{-E_\gamma(t_x-T)} \right)M_W^{\mu\nu}(t_x,t;\boldsymbol{k},\boldsymbol{p})\,.\label{eq:Cmunudef}
\eea
On a lattice with a large but finite temporal extent $T$, the required matrix element can be obtained from the first term on the top line of Eq.\,(\ref{eq:Cmunudef}). This is illustrated in the left-hand diagram of Fig.\,\ref{fig:correlator} and it should be remembered that $t_x$ can also be larger than $t$. The second term on the second line of Eq.\,(\ref{eq:Cmunudef}) represents the time-reversed process (we discuss the properties of the matrix elements under time reversal at the end of this section) and is illustrated in the right-hand diagram of Fig.\,\ref{fig:correlator} and again it should be remembered that $t_x$ can also be smaller than $t$. 
The second term on the top line of Eq.\,(\ref{eq:Cmunudef})   represents, on a periodic lattice of finite temporal extent, the  ordering where the electromagnetic current acts at an earlier time than the meson source  that,  in the reduction formula to create an initial meson state, should be asymptotically far in the past.  Indeed,
the contribution of this term disappears in the limit $T \to \infty$. 
On the lattices used here however,  we have found that its inclusion corrects sizeable finite $T$ effects and improves the quality of the numerical fits of $C_W^{\mu\nu}(t,E_\gamma,\bs{k},\bs{p})$. 
Similarly, the first term on the second line of Eq.\,(\ref{eq:Cmunudef}) represents, for the time-reversed process, the electromagnetic currents acting at a time larger  than the meson source. Its contribution also disappears in the infinite $T$ limit, but its inclusion improves the quality of the fit of  
$C_W^{\mu\nu}(t,E_\gamma,\bs{k},\bs{p})$.


\begin{figure}[t]
    \begin{center}
    \includegraphics[width=0.8\textwidth]{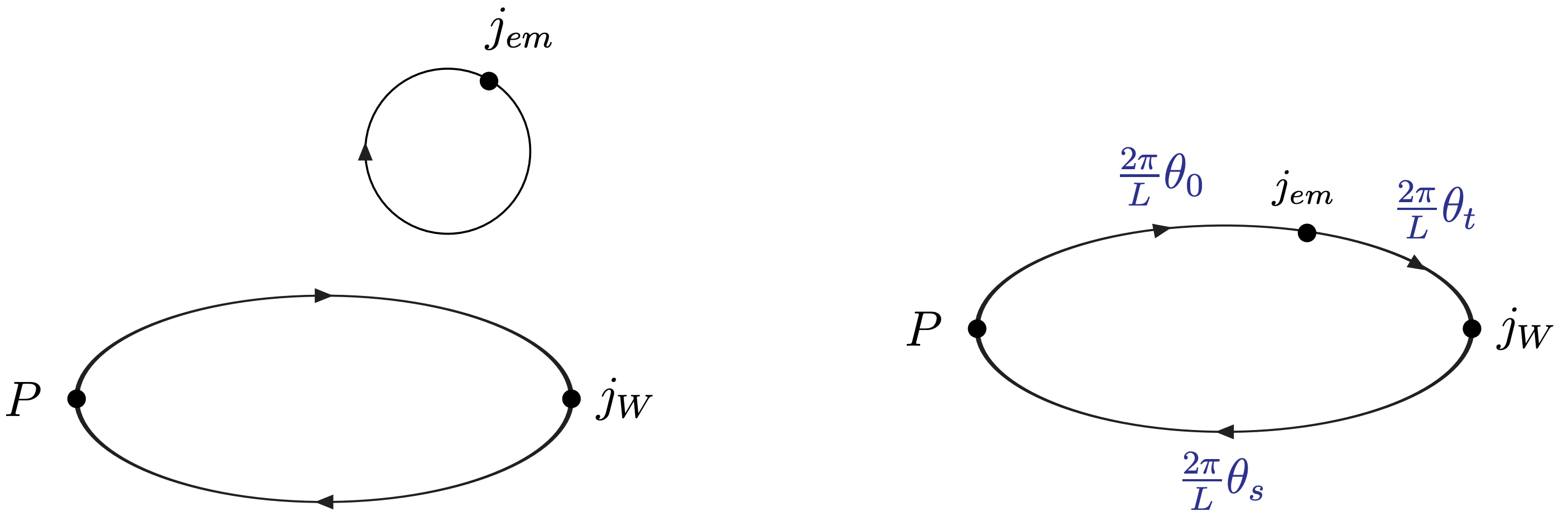}
    \end{center}
    \caption{\it The diagram on the left represents the contributions to the correlation functions arising from the emission of the photon by the sea quarks. In our numerical simulations we work in the electroquenched approximation and neglect such diagrams. The diagram on the right explains our choice of the spatial boundary conditions, which allow us to set arbitrary values for the meson and photon spatial momenta.
    The spatial momenta of the valence quarks, modulo $2\pi/L$, in terms of the twisting angles are as indicated. Each diagram implicitly includes all orders in QCD.
    \hspace*{\fill}
    \label{fig:disctheta}}
    \end{figure}

Fig.\,\ref{fig:disctheta} contains two diagrams presented to illustrate two important points concerning our numerical calculation of the correlation functions  and of the form factors. The diagram in the left panel shows a {\it quark-disconnected} contribution to the correlation function originating from the possibility that the virtual photon is emitted from sea quarks. In this paper we use the so-called electroquenched approximation in which the sea-quarks are electrically neutral. In practice this means that we have neglected the contributions represented by the diagram in the left panel of Fig.\,\ref{fig:disctheta}. We note that the contribution of these diagrams vanishes in the limit of exact $SU(3)$ flavour symmetry.

The quark-connected diagram in the right panel of Fig.\,\ref{fig:disctheta} is shown in order to explain the strategy we have used to set the values of the spatial momenta. We have exploited the fact that by working within the electroquenched approximation it is possible to choose arbitrary values of the spatial momenta by using different spatial boundary conditions for the quark fields\,\cite{de_Divitiis_2004}. More precisely, we set the spatial boundary conditions for the ``spectator" quark such that 
    \bea
    \psi(x+\boldsymbol{n} L)=\exp(2\pi i \boldsymbol{n} \cdot \boldsymbol{\theta_s}) \psi(x)\, , \label{eq:twist} 
    \eea
    where $\bs{n}$ is a three-vector of integers and $\bs{\theta_s}$ is a three-vector of angles. For the temporal direction we employ anti-periodic boundary conditions. For each quark flavour $f$, we impose different boundary conditions on $q_f$ and $\bar{q}_f$, the two component fields of $J_f^\mu$. This is possible at the price of accepting violations of unitarity that are exponentially suppressed with the volume \cite{Boyle:2007wg,Sachrajda:2004mi}. By setting the boundary conditions as illustrated in the figure we have thus been able to choose arbitrary (non-quantised) values for the meson and photon spatial momenta
    \begin{flalign}
    \boldsymbol p = \frac{2\pi}{L}\left( \boldsymbol \theta_0-\boldsymbol \theta_s\right)\;,
    \qquad
    \boldsymbol k = \frac{2\pi}{L}\left( \boldsymbol \theta_0-\boldsymbol \theta_t\right)\;,
    \label{eq:momenta}
    \end{flalign}
    by tuning the real three-vectors $\boldsymbol \theta_{0,t,s}$. We find that the most precise results are obtained with small values of $|\boldsymbol{p}|$ and in particular with $\boldsymbol p=\boldsymbol 0$.
        
In order to show that it is possible to extract the hadronic matrix element in Eq.\,(\ref{tensor}) from the function in Eq.\,(\ref{eq:correlator22}) we perform a spectral decomposition of $C_W^{\mu\nu}(t)$.
On the assumption that the inequalities in Eqs.\,(\ref{condition1}) and (\ref{condition2}) are satisfied, we derive the relation
\begin{widetext}

\bea\label{C-H}
& &C_W^{\mu\nu}(t,E_\gamma,\boldsymbol{k},\boldsymbol{p})=\nonumber\\
& &\Theta(T/2-t)\,\frac{e^{-t(E-E_\gamma)}\, \langle P\vert P\vert  0\rangle}{2E}H_W^{\mu\nu}(k,\boldsymbol p)+\Theta(t-T/2)\,  \frac{e^{-(T-t)\, (E-E_\gamma)}\langle 0\vert  P\vert P\rangle }{2E}\left[H_W^{\mu\nu}(k,\boldsymbol p)\right]^\dagger+\ .\ .\ .\\ \nonumber
\eea
\end{widetext}
where $H_W^{\mu\nu}(k,\boldsymbol p)$ is the \emph{physical} matrix element defined in Eq. (\ref{tensor}) and the dots represent the subleading exponentials, suppressed as $e^{-\Delta E t}$ or $e^{-\Delta E(T-t)}$, where $\Delta E$ can be either $E_n+E_\gamma-E$ or $E_n-E_\gamma$ .
We also see that when $t>T/2$ the correlator represents the time-reversal of the original process.
It is useful to note that, in order to separate the axial and vector form factors, it is enough to compute separately the correlation functions corresponding to the vector, $C^{\mu\nu}_V(t, E_\gamma,\boldsymbol k,\boldsymbol p)$, and the axial, $C^{\mu\nu}_A(t, E_\gamma,\boldsymbol k,\boldsymbol p)$, components of the weak current. Moreover, from the properties
\bea
\left[H^{\mu\nu}_A(k,p)\right]^\dagger=H_A^{\mu\nu}(k,p),\quad \left[H_V^{\mu\nu}(k,p)\right]^\dagger=-H_V^{\mu\nu}(k,p)
\eea
we deduce the following properties of the corresponding correlation functions under time reversal:
\bea
C^{\mu\nu}_{A}\left(T-t, T/2, E_\gamma, \boldsymbol{k}, \boldsymbol{p}\right)=C^{\mu\nu}_{A}\left(t, T/2, E_\gamma, \boldsymbol{k}, \boldsymbol{p}\right),\quad C^{\mu\nu}_{V}\left(T-t, T/2, E_\gamma, \boldsymbol{k}, \boldsymbol{p}\right)=-C^{\mu\nu}_{V}\left(t, T/2, E_\gamma, \boldsymbol{k}, \boldsymbol{p}\right)\,.
\eea 
We use these time reversal properties of the lattice correlators, to either symmetrize or anti-symmetrize the correlators between the two halves $[0, T/2]$ and $[T/2,T]$ of the lattice and then we will work just within the first half of the lattice time-extent, defining
\bea\label{eq:lat_hadr_tens}
H^{\mu\nu}_L(t,k,\boldsymbol p)=\frac{2E}{e^{-t(E-E_\gamma)}\mel{P}{P}{0}} C_W^{\mu\nu}(t,E_\gamma,\boldsymbol{k},\boldsymbol{p})= H^{\mu\nu}_W(k,\boldsymbol p)+...
\eea
where the subscript $L$ stands for ``lattice" and the ellipsis represents the sub-leading exponentials. In addition to decreasing the statistical error, averaging the correlation function between the two halves of the lattice in this way ensures $O(a)$improvement, i.e. the reduction of the discretisation errors to ones of $O(a^2)$.

In this section we have shown how to obtain the hadronic tensor from lattice correlation functions and we now proceed to discuss the extraction of all the structure-dependent hadronic form factors.
\section{Extraction of the Structure-Dependent Form Factors}\label{sdff}
As already stated above, the axial and vector part of the hadronic tensor can be evaluated separately in order to determine the corresponding form factors. In our numerical study we choose the meson to be at rest, $\bs{p}=\bs{0}$ (the correlation functions are less noisy in this case) and the spatial momentum of the photon to be in the $z$-direction, $\bs{k}=(0,0,k_z)$. 
The form-factors depend on two independent variables which can be chosen to be the invariants $k^2$, where $k$ is the four-momentum of the photon, and $q^2\equiv(p-k)^2$. In section\,\ref{ffnum} we present our results in terms of the dimensionless variables $x_k$ and $x_q$ defined in Eq.\,(\ref{eq:xkqdef}) terms of $k^2$ and $q^2$. In this section however, in which we discuss the extraction of the form factors from correlation functions computed in the frame defined above, it is more transparent to present the discussion with $k^2$ and $k_z$ as the independent variables, together with the energy of the photon $E_\gamma$ given by $E_\gamma^2=k^2+k_z^2$.

In the rest frame of the meson and with $\bs{k}=(0,0,k_z)$, the only non-zero elements of the vector component of the hadronic tensor, $H_V^{\mu\nu}$, are
$H^{12}_V$ and $H^{21}_V$ which are related to the vector form factor $F_V$ by
\bea
H^{12}_V=-H_V^{21}=iF_Vk_z\,.
\eea

The axial component of the hadronic tensor, $H_A^{\mu\nu}$, is parametrised by the SD form factors $F_{A}$, $H_{1}$ and $H_{2}$, and by the meson decay constant $f_P$.
In the reference frame defined above, the non-zero elements of $H_A^{\mu\nu}$ are given by
%
\bea
\label{eq:hadronic_tensor}
H_A^{00}&=&-H_1\frac{k_z^2}{m_P}-H_2\frac{k_z^2\left(m_P-E_\gamma\right)}{2m_PE_\gamma-k^2}-F_A\frac{k_z^2}{m_P}+f_P\frac{2m_P^2-m_PE_\gamma+k_z^2}{2m_PE_\gamma-k^2}\,,\\ [2ex]
H_A^{03}&=&-H_1\frac{E_\gamma k_z}{m_P}+H_2\frac{k_z(E_\gamma^2-k^2)}{2m_PE_\gamma-k^2}-F_A\frac{\left(m_P-E_\gamma\right)k_z}{m_P}-f_P\frac{k_z\left(2m_P-E_\gamma\right)}{2m_PE_\gamma-k^2}\,,
\label{eq:H03}\\[2ex]
H^{30}_A&=&-H_1\frac{E_\gamma k_z}{m_P}-H_2\frac{k_zE_\gamma\left(m_P-E_\gamma\right)}{2m_PE_\gamma-k^2}+F_A\frac{k_zE_\gamma}{m_P}-f_P\frac{k_z\left(m_P-E_\gamma\right)}{2m_PE_\gamma-k^2}\,,\\[2ex]
\label{33}
H^{33}_A&=&-H_1\frac{E_\gamma^2}{m_P}+H_2\frac{E_\gamma k_z^2}{2m_PE_\gamma-k^2}-F_A\frac{E_\gamma\left(m_P-E_\gamma\right)}{m_P}	-f_P\frac{E_\gamma\left(2m_P-E_\gamma\right)}{2m_PE_\gamma-k^2}\,,\\[2ex]
\label{11}
H^{11}_A&=&H^{22}_A=-H_1\frac{k^2}{m_P}-F_A\frac{\left(m_PE_\gamma-k^2\right)}{m_P}-f_P\,.
\eea

Here and in the following we use continuum notation for the four-vectors but in lattice computations, in order to reduce the discretization uncertainties, the energy and momentum 
carried by the electromagnetic current should be understood by the following replacements: 
\bea
k_z\to \hat{k}_z=\frac{2}{a}\sin\left(\frac{ak_z}{2}\right)\,,\quad E_\gamma=\frac{2}{a}\sinh^{-1}\left[\frac{a}{2}\sqrt{
\hat{k}_z^2+\left(\frac2{a}\sinh\left(\frac{a\sqrt{k^2}}{2}\right)\right)^{\!\!2}}\,\,\right]
\eea
where $k_z$ and $\sqrt{k^2}$ are the continuum, physical values for the photon's spatial momentum (which here is directed along the $z$-axis) and for the photon's virtuality respectively.\\

To determine the SD axial form factors from knowledge of the non-zero components of $H_A^{\mu\nu}$, it is necessary to subtract the point-like terms proportional to $f_P$. From the previous equations, it follows that the point-like terms become dominant in the infrared limit, $k \to 0$, where the SD part of the hadronic tensor vanishes. This is expected, since soft photons cannot probe the internal structure of the meson. However, this poses the problem for the numerical evaluation of the SD form factors at small $k^2$, that $\mathcal{O}(a^{2})$ discretization effects in the subtraction of the point-like contribution result in enhanced artefacts in the determined  values of the SD form factors. Moreover, these artefacts diverge as $k \to 0$. This problem has already been encountered in our previous work on  $P\to\ell\bar\nu_\ell\gamma$ decays\,\cite{Desiderio2020}, where it was found that performing the subtraction using the value of $f_{P}$ extracted from two-point correlation functions results in unphysically large values of $F_{A}$ in the soft-photon limit. In the same paper, we proposed a solution to this problem.  We showed that by exploiting the electromagnetic Ward Identity in the lattice theory, the subtraction of the point-like contribution can be performed non-perturbatively to all orders in the lattice spacing, thus avoiding infrared-divergent  
lattice artefacts in the resulting SD form factors.
In particular, we demonstrated that, for the diagonal spatial components of the lattice correlation function, which are smooth in the limit $k\to 0$, this can be achieved by using the values of $f_P$ obtained from the same components evaluated at zero photon momentum\,\cite{Desiderio2020}.


A similar situation occurs also when the final-state photon is virtual, albeit in this case the lepton masses provide a energy-momentum cut-off for the photon. Proceeding in a similar way, we define the subtracted quantities for the diagonal components as follows:
\bea\label{eq:sub_diag_comp}
\tilde{H}_{A}^{33}(k_{z},k^{2}) &\equiv& H_{A}^{33}(k_{z},k^{2})-H_{A}^{33}(0,0)\frac{E_\gamma\left(2m_P-E_\gamma\right)}{2m_PE_{\gamma} -k^2 }=-H_1\frac{E_\gamma^2}{m_P}+H_2\frac{E_\gamma k_z^2}{2m_PE_\gamma-k^2}-F_A\frac{E_\gamma\left(m_P-E_\gamma\right)}{m_P} \nonumber \\
\tilde{H}_{A}^{11}(k_{z},k^{2}) &\equiv& H_{A}^{11}(k_{z},k^{2}) - H_{A}^{11}(0,0)=-H_1\frac{k^2}{m_P}-F_A\frac{\left(m_PE_\gamma-k^2\right)}{m_P}\,.\label{eq:fPsubtraction}
\eea
Unfortunately the same procedure cannot be used for the other components. The reason for this is that, 
in the limit $k\to 0$, the ``excited" state consisting of a meson $P$ with momentum $-\boldsymbol{k}$ and a photon with energy $E_\gamma$ becomes degenerate with the ``ground" state of the meson $P$ at rest. In the $k\to 0$ limit, the off-diagonal components, $C_{A}^{30}$ and $C_{A}^{03}$ go to zero; the contribution of the $P+\gamma$ state cancels that of the ground state (we refer to the Appendix.~\ref{kto0} for a detailed discussion on this point). These components at zero photon momentum cannot therefore be used to subtract the contribution proportional to $f_P$.
Instead we define a linear combination of the two off-diagonal components, which in the continuum cancels the point-like term proportional to $f_P$, that is:
\begin{equation}
H_{A}^{[3,0]}(k_{z},k^{2}) \equiv H^{30}_{A}(k_{z},k^{2}) - H^{03}_{A}(k_{z},k^{2})\,\left( \frac{m_P- E_{\gamma}}{2m_P-E_{\gamma}}   \right)=-H_1\frac{E_{\gamma} k_z }{2 m_P-E_{\gamma}}-H_2\frac{k_{z} (m_P-E_{\gamma})}{2 m_P-E_{\gamma}}+F_A\frac{ k_{z}  m_P}{2 m_P-E_{\gamma}}\,.\label{eq:H30m03}
\end{equation}
We have verified that the difference in Eq.(\ref{eq:H30m03}) does not give rise to enhanced unphysical infrared effects from the residual discretization errors.
In the present study of $K\to l\,\nu_l\,l'^+\,l'^-$ decays, as mentioned above, we have an infrared cut-off on the photon virtuality $k^2$, given either by the non-negligible muon mass or by the experimental cut on the two-electron invariant mass $m_{ee}=\sqrt{k^2}>145\ \textrm{MeV}$  \cite{Poblaguev_2002}. 
Above these cut-offs we observe a smooth behaviour of the form factors as a function of the photon's momentum without any anomalous increase in the infrared region.

An alternative possibility, one which we have not explored in this study, would have been to compute the correlation function with the meson in motion ($\bs{p}\neq\bs{0}$) and to use different components of the correlation functions to extract the form factors.

Once we computed three independent linear combinations of the three axial form factors using lattice QCD, the form factors themselves are obtained by inverting the matrix of coefficients. Specifically, our estimators of the axial form factors, $\bar{H}_{1}(t, k^2,k_z), \bar{H}_{2}(t,k^2,k_z)$ and  $\bar{F}_{A}(t,k^2,k_z)$, are obtained from the axial component of the lattice tensor $H^{\mu\nu}_{L,A}(t,k) \equiv H^{\mu\nu}_{L,A}(t,k,\bs{0})$ of Eq.\,(\ref{eq:lat_hadr_tens}) as follows:
\begin{align}
\label{eq:est_def}
\begin{pmatrix} \bar{H}_{1}(t,k^2,k_z) \\[10pt] \bar{H}_{2}(t,k^2,k_z) \\[10pt] \bar{F}_{A}(t,k^2,k_z)  \end{pmatrix}\equiv ~Z(t)~
\left(
\begin{array}{ccc}
 -\frac{E_\gamma k_z}{2 m_P-E_\gamma} & -\frac{k_z (m_P-E_\gamma)}{2 m_P-E_\gamma} & \frac{k_z m_P}{2 m_P-E_\gamma} \\
 -\frac{E_\gamma^2+k^2}{m_P} & \frac{E_\gamma k_z^2}{2 E_\gamma m_P-k^2} & \frac{E_\gamma^2-2 E_\gamma m_P+k^2}{m_P} \\
 \frac{k^2-E_\gamma^2}{m_P} & \frac{E_\gamma k_z^2}{2 E_\gamma m_P-k^2} & \frac{E_\gamma^2-k^2}{m_P} \\
\end{array}
\right)^{-1}
\begin{pmatrix} H^{[3,0]}_{L,A}(t,k) \\[10pt] \tilde{H}^{33}_{L,A}(t,k) + \tilde{H}^{11}_{L,A}(t,k) \\[10pt] \tilde{H}^{33}_{L,A}(t,k) - \tilde{H}^{11}_{L,A}(t,k) \end{pmatrix} 
\end{align}
where $Z(t)$ is the factor relating the matrix element of the bare local axial current, for a meson at rest, $f_P^{\mathrm{bare}} m_P$, to the corresponding  physical matrix element (up to terms of $O(a^2)$).
With twisted-mass fermions at maximal twist there is an exact PCVC relation which ensures that the physical decay constant can be obtained from two-point correlation functions of bare local pseudoscalar operators \cite{Jansen:2003ir,Shindler:2007vp}; this is denoted here by $f_P^{2\mathrm{pt}}$.  Thus 
\begin{equation}
Z(t)=\frac{f_P^{2\mathrm{pt}}}{f_P^{\mathrm{bare}}}= \frac{-2f_{P}^{2\mathrm{pt}}}{(H_{L,A}^{11}(t,0)+H_{L,A}^{22}(t,0))}
\end{equation}
 and at sufficiently large $t$, $Z(t)$ is independent of $t$.
We recall that the diagonal components $\tilde{H}^{33}_{L,A}$ and $\tilde{H}^{11}_{L,A}$ are defined in Eq.\,(\ref{eq:sub_diag_comp}) after the subtraction of the 
point-like contributions proportional to $H_{A,L}^{33}(t,0)= H_{A,L}^{11}(t,0)= H_{A,L}^{22}(t,0)$. 

At large times $t/a \gg 1$ but with $t/a \ll T/2$, the estimators $\bar{H}_{1}(t,k^2,k_z), \bar{H}_{2}(t,k^2,k_z)$ and $\bar{F}_{A}(t,k^2,k_z)$, tend to the corresponding form factors. In the limit $k_{z} \to 0$, two of the components of the vector on the right-hand side of Eq.\,(\ref{eq:est_def}), $H_{L,A}^{[3,0]}(t,k)$ and $\tilde{H}_{L,A}^{33}(t,k)-\tilde{H}_{L,A}^{11}(t,k)$, both go to zero for all values of $k^2$, see Eqs.\,(\ref{eq:H03})\,-
(\ref{eq:H30m03}).
This fact can be used to define equivalent estimators of the form factors, obtained by making the following replacement(s) in Eq.\,(\ref{eq:est_def}):
\begin{align}
H^{[3,0]}_{L,A}(t,k)&\to H^{[3,0]}_{L,A}(t,k)  - H^{[3,0]}_{L,A}(t,(\sqrt{k^{2}}, \boldsymbol{0}))
\end{align}
and/or
\begin{align}
\tilde{H}^{33}_{L,A}(t,k) - \tilde{H}^{11}_{L,A}(t,k) &\to \left(\tilde{H}^{33}_{L,A}(t,k) - \tilde{H}^{33}_{L,A}(t,(\sqrt{k^{2}}, \boldsymbol{0})) \right) - \left(\tilde{H}^{11}_{L,A}(t,k)- \tilde{H}^{11}_{L,A}(t,(\sqrt{k^{2}},\boldsymbol{0}))\right)~.
\end{align}
The correlated subtraction of the contribution coming from the kinematic point with the same value of  $k^{2}$ but zero photon spatial momentum $k_z$, can reduce the statistical noise of the estimators and improve the corresponding plateaux and below we have used this freedom to improve the resulting accuracy. The amount of improvement depends on the kinematic point and on the form factor being considered.

Finally,  for the vector form factor $F_{V}$, we define the following estimator: 
\begin{align}
\bar{F}_{V}(t,k^2,k_z) = \frac{Z_{A}}{Z_{V}}\,Z(t)\, \frac{ H_{L,V}^{12}(t,k) - H_{L,V}^{2,1}(t,k)}{2ik_{z}}~,   
\end{align}
which again for $t/a \gg 1$ and $t/a \ll T/2$ tends to $F_{V}$. The ratio of the axial ($Z_{A}$) and vector ($Z_{V}$) renormalisation constants is needed to obtain the properly renormalised value of $F_{V}$ when using twisted mass fermions.

Having explained our procedure for extracting the SD form factors from three-point lattice correlation functions, we now proceed to presenting our numerical results.

\section{Numerical results for the form factors}\label{ffnum}
In this section we implement the procedure developed in the preceding  sections to study $K\to \ell\,\nu_\ell\,\ell^{\prime+}\ell^{\prime-}$ decays numerically. The simulations have been performed on the $A40.32$ ensemble generated by the ETMC\,\cite{Carrasco2014} with $N_{f}=2+1+1$ dynamical quark flavours, a space-time volume of $32^3\times 64$ and a lattice spacing of $a= 0.0885(36)\,{\rm fm}$. 
The analysis was performed on 100 gauge configurations.
The light quarks are heavier than the corresponding physical ones and correspond to
$m_{\pi}\simeq 320\,{\rm MeV}$ and $m_{K}\simeq 530\,{\rm MeV}$. We used smeared interpolating sources for the kaon field, obtained applying 128 steps of Gaussian smearing with step-size parameter $\epsilon=0.1$. Moreover, we used four stochastic sources on each time slice when inverting the Dirac operator.
On this ensemble the values of the two renormalisation constants are $Z_{A}= 0.731(8)$ and $Z_{V}= 0.587(4)$\,\cite{Carrasco2014}.

Below we will compare our results for the form factors and branching ratios with those determined in experiment\,\cite{Poblaguev_2002} and chiral perturbation theory. While these comparisons are interesting and instructive, it must be remembered that our computations were performed with unphysical quark masses, at a single value of the lattice spacing and on a single volume\,\footnote{We also compare our results with a previous lattice computation\,\cite{xu}, which was performed on a $24^3\times 48$ lattice with $a\simeq 0.093$\,fm with quark masses corresponding to $m_\pi\simeq352$\,MeV and $m_K\simeq 506$\,MeV.}.
Until the corresponding systematic uncertainties are studied in the future, the comparison with the experimental measurements may be indicative, but cannot be considered definitive.

As already outlined in Sec.~\ref{lattcorr}, we used twisted boundary conditions in order to evaluate the hadronic tensor for a range of values of the photon's spatial momentum $\boldsymbol{k}$. To probe the region of the phase-space relevant for the four $K\to \ell\,\nu_\ell\,\ell^{\prime+}\ell^{\prime-}$ decay channels, with $\ell,\ell' = e,\mu$, we evaluate the Euclidean three-point functions $C^{\mu\nu}(t,E_{\gamma}, \boldsymbol{k}, \boldsymbol{p})$ for fifteen different values of $(E_{\gamma}, \boldsymbol{k})$, with $\boldsymbol{k}= (0,0,k_{z})$, and restricted our analysis to the kaon rest frame $\boldsymbol{p}= 0$. We find it convenient to parametrize the phase space in terms of the two dimensionless parameters $x_{k}$ and $x_{q}$, defined as
\begin{equation}
\label{eq:xkqdef} 
x_k \equiv \sqrt{\frac{k^{2}}{m_{K}^{2}}}\, , \qquad x_q \equiv \sqrt{\frac{q^{2}}{m_{K}^{2}}}\,, 
\end{equation}
where $q$ is the four-momentum of the lepton-neutrino pair created by the weak Hamiltonian.
In terms of $x_{k}$ and $x_{q}$ the 
photon's four-momentum, $(E_\gamma,0,0,k_z)$, (in the kaon's rest frame) is given by 
\begin{equation}
E_{\gamma}  = \frac{m_{K}}{2}(1 + x_{k}^{2}- x_{q}^{2})\,,\qquad
k_z = \frac{m_{K}}{2}\sqrt{ (1-x_k^{2}-x_q^{2})^{2}- 4x_k^{2}x_q^{2} }\,.
\end{equation}
The range of values of $x_{k}$ and $x_{q}$ is given in terms of the 
lepton masses, $m_{\ell}$ and $m_{\ell'}$, by
\begin{equation}
\label{xkq_range}
\frac{m_\ell}{m_K}\,\leq\,x_q\,\leq\,1-x_k\,;\qquad
\frac{2m_{\ell^\prime}}{m_K}\,\leq\,x_k\,\leq\,1-\frac{m_\ell}{m_K},
\end{equation}
so that the phase space has a triangular shape in the $x_{k}$\,-\,$x_{q}$ plane.\\

\begin{figure}[t]
	\centering
	\includegraphics[scale=0.5]{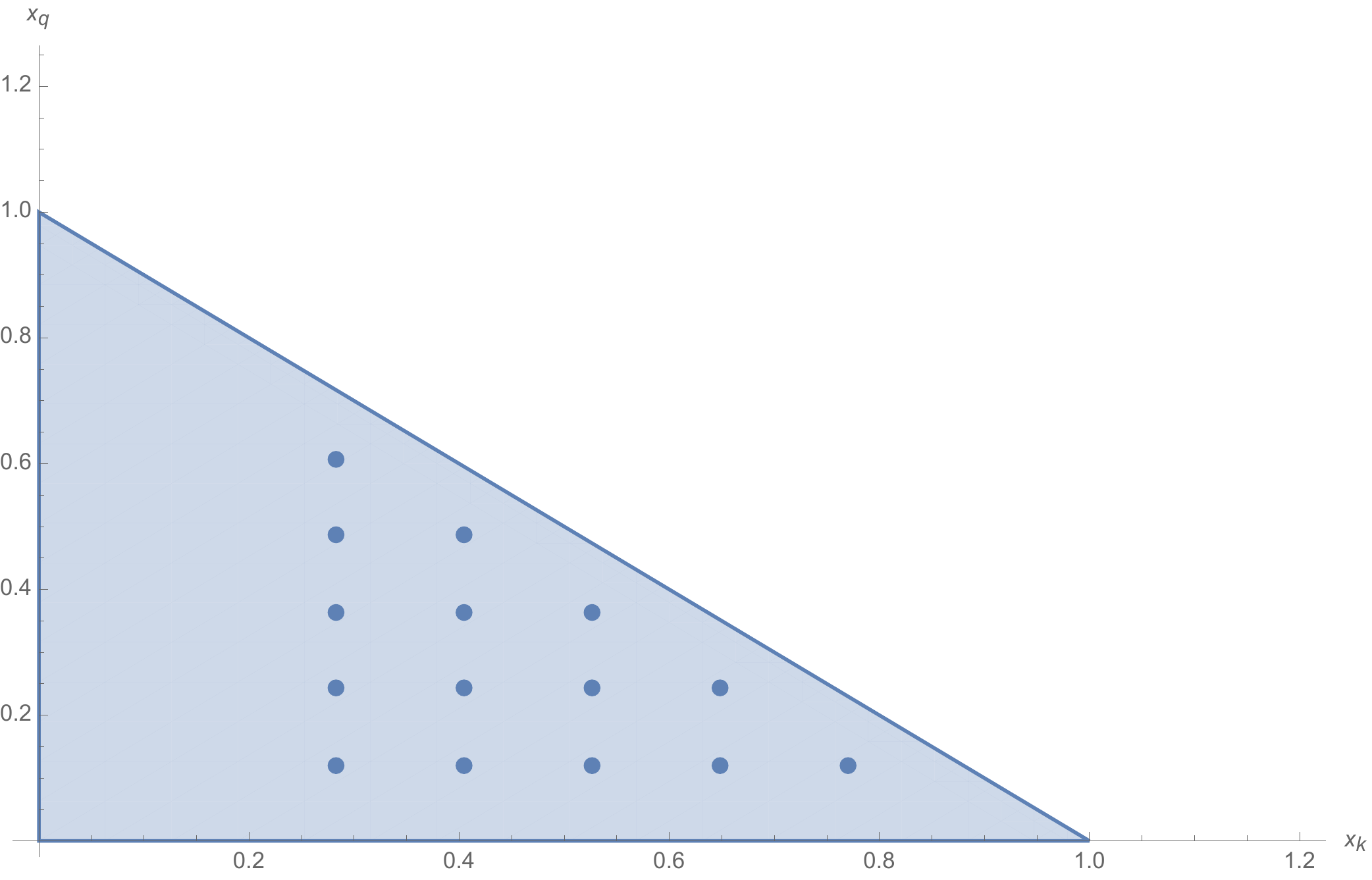}
	\caption{\it The shaded area represents the range of allowed physical values of $(x_k,x_q)$ when neglecting lepton masses, so that $0<x_k<1$ and $0<x_q<1-x_k$. The points correspond to the 15 choices of $(x_k,x_q)$ used in this analysis.}
	\label{phsp_plot}
\end{figure}

In Fig.\,\ref{phsp_plot} we show the positions of the fifteen simulated kinematic configurations, which we take as equally spaced in the $x_{k}$\,-\,$x_{q}$ plane. For completeness, the  corresponding numerical values of $x_{k}$ and $x_{q}$ are reported in Tab.\,\ref{tab:num_val_xk_xq}. It should be noted that our computations are limited to $x_{k}\geq 0.28$. This choice is appropriate to describe both the cases in which a $\mu^{+}\mu^{-}$ or a $e^{+}e^{-}$ pair is produced in the radiative decay of the kaon. Indeed, in the first case the lowest allowed value of $x_{k}$ is given by $2m_{\mu}/m_{K} \simeq 0.428$, while for decays in which an $e^+e^-$ pair is produced, although very low values of $x_{k}$ are kinematically allowed, the  experimental branching ratios have been determined with values 
of the electron-positron invariant mass $\sqrt{k^{2}} > 145\,{\rm MeV}$ $(x_{k} > 0.294)$ for $K^+\to \mu^+\,\nu_\mu\,e^+\,e^-$ decays and
$\sqrt{k^{2}} > 150\,{\rm MeV}$ $(x_{k} >0.304)$ for $K^+\to e^+\,\nu_e\,e^+\,e^-$ decays\,\cite{Poblaguev_2002}.

\begin{center}
\begin{table}[]
    \begin{tabular}{c | c c c  c c c c c c c c c c c c c}
    \hline
     kinematics & 1 & 2 & 3 & 4 & 5 & 6 & 7 & 8 & 9 & 10 & 11 & 12 & 13 & 14 & 15 \\
    \hline
    $ x_{k}$ & 0.28 & 0.28 & 0.28 & 0.28 & 0.28 & 0.41 & 0.41 & 0.41 & 0.41 & 0.53 & 0.53 & 0.53 & 0.65 & 0.65 & 0.77 \\
    \hline
    $ x_{q}$ & 0.12 & 0.24 & 0.36 & 0.48 & 0.61 & 0.12 & 0.24 & 0.36 & 0.48 & 0.12 & 0.24 & 0.36 & 0.12 & 0.24 & 0.12 \\
    \hline
    \end{tabular}
    \caption{\small\it Table of the values of $x_{k}$ and $x_{q}$ corresponding to the fifteen simulated kinematic points.}
    \label{tab:num_val_xk_xq}
\end{table}
\end{center}

\begin{figure}[tp]
       \subfloat{%
       \includegraphics[scale=0.32]{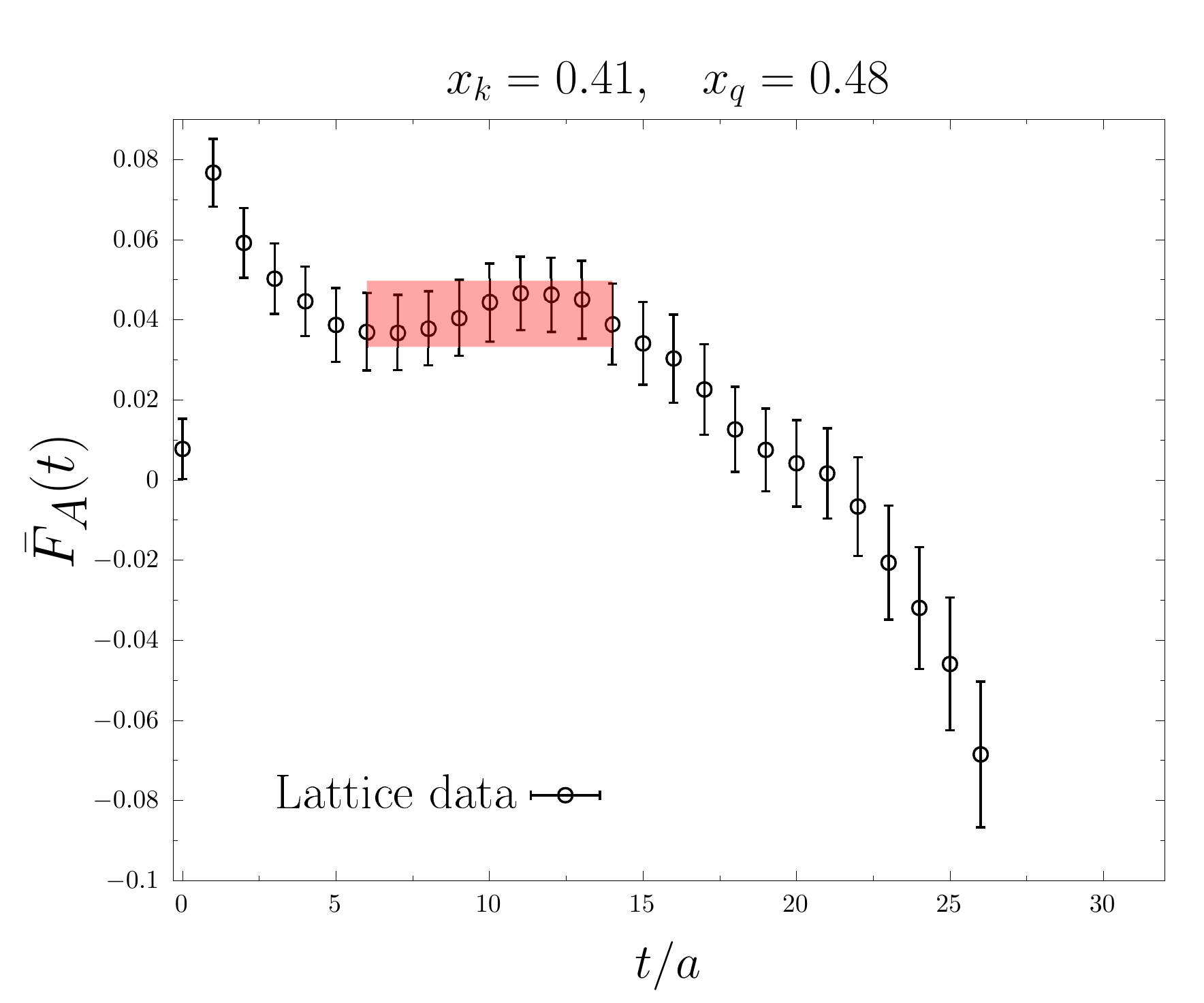}}
     \subfloat{%
       \includegraphics[scale=0.32]{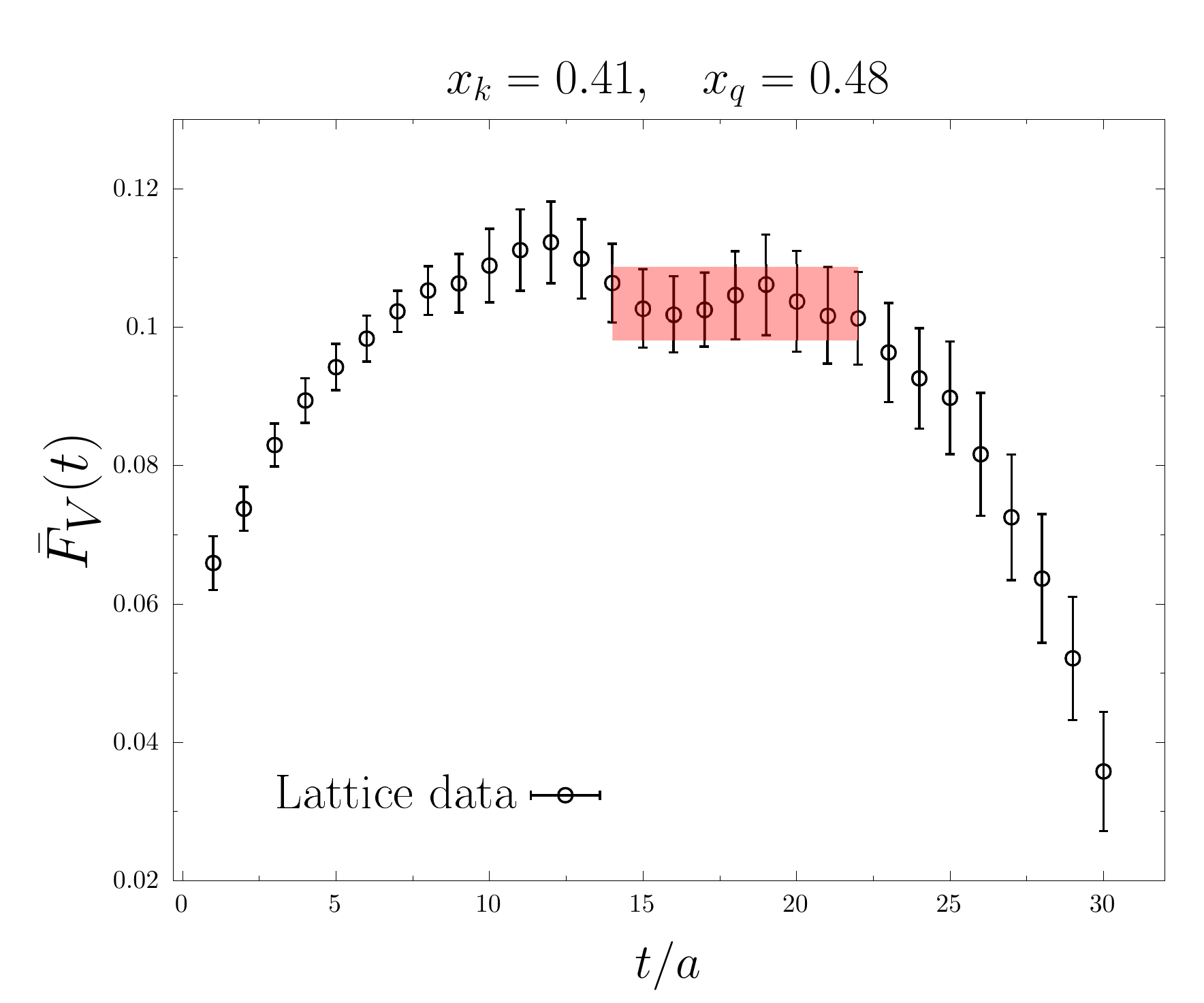}} \\
     \subfloat{%
       \includegraphics[scale=0.32]{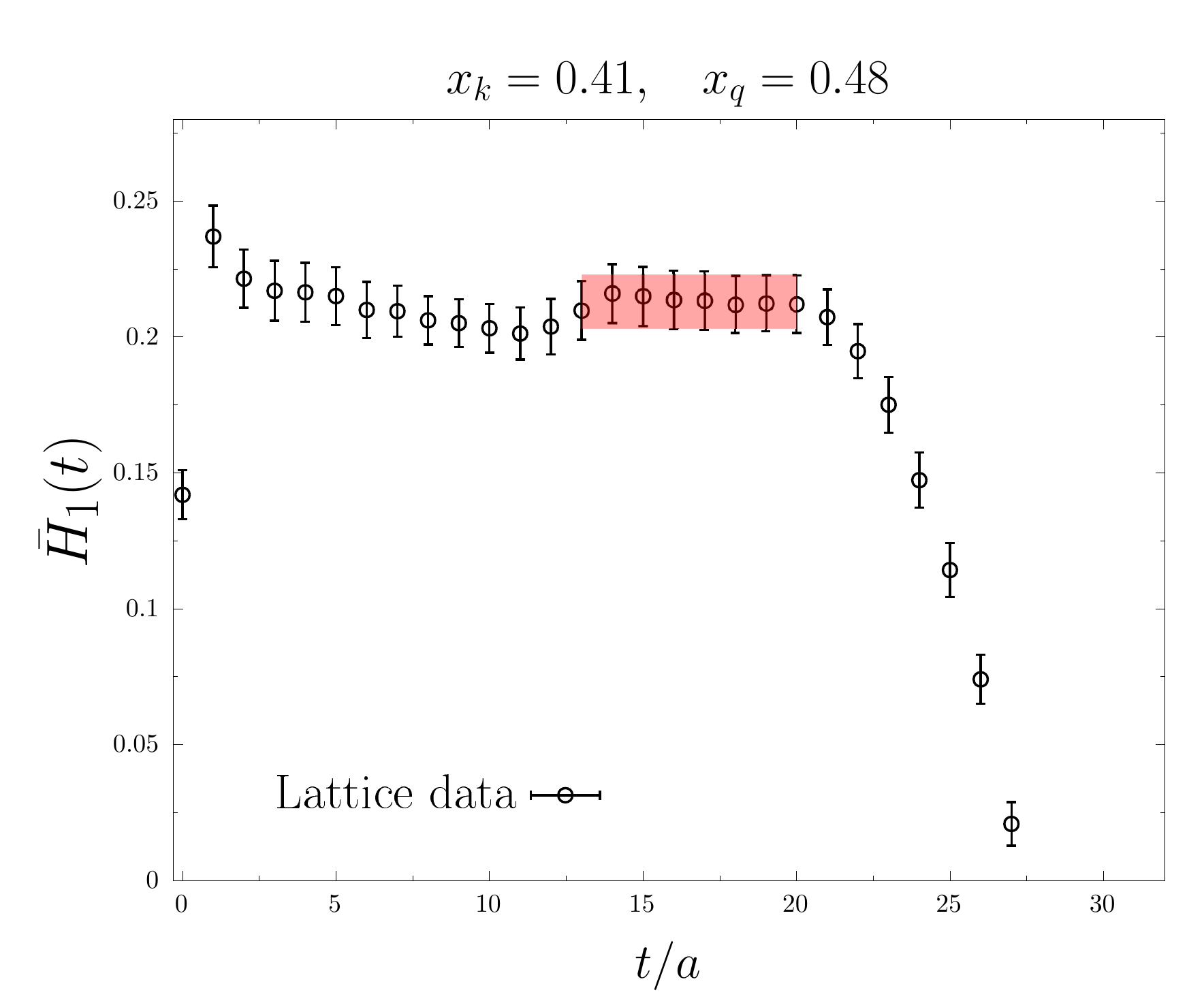}}
     \subfloat{%
       \includegraphics[scale=0.32]{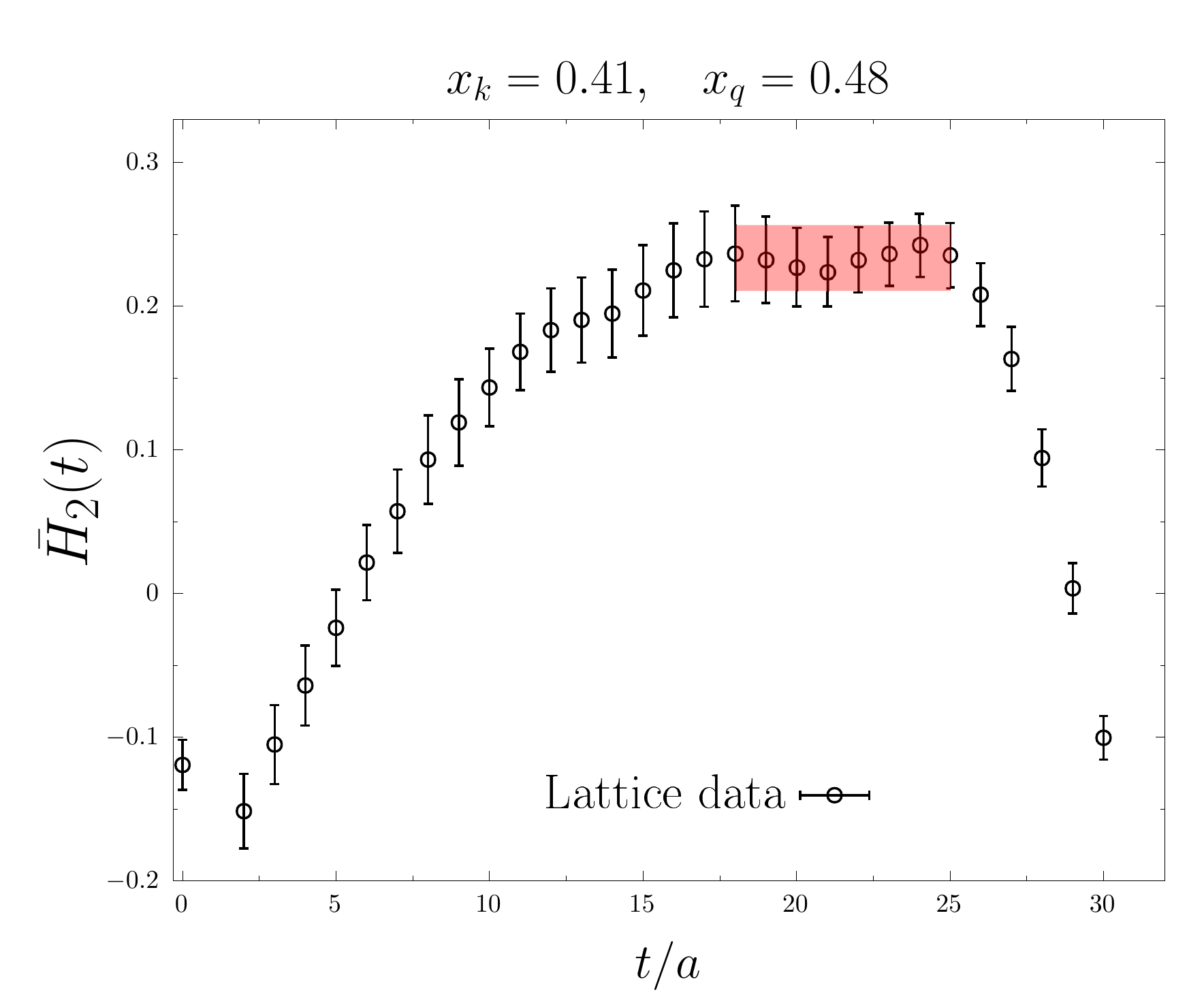}} 
	\caption{\it Extraction of the form factors $F_{A}, F_{V},H_{1}, H_{2}$ from the plateaux of the corresponding estimator. The data correspond to $x_{k}=0.41$ and $x_{q}=0.48$.}
	\label{xk_0.4064_xq_0.4845}
\end{figure}

\begin{figure}[tp]
       \subfloat{%
       \includegraphics[scale=0.32]{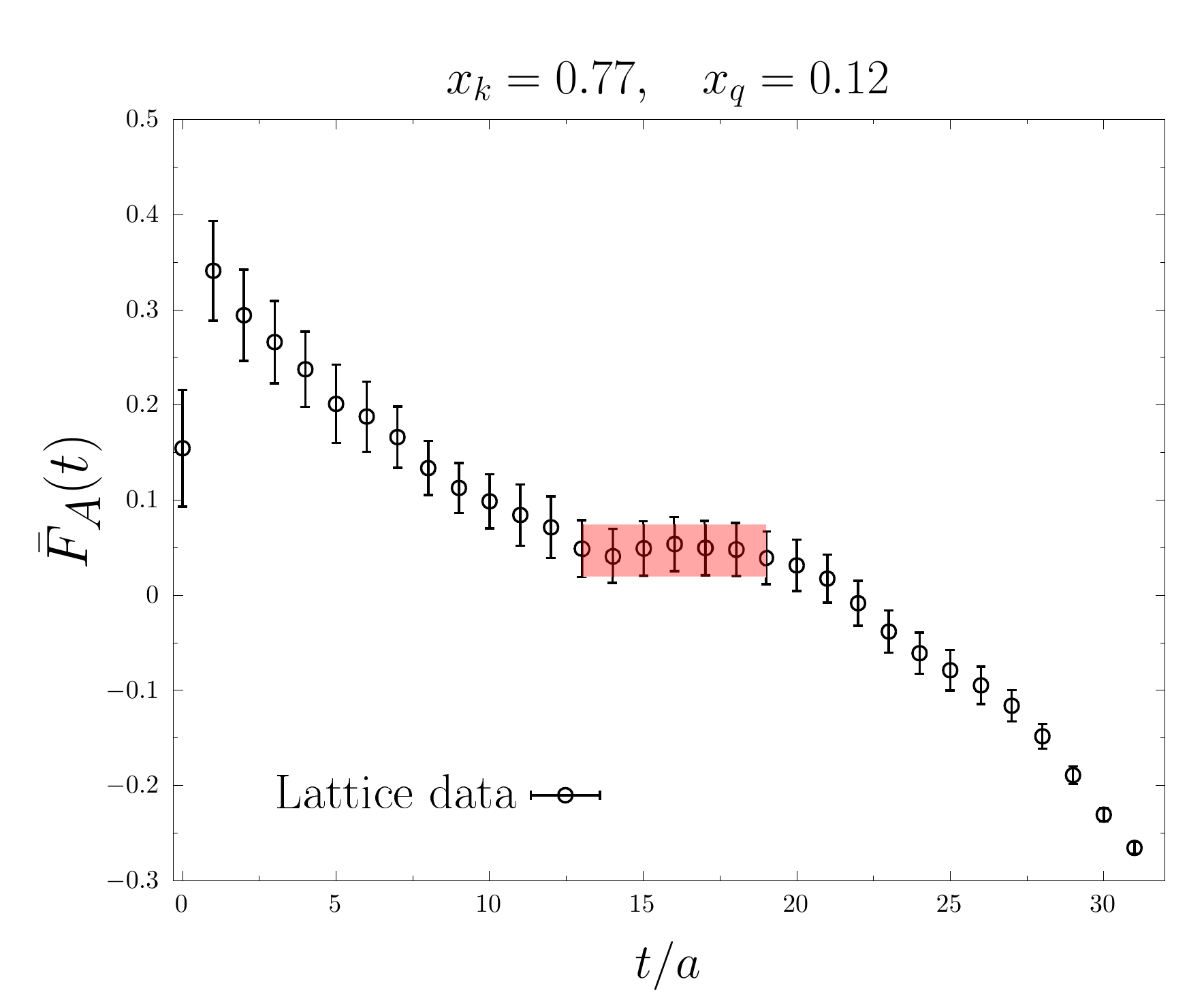}}
     \subfloat{%
       \includegraphics[scale=0.32]{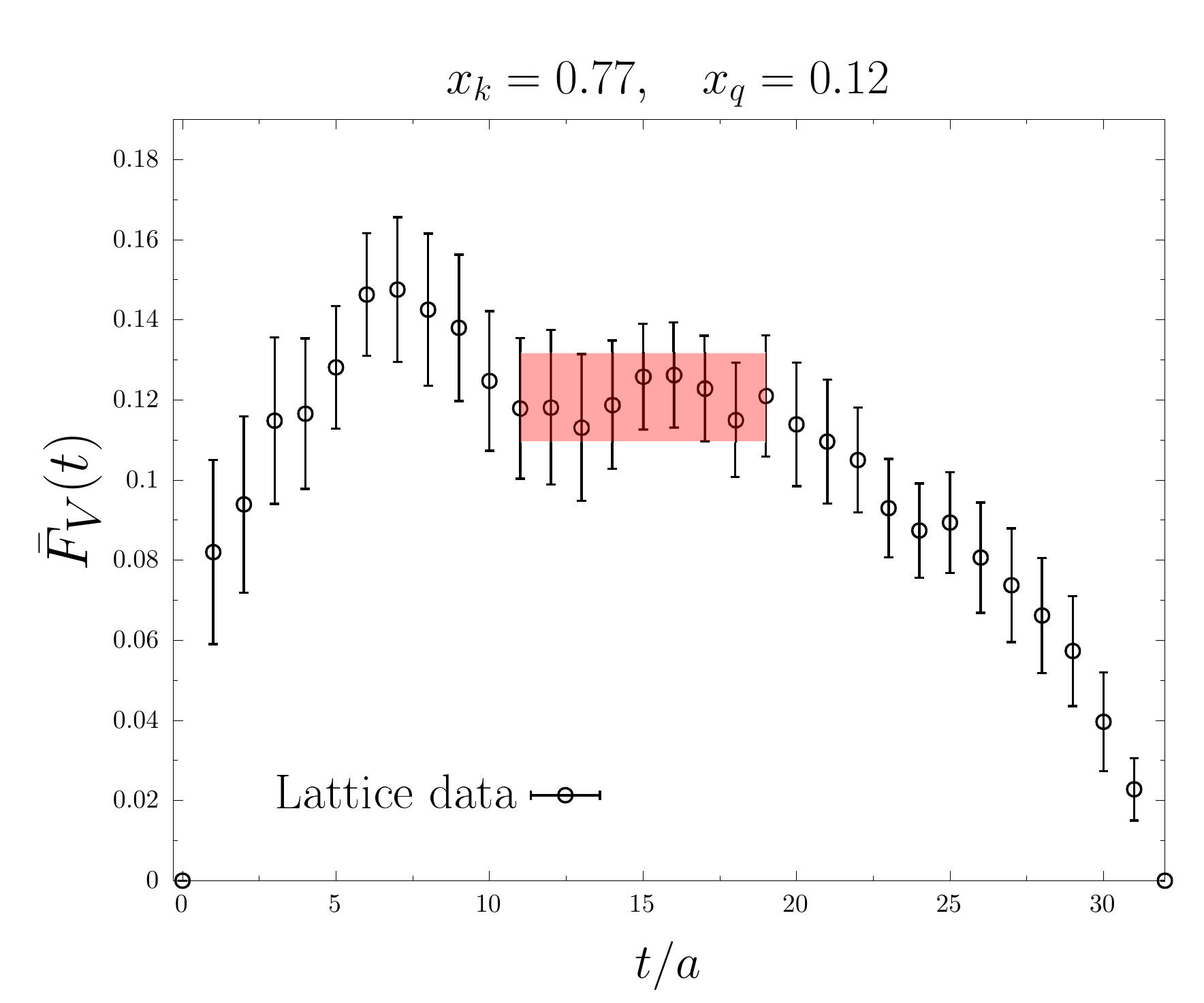}} \\
     \subfloat{%
       \includegraphics[scale=0.32]{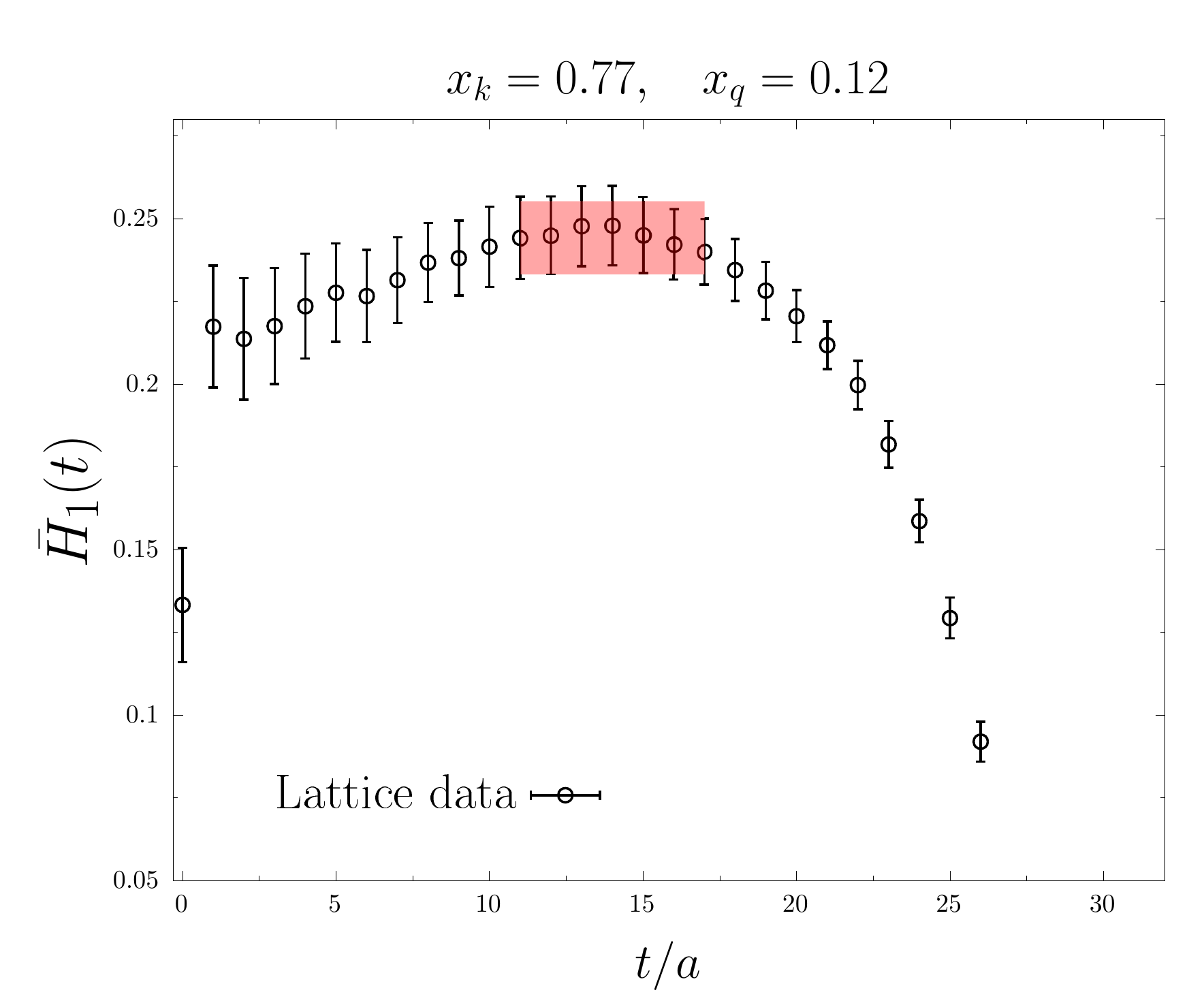}}
     \subfloat{%
       \includegraphics[scale=0.32]{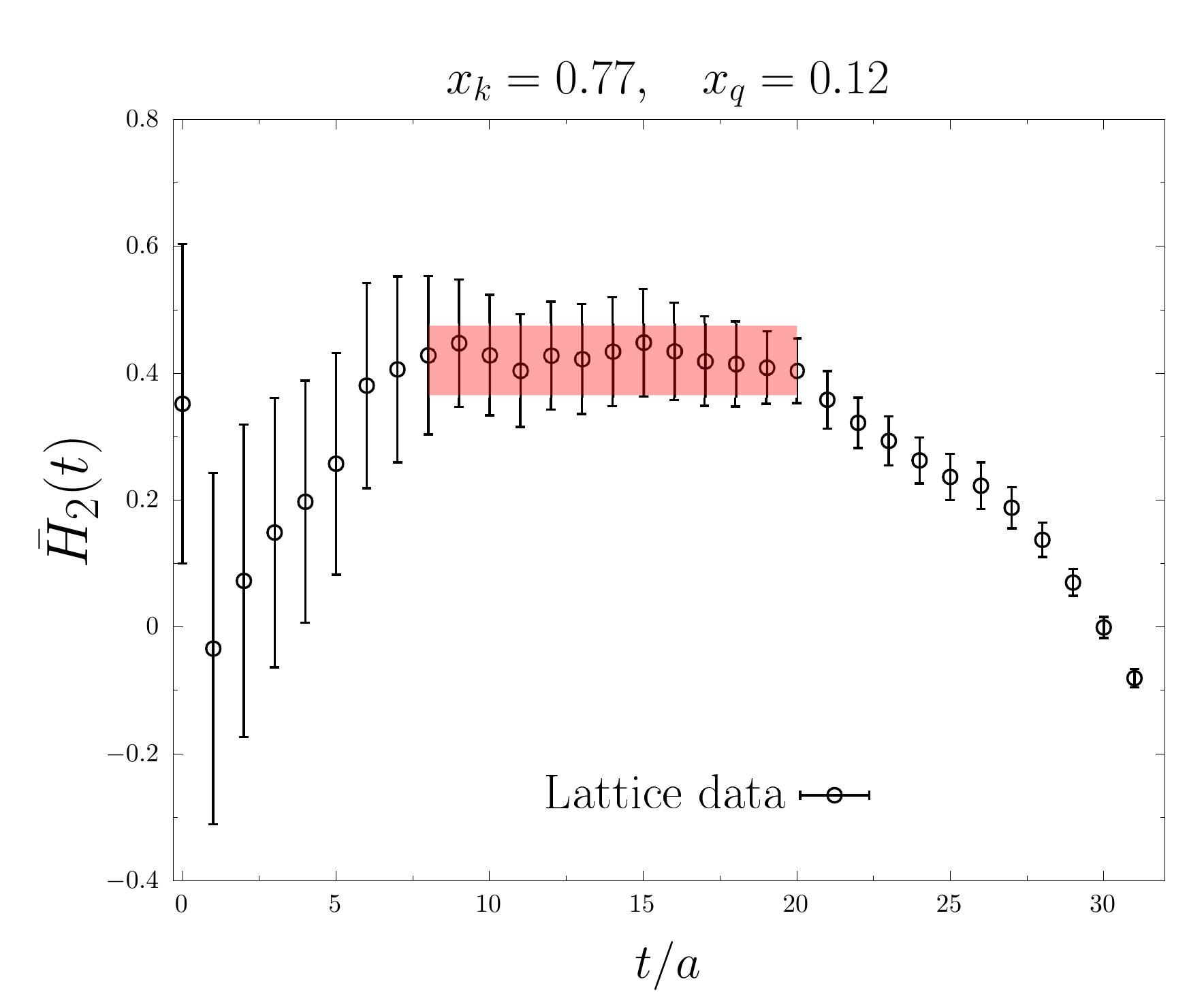}} 
	\caption{\it Extraction of the form factors $F_{A}, F_{V},H_{1}, H_{2}$ from the plateaux of the corresponding estimator. The data correspond to $x_{k}=0.77$ and $x_{q}=0.12$.}
	\label{xk_0.7722_xq_0.1185}
\end{figure}

\begin{figure}[t]
       \subfloat{%
       \includegraphics[scale=0.32]{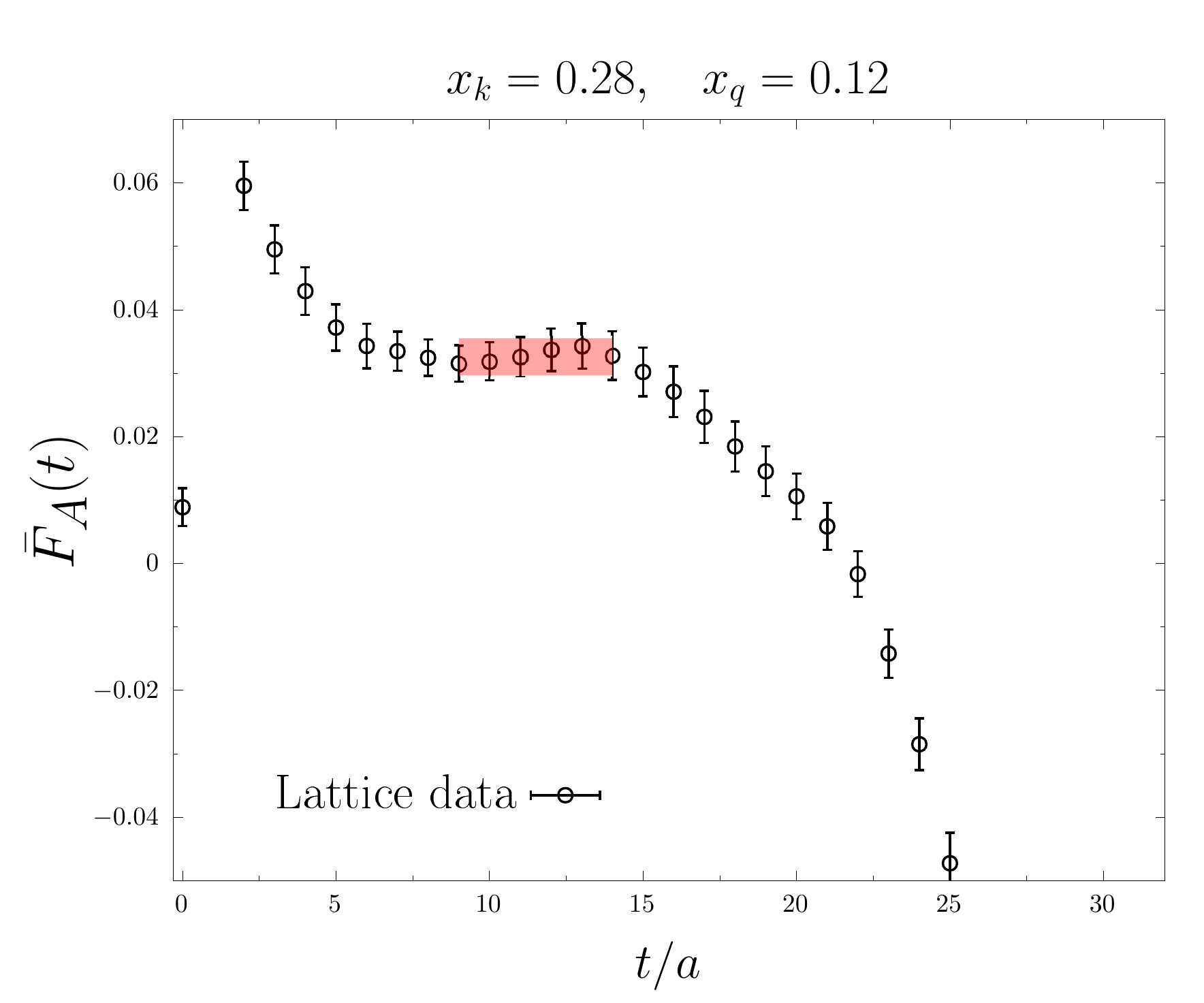}}
     \subfloat{%
       \includegraphics[scale=0.32]{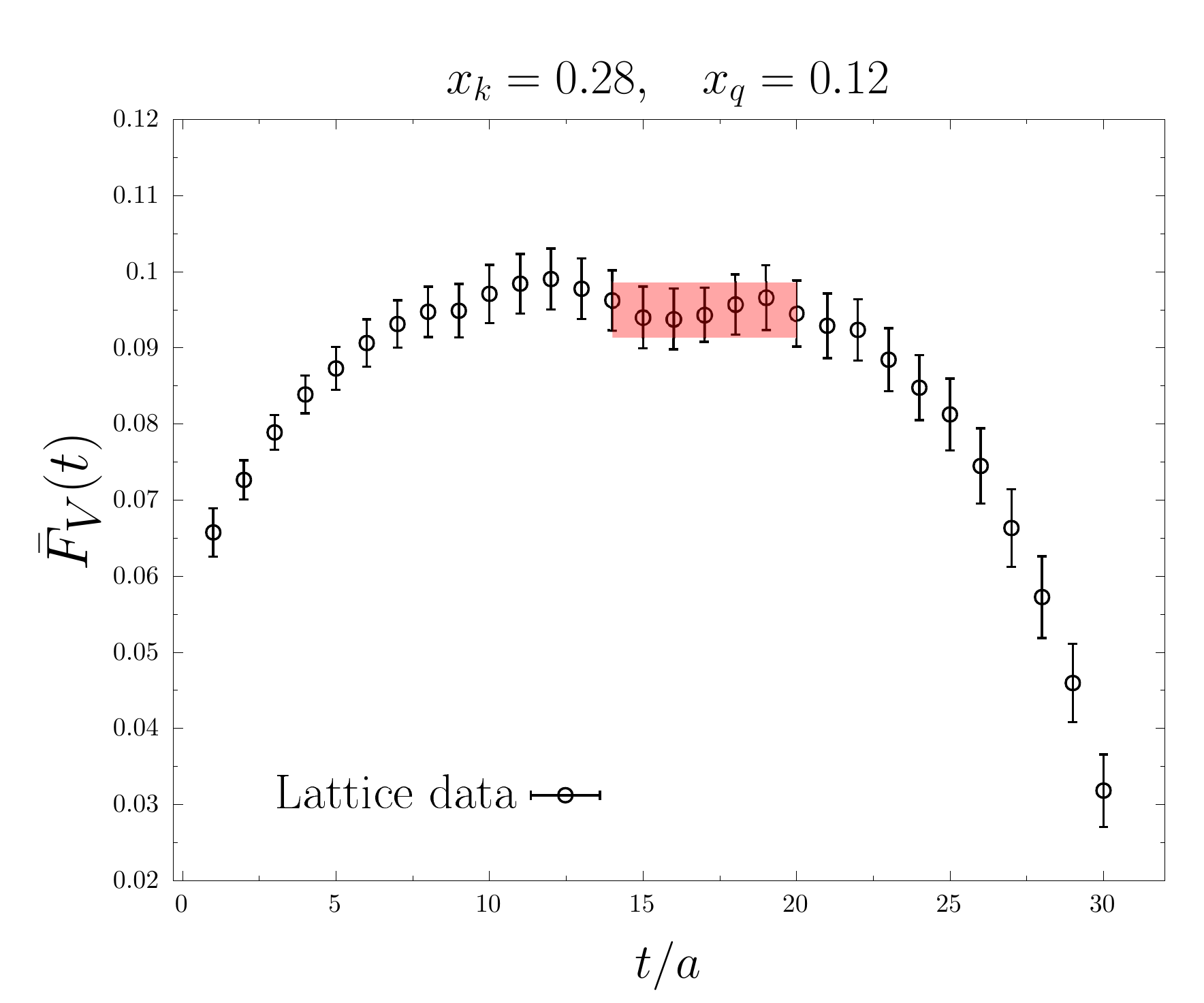}} \\
     \subfloat{%
       \includegraphics[scale=0.32]{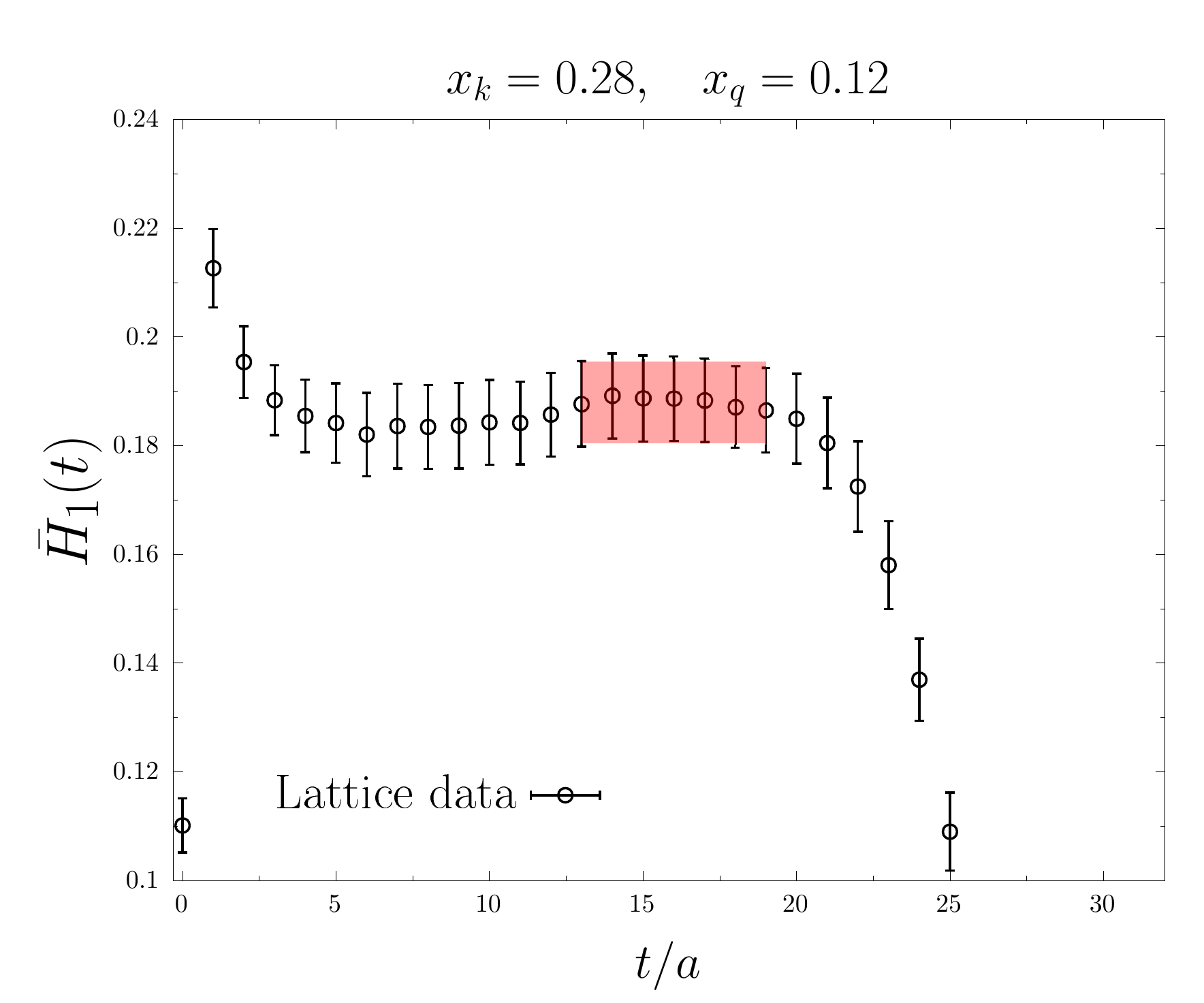}}
     \subfloat{%
       \includegraphics[scale=0.32]{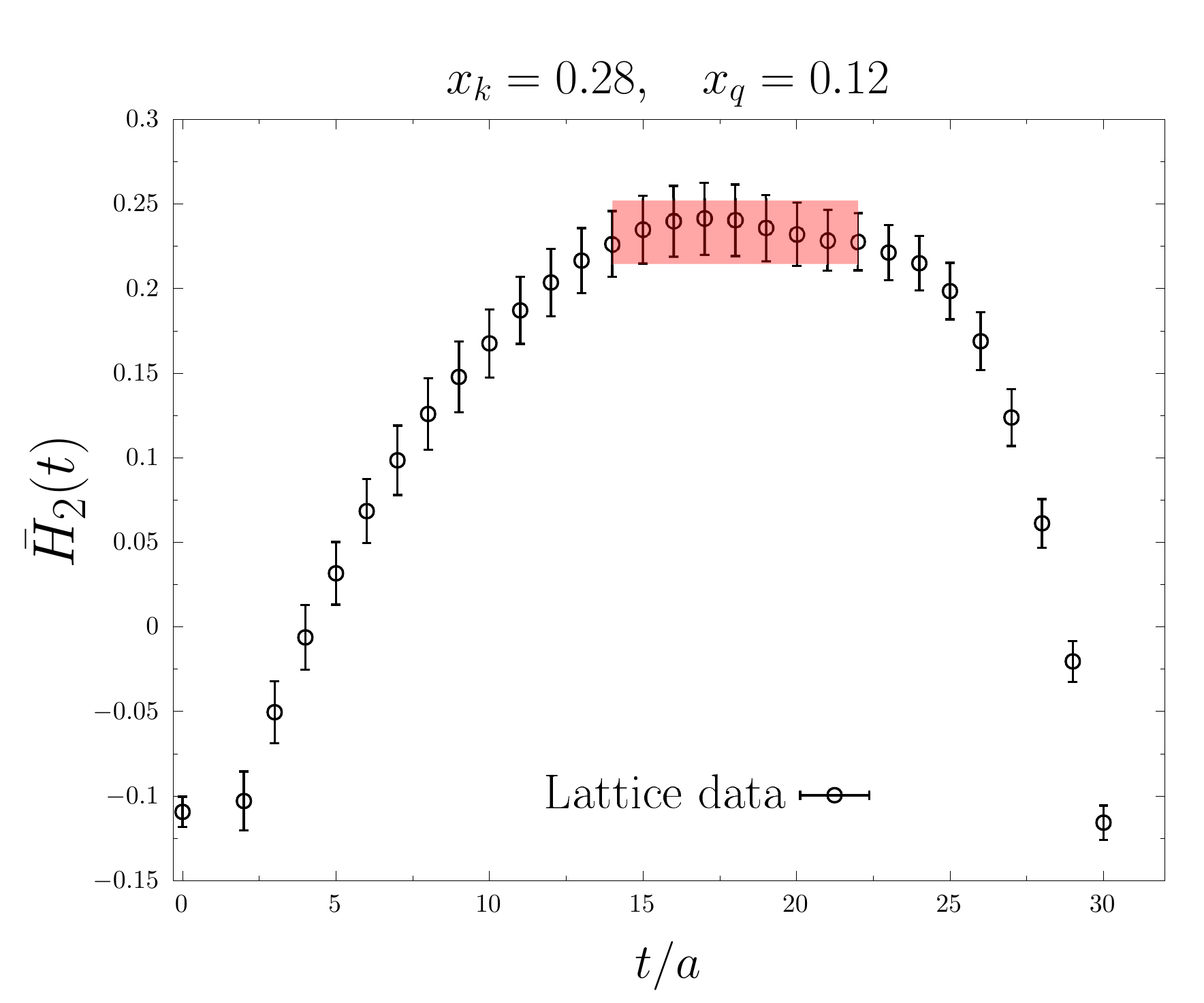}} 
	\caption{\it Extraction of the form factors $F_{A}, F_{V},H_{1}, H_{2}$ from the plateaux of the corresponding estimator. The data correspond to $x_{k}=0.28$ and $x_{q}=0.12$.}
	\label{xk_0.2845_xq_0.1153}
\end{figure}

In Figs.\,\ref{xk_0.4064_xq_0.4845},~\ref{xk_0.7722_xq_0.1185} and \ref{xk_0.2845_xq_0.1153}, we present the estimators $\bar{H}_{1}(t, x_{k}, x_{q})$, $\bar{H}_{2}(t, x_{k}, x_{q})$, $\bar{F}_{A}(t, x_{k}, x_{q})$ and $\bar{F}_{V}(t, x_{k}, x_{q})$ for selected values of $x_{k}$ and $x_{q}$. In each figure, the shaded region indicates the result of a constant fit in the corresponding time interval. Figs.\,\ref{xk_0.4064_xq_0.4845} and \ref{xk_0.7722_xq_0.1185} illustrate the feature that 
for kinematics corresponding to small values of $k_z$ (i.e. when $x_q+x_k\simeq 1$) the estimator of the axial form factor $F_A$ becomes somewhat noisy leading to increased uncertainties in its determination.  On the other hand, for other values of $(x_k,x_q)$ and for all other form factors, the precision achieved is very good and typically of the order of five to ten percent.

In order to evaluate the decay rate, we fit the lattice form factors, using two different ansatzes to describe their dependence on $x_{k}$ and $x_{q}$. The first is a simple polynomial in $x_{k}^{2}$ and $x_{q}^{2}$ given by
\begin{align}
F_{\mathrm{poly}}(x_k,x_q)=a_0+a_kx_k^2+a_qx_q^2\,,\label{poly} 
\end{align}
where $a_{0}, a_{k}$ and $a_{q}$ are free fitting parameters. We find that this simple form represents our data very well and the corresponding results presented below are obtained using Eq.\,(\ref{poly}). However, we have also performed fits using ansatzes which include additional terms which are quartic in $x_k$ and $x_q$, i.e. terms proportional to $x_k^2x_q^2$, $x_k^4$ and $x_q^4$. We find that including all or some of such terms does not improve the fits, generally results in an overfit of our data and only negligibly changes the results for the form factors and decay rates. This is not surprising as the 15 points in the $(x_k,x_q)$ plane at which we compute the form factors (see Tab.\,\ref{tab:num_val_xk_xq}) cover well the kinematic regions studied in the E865 experiment\,\cite{Poblaguev_2002,Ma_2006} to which we compare our results in Sec.\,\ref{brnum}.  We therefore require only minor interpolations of our results to be able to perform the phase-space integrations.

\begin{figure}[p]
       \subfloat{%
       \includegraphics[scale=0.35]{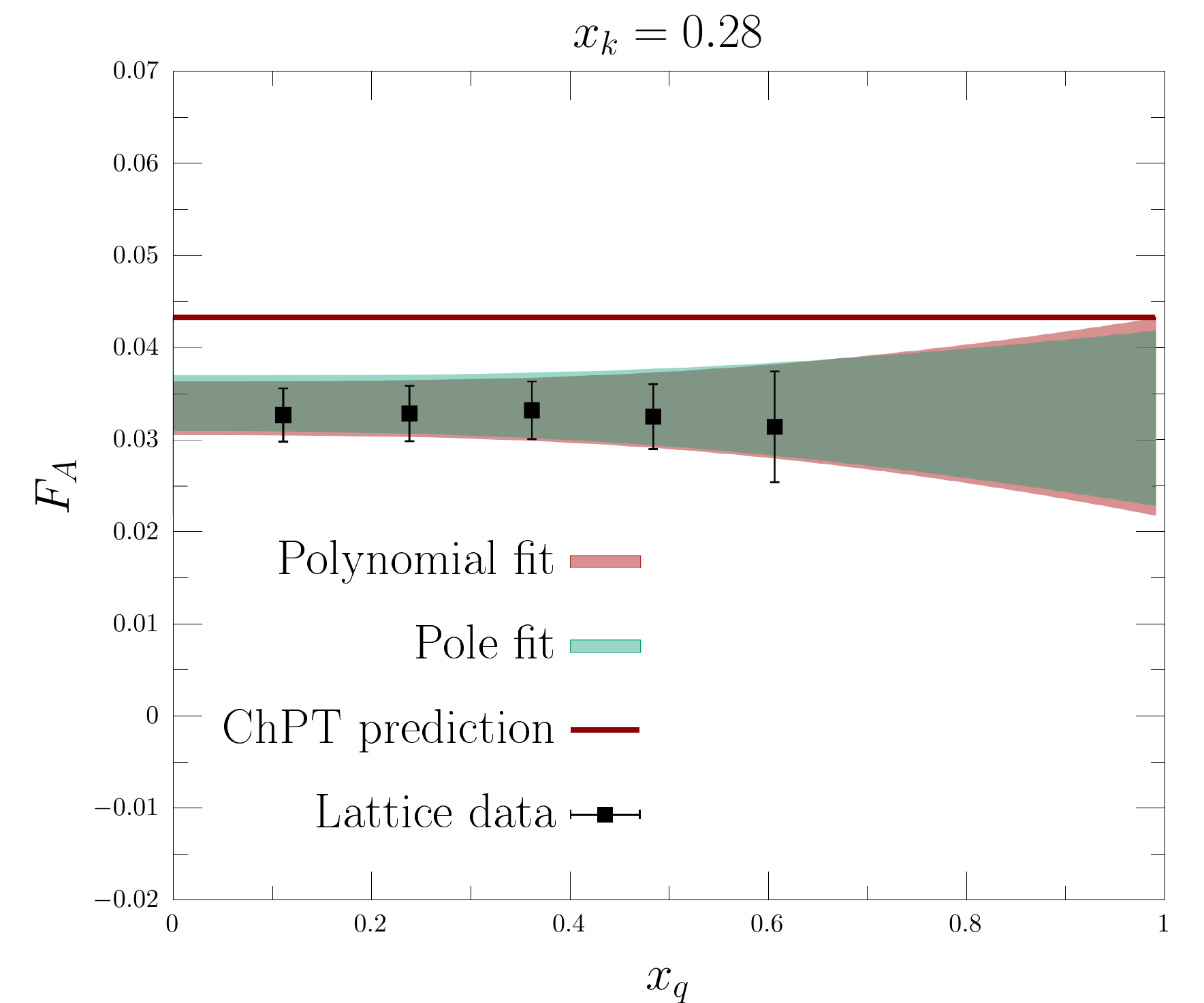}}
     \subfloat{%
       \includegraphics[scale=0.35]{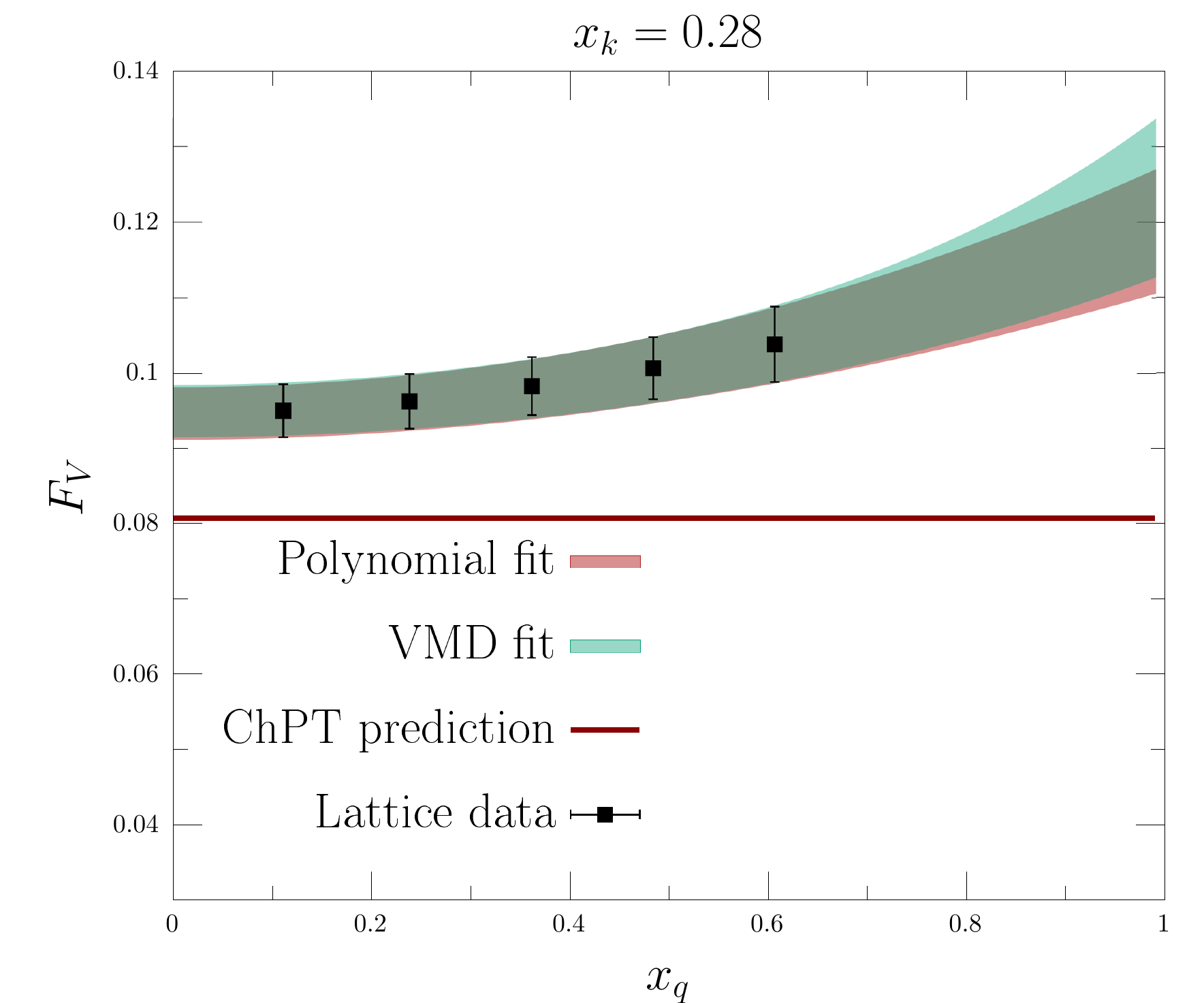}} \\
     \subfloat{%
       \includegraphics[scale=0.35]{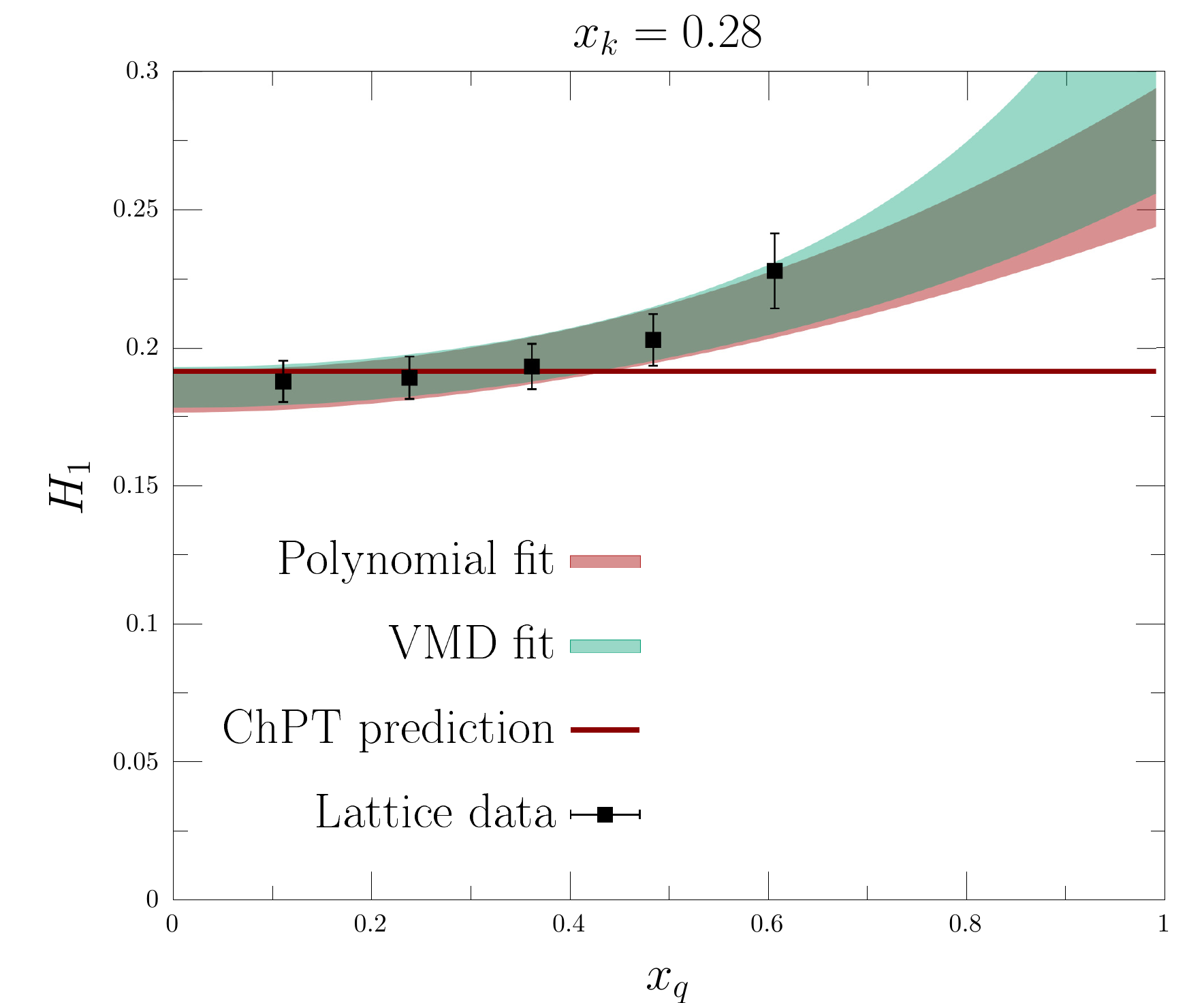}}
     \subfloat{%
       \includegraphics[scale=0.35]{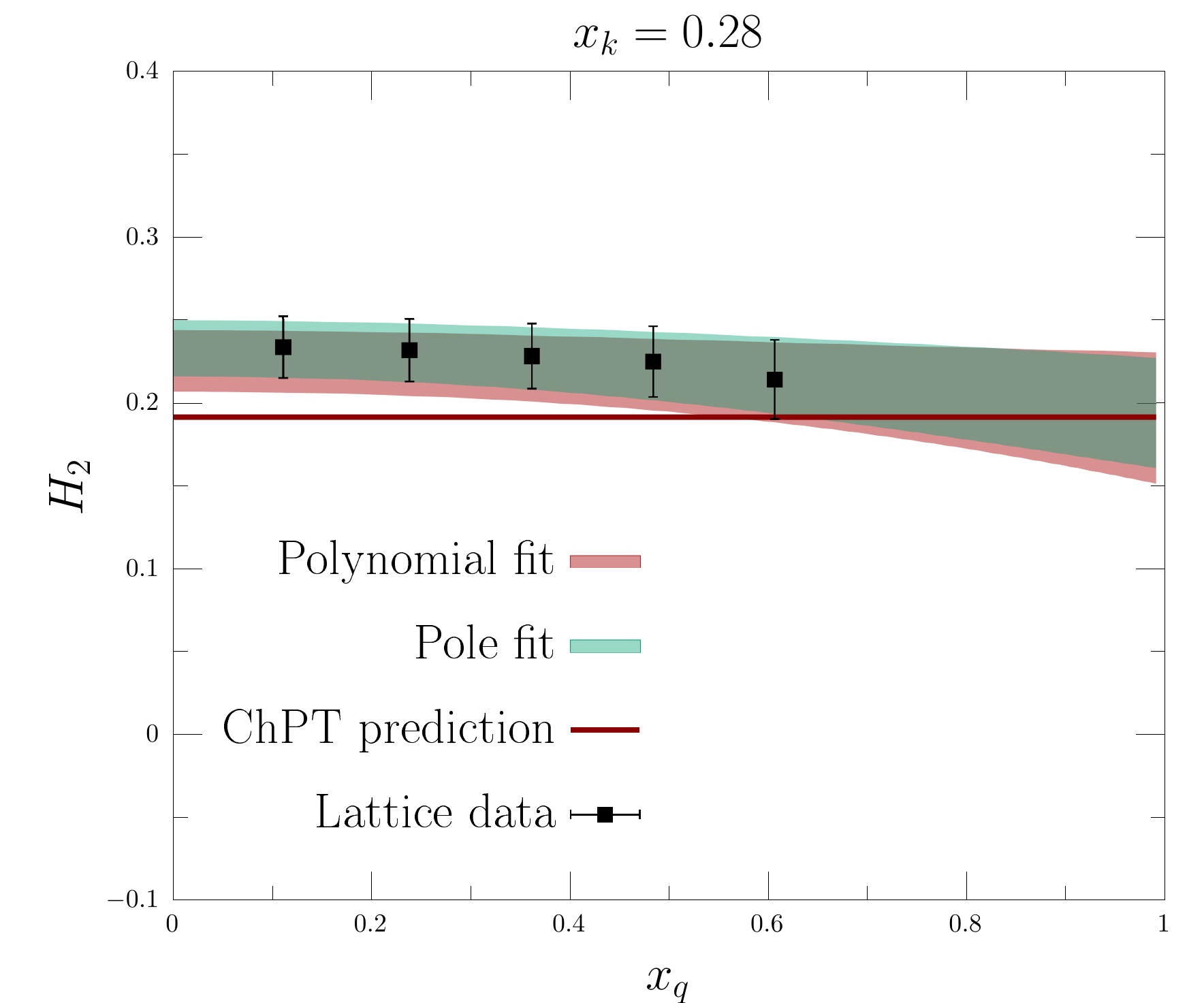}} \\
       \subfloat{%
       \includegraphics[scale=0.35]{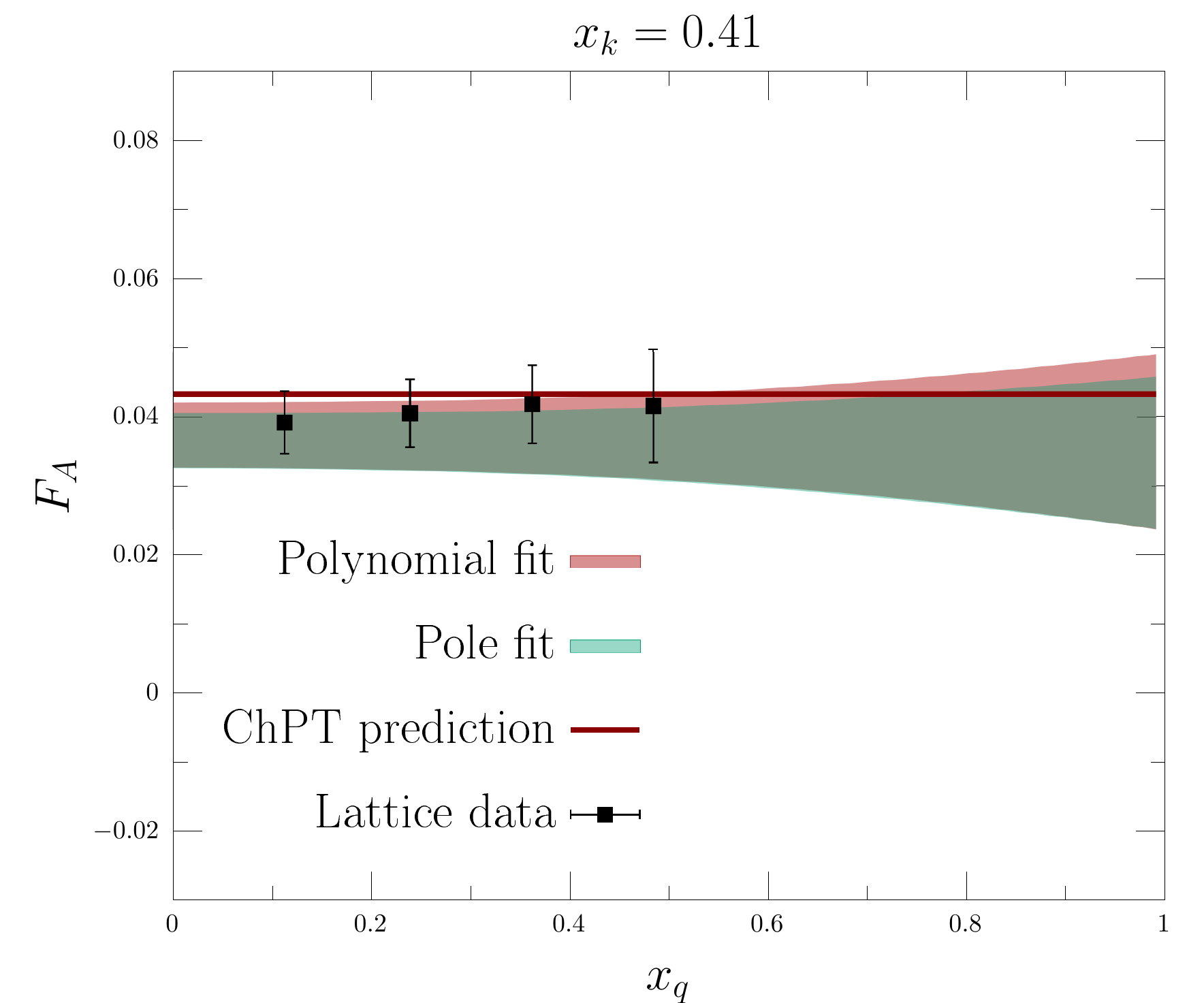}}
     \subfloat{%
       \includegraphics[scale=0.35]{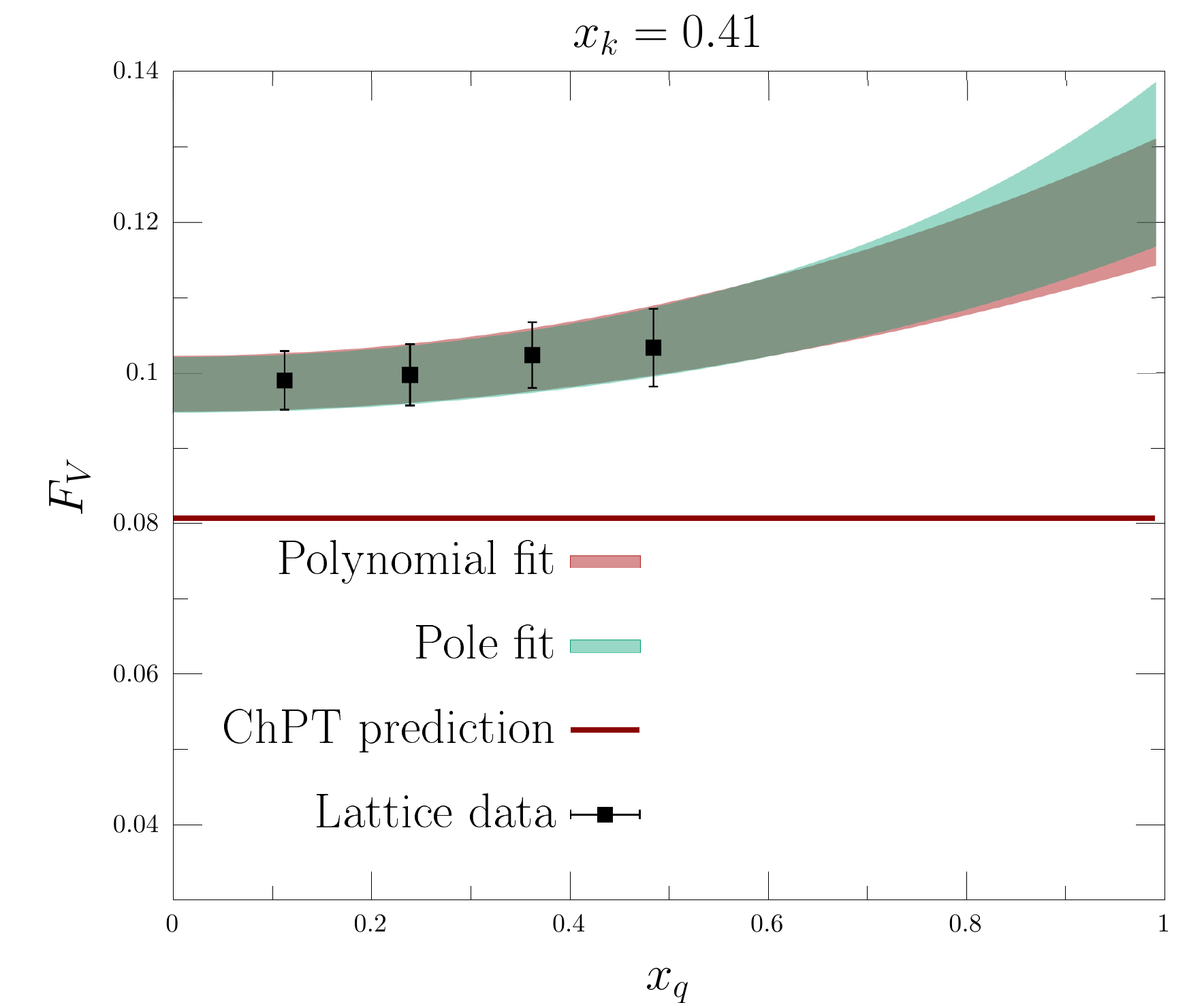}} \\
     \subfloat{%
       \includegraphics[scale=0.35]{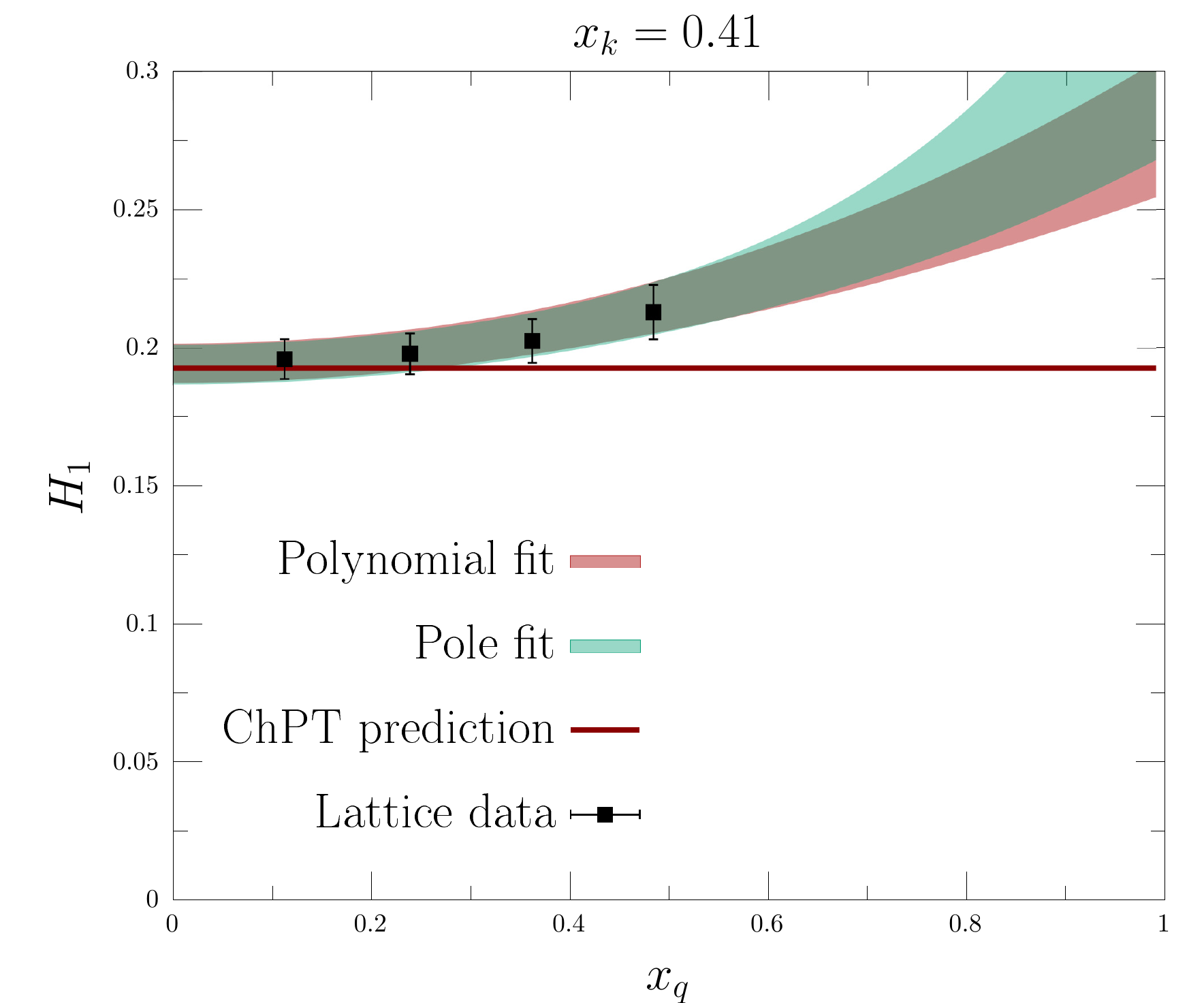}}
     \subfloat{%
       \includegraphics[scale=0.35]{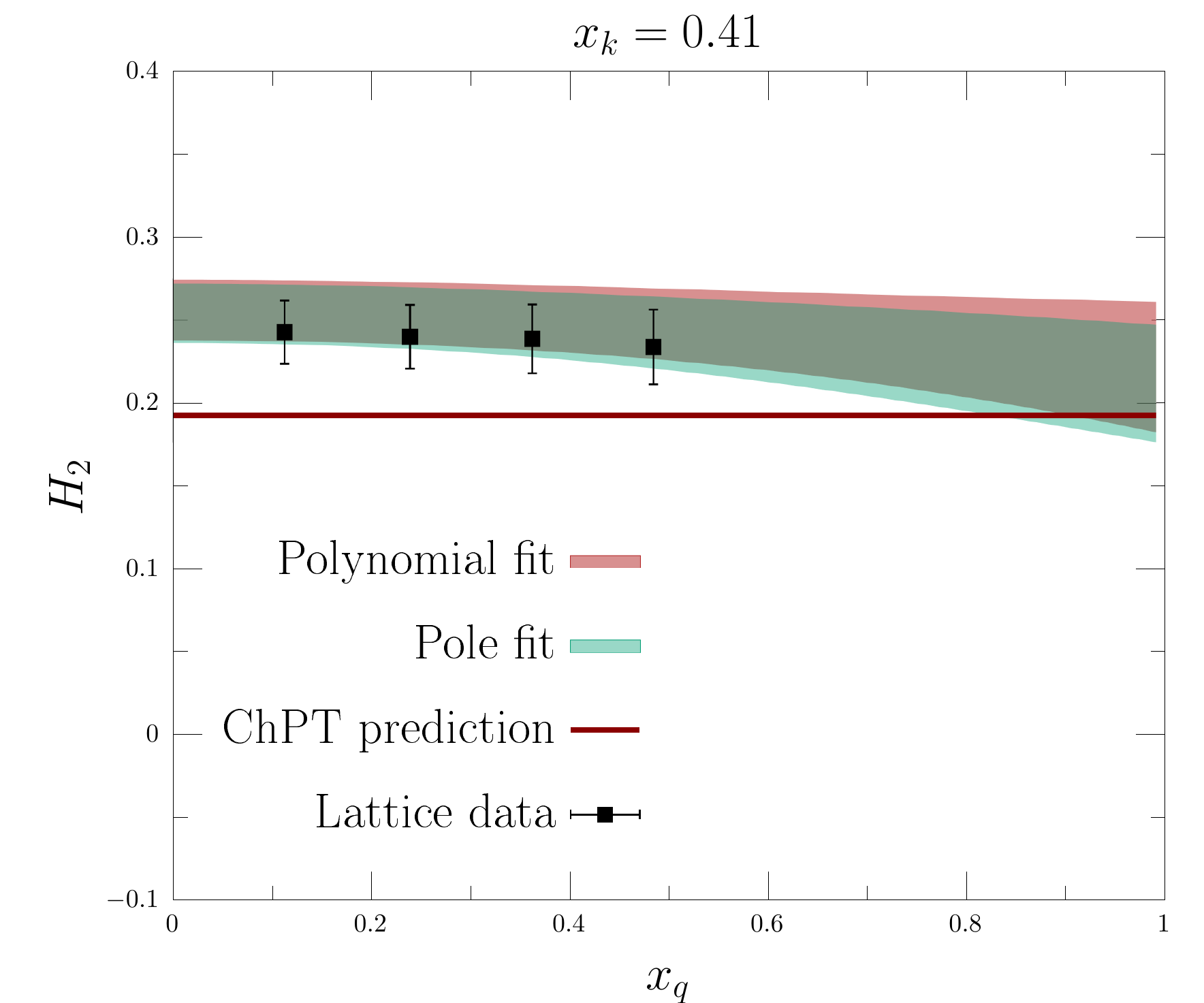}} 
       \caption{\it The fitting functions corresponding to the polynomial and the pole-like fits of Eqs.\,(\ref{poly}) and(\ref{pole}) are plotted, along with the lattice data, as function of $x_{q}$ and at a fixed value of $x_k=0.28$ (panels 1-4) and $x_k=0.41$ (panels 5-8) . The red line corresponds to the 1-loop ChPT prediction with $F=f_K/\sqrt{2}$.} 
	\label{xk_0.2845_0.4064}
\end{figure}

\begin{figure}[p]
       \subfloat{%
       \includegraphics[scale=0.35]{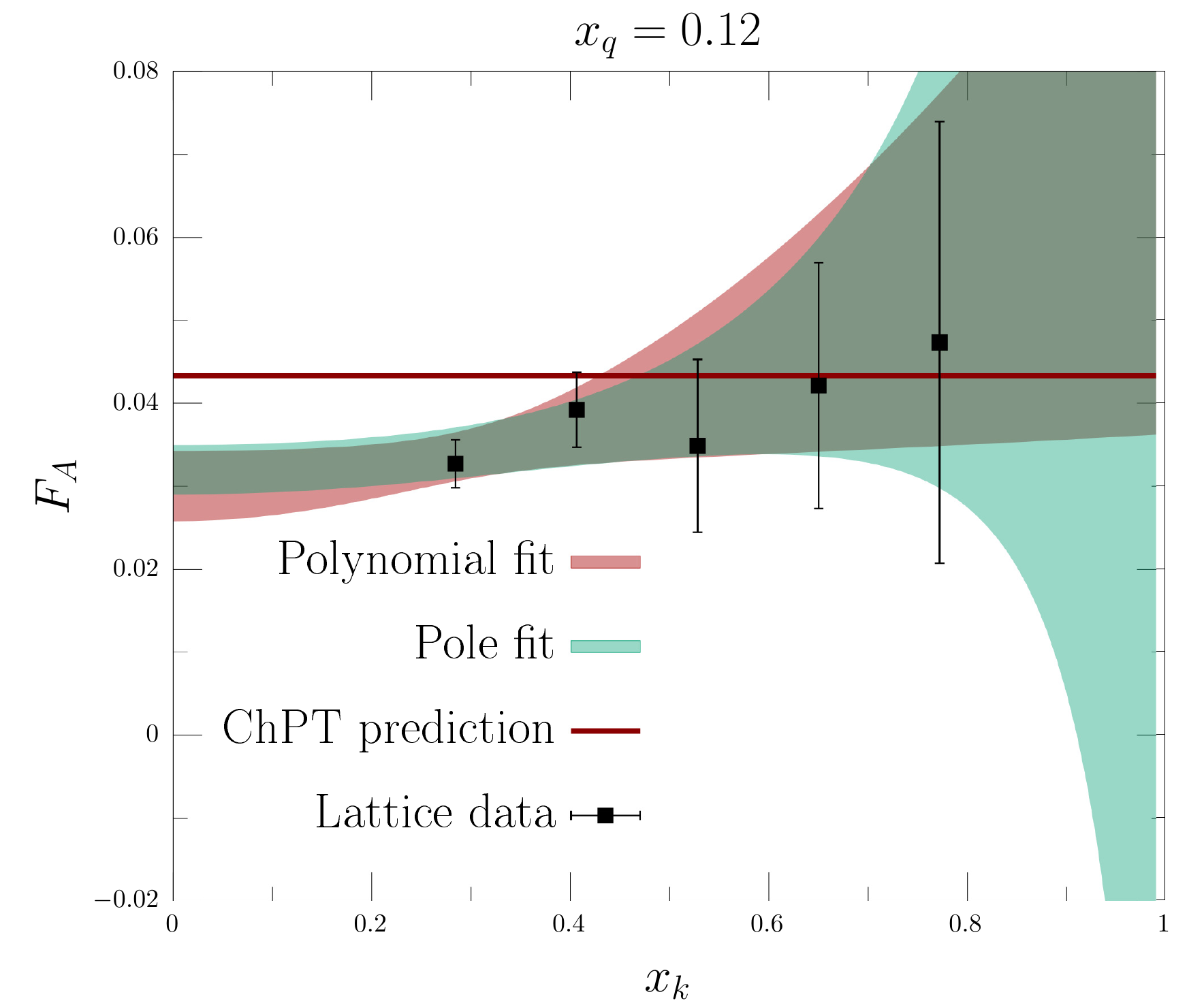}}
     \subfloat{%
       \includegraphics[scale=0.35]{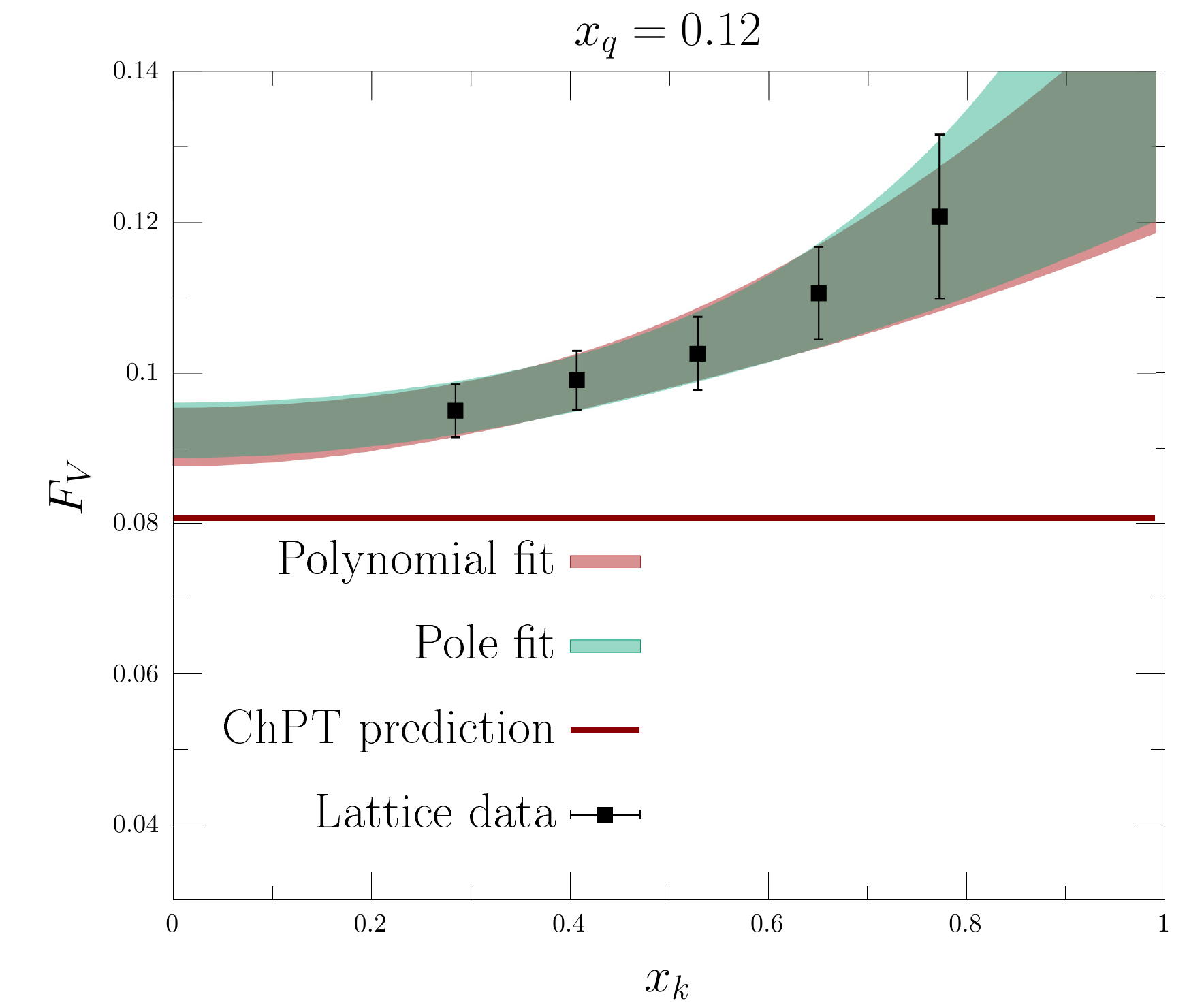}} \\
     \subfloat{%
       \includegraphics[scale=0.35]{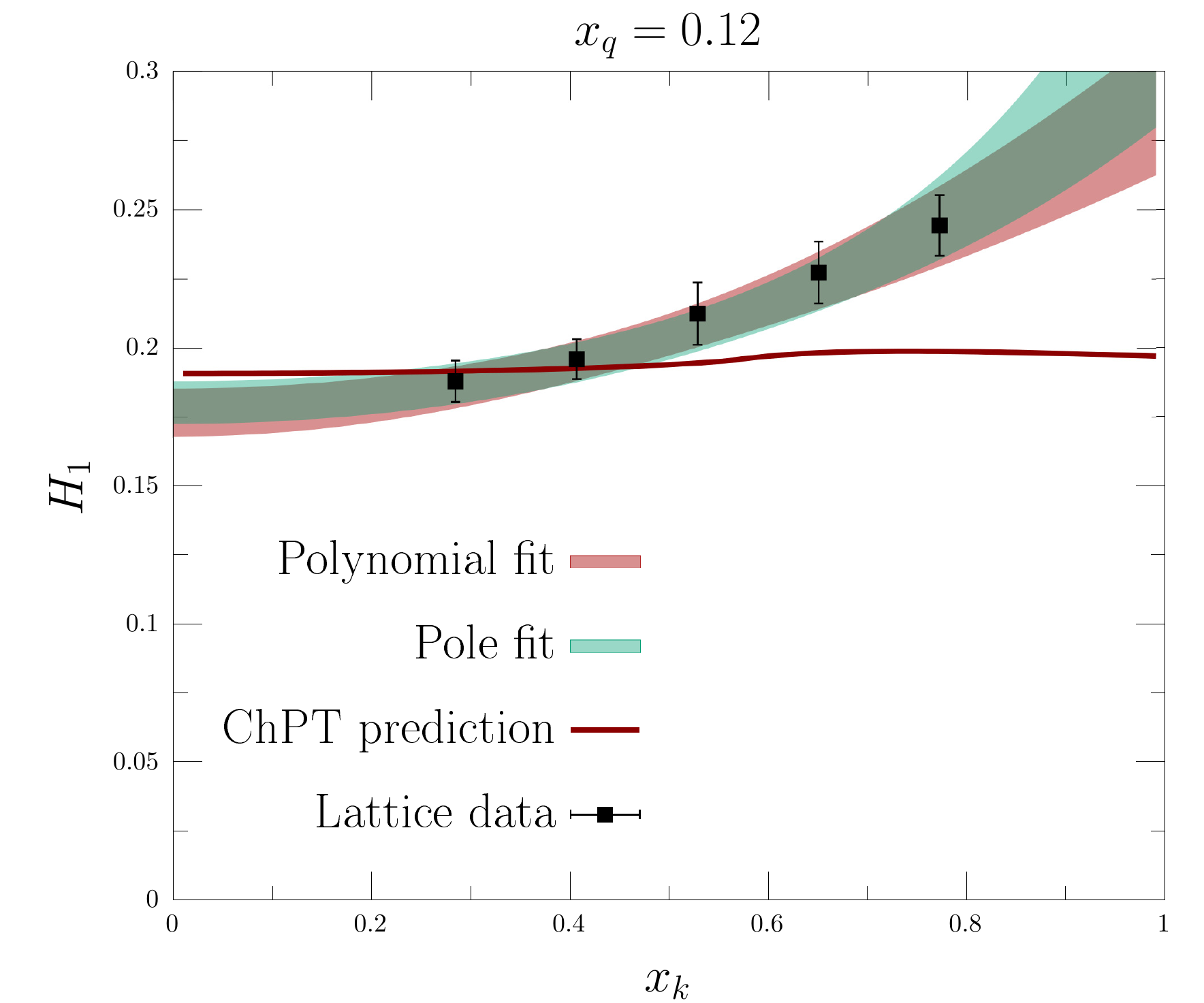}}
     \subfloat{%
       \includegraphics[scale=0.35]{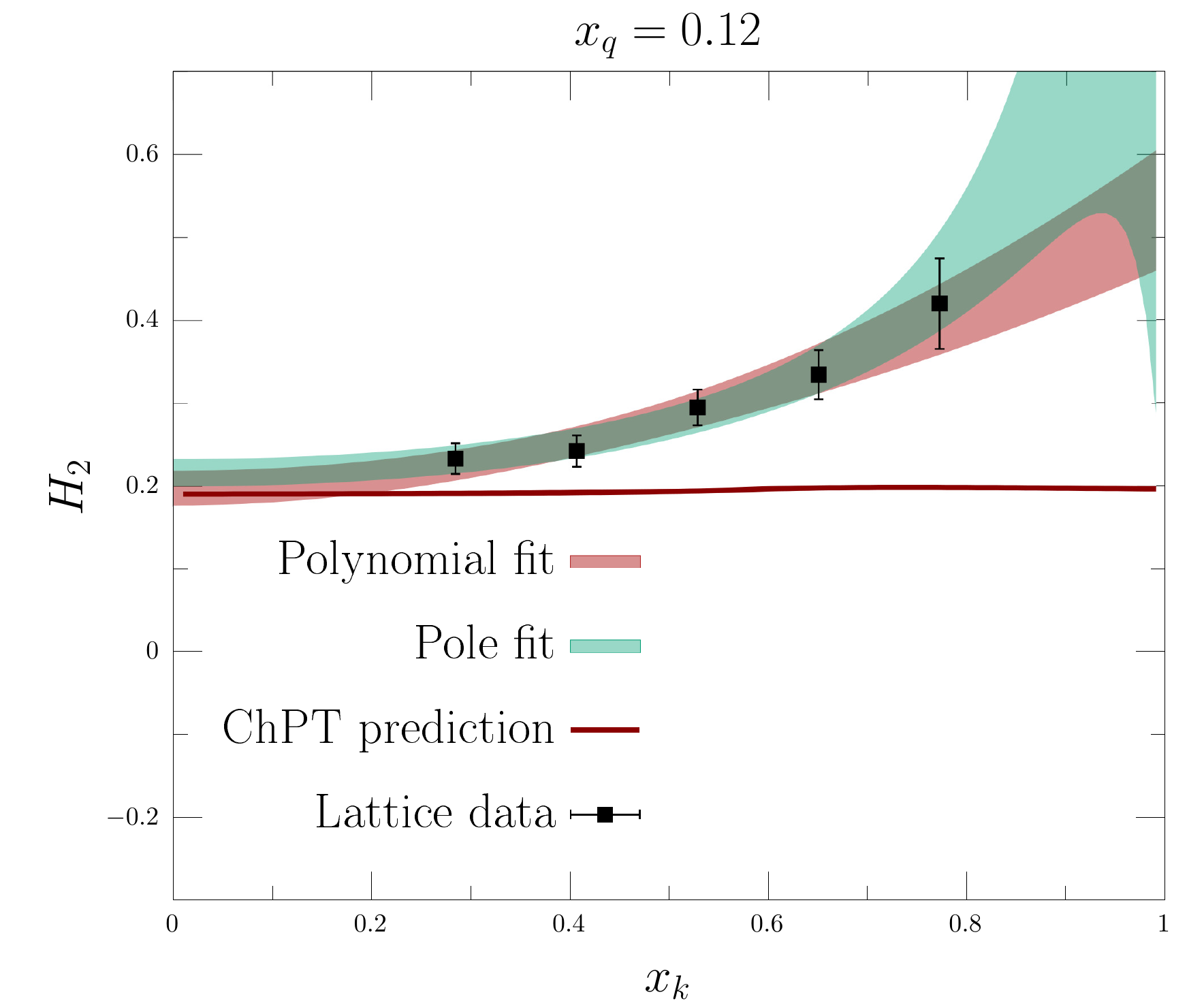}} \\
       \subfloat{%
       \includegraphics[scale=0.35]{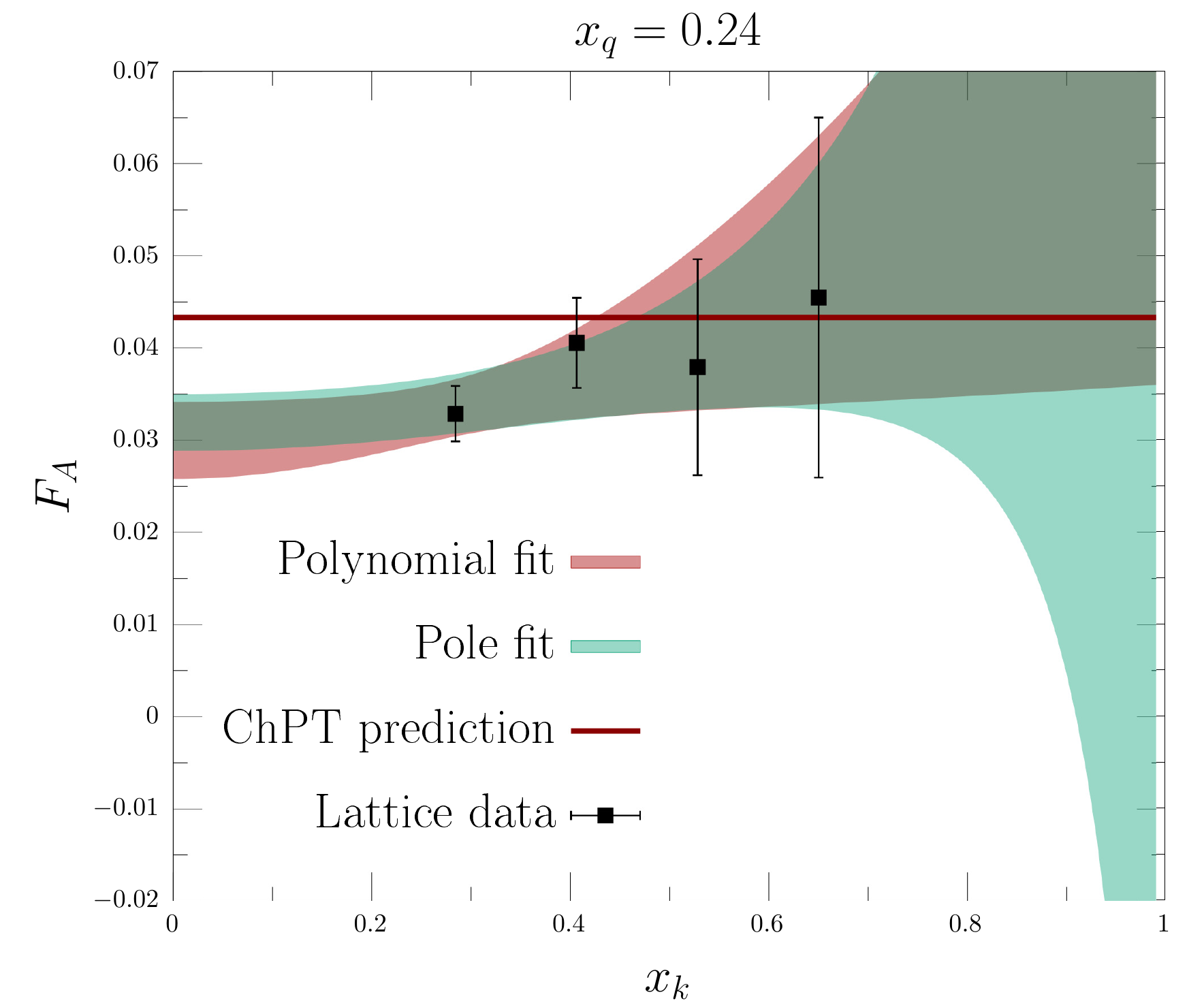}}
     \subfloat{%
       \includegraphics[scale=0.35]{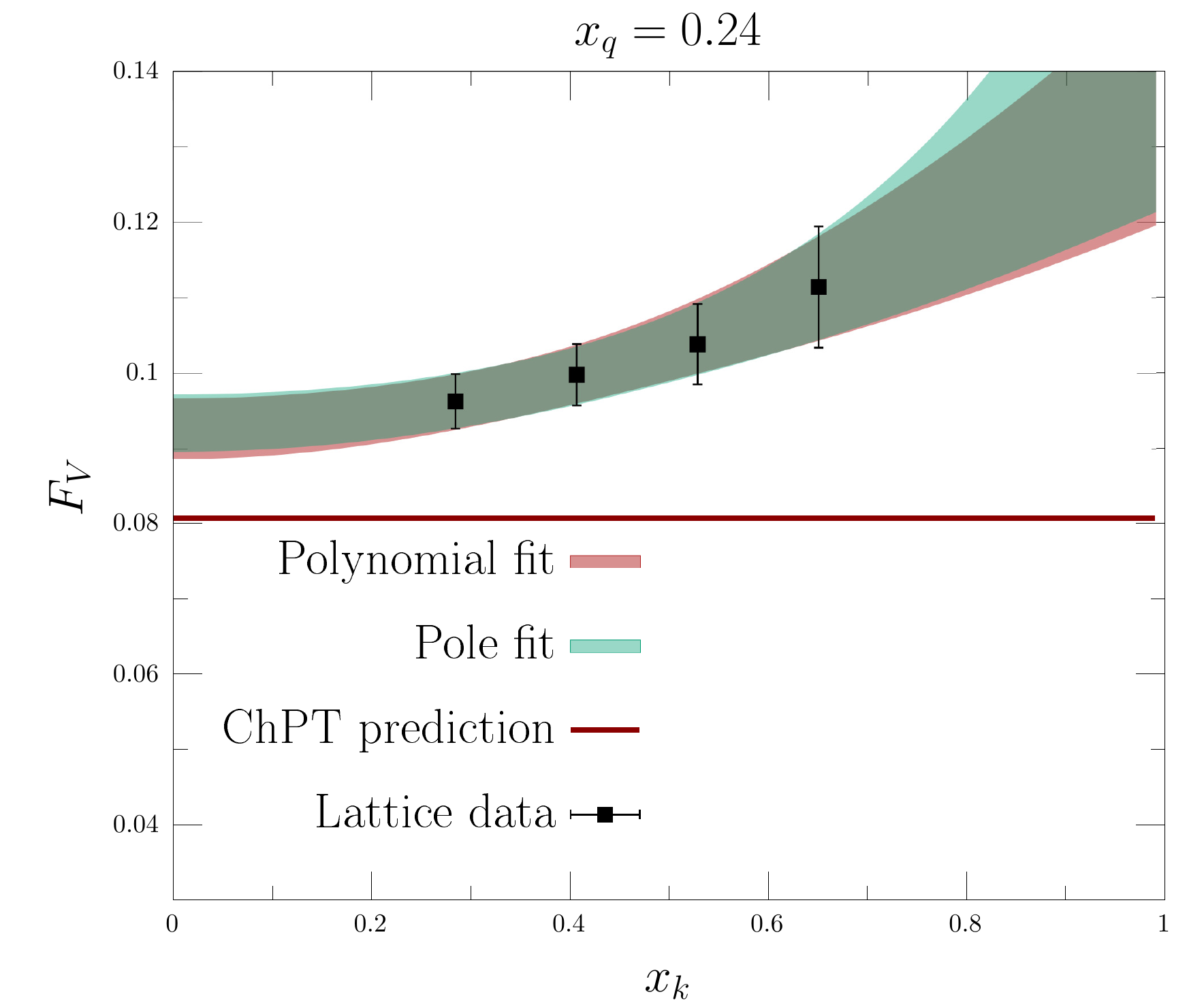}} \\
     \subfloat{%
       \includegraphics[scale=0.35]{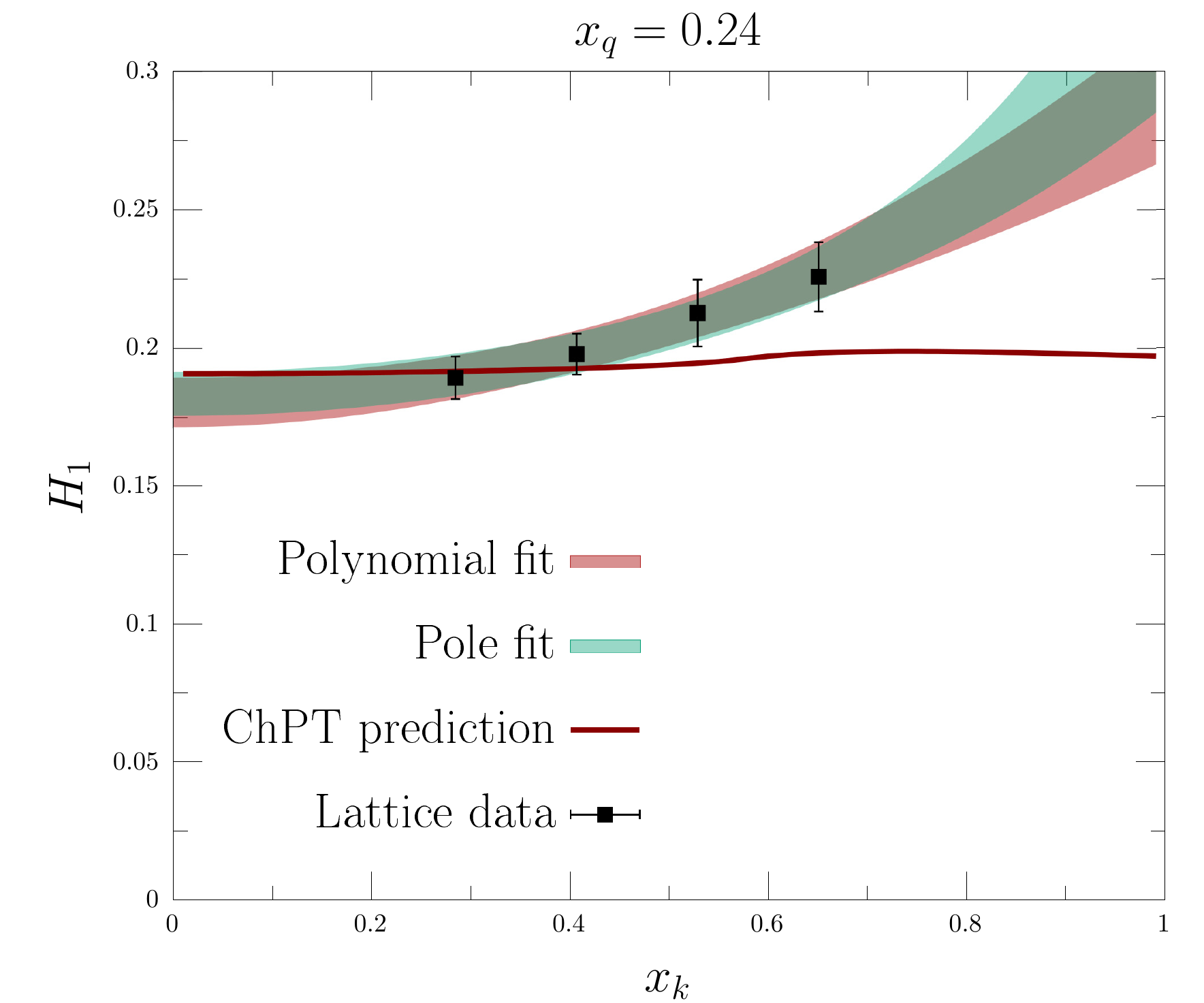}}
     \subfloat{%
       \includegraphics[scale=0.35]{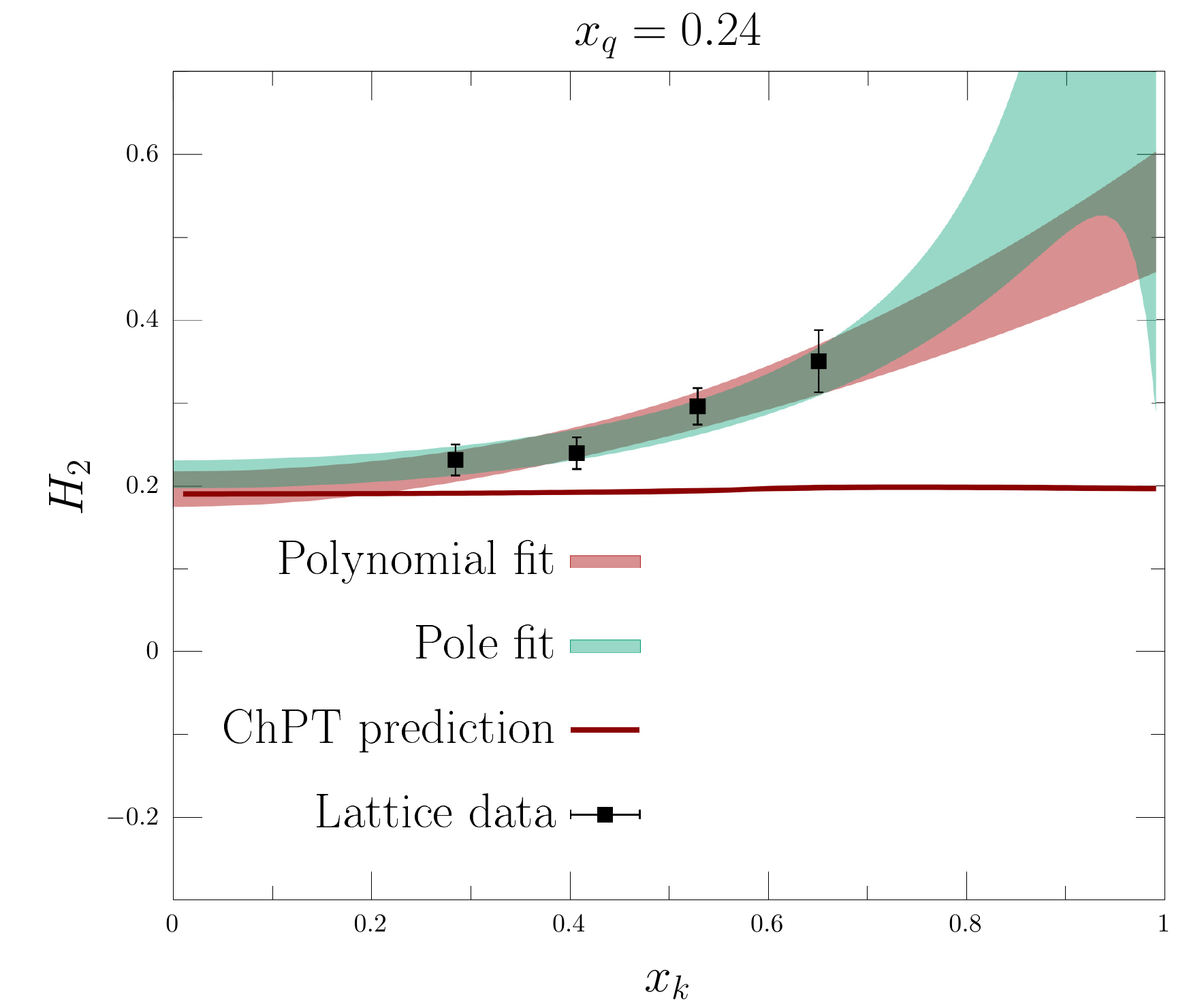}} 
       \caption{\it The fitting functions corresponding to the polynomial and the pole-like fits of Eqs.\,(\ref{poly}) and(\ref{pole}) are plotted, along with the lattice data, as function of $x_{k}$ and at a fixed value of $x_q=0.12$ (panels 1-4) and $x_q=0.24$ (panels 5-8). The red line corresponds to the 1-loop ChPT prediction with $F=f_K/\sqrt{2}$.}
	\label{xq_0.12_0.24}
\end{figure}

The second ansatz has a pole-like structure of the form
\begin{align}
F_{\mathrm{pole}}(x_k,x_q)=\frac{A}{\left(1-R_k x_k^2\right)\left(1-R_q x_q^2\right)}\label{pole}\,, 
\end{align}
where again $A, R_{k}$ and $R_{q}$ are free fitting parameters. The resulting fitting curves, along with the lattice data and with the ChPT prediction, are shown, for all four form factors $H_{1}, H_{2}, F_{A}$ and $F_{V}$, in the panels of Figs.\,\ref{xk_0.2845_0.4064}\,-\,\ref{xq_0.12_0.24} , as a function of $x_{q}$ at fixed $x_{k}$, and vice-versa. The quality of the fit in all cases is very good, with the reduced $\chi^{2}$ always smaller than one. The parameters of both the polynomial and pole-like fits are collected in Tab.\,\ref{fitpar}. 


The $1$-loop ChPT predictions for the form factors $H_{1}$, $H_{2}$, $F_{A}$ and $F_{V}$ of the $K^{+}$-meson read\,\cite{Bijnens:1994me}
\begin{eqnarray}
	\label{eq:chpt_pred}
	F_{V} &=& \frac{m_{K}}{4\sqrt{2}\pi^{2}F}\,, \qquad\qquad\qquad\qquad\qquad
	F_{A} = \frac{4\sqrt{2}\,m_{K}}{F}(L_{9}^{r}+L_{10}^{r})\,,\nonumber\\
	H_{1}(k^{2}) &=& 2f_{K}m_{K}\frac{\left( F_{V}^{K}(k^{2})-1  \right)}{k^{2}}\,, \qquad
	H_{2}(k^{2}) = 2f_{K}m_{K}\frac{\left( F_{V}^{K}(k^{2})-1  \right)}{k^{2}},
\end{eqnarray}
where $F$ is the ChPT leading-order  low-energy constant (LEC), while $L_{9}^{r}$ and $L_{10}^{r}$ are ChPT LECs at next-to-leading order.
The 1-loop ChPT prediction for the kaon electromagnetic form factor $F_{V}^{K}(k^{2})$ (which depends on $F$ and $L_{9}^{r}$) can be found in Ref.\,\cite{Bijnens:1994me}. In the following we will evaluate the ChPT predictions for the form factors using the physical charged kaon and pion masses and setting either $F=f_\pi/\sqrt{2}$ or $F=f_K/\sqrt{2}$, where $f_\pi$ and $f_K$ are the physical values of these decays constants. We label these two determinations as $\mathrm{ChPT}(f_\pi)$ and $\mathrm{ChPT}(f_K)$, respectively.
For the LECs $L_{9}^{r}$ and $L_{10}^{r}$, we use the values
\bea
L_{9}^{r} &=& 6.9\times 10^{-3}~\,\,\,\qquad L_{10}^{r}= -5.2\times 10^{-3}\,,
\eea
taken from~\cite{LEC}.
Notice that at this order, the $1$-loop prediction for the form factors does not depend on $q^{2}$. Moreover, the dependence on the virtuality $k^{2}$ is very mild as well, and only enters the prediction for the form factors $H_{1}$ and $H_{2}$, through the higher order corrections to the linear parametrization for the electromagnetic form factor
\begin{align}\label{eq:FVlinear}
	F_{V}^{K}(k^{2}) = 1+ \frac{\langle r_{K}^{2}\rangle}{6}k^{2} + \mathcal{O}(k^{4})\,,
\end{align}
where $\langle r_{K}^{2}\rangle$ is the kaon's mean-square radius.
From the figures it can be seen that our results are reasonably consistent with the ChPT prediction. However, note that, at NLO, ChPT does not include any momentum dependence of the form factors and so the comparison should be made with their values at $x_k=0$ and $x_q=0$, that is with the parameters $a_0$ or $A$. Moreover, even at zero momentum transfer,  on the one hand we do not expect the NLO ChPT prediction to be exact, while on the other hand our lattice estimates are affected by systematic errors, such as the continuum, chiral and infinite-volume extrapolations, that will be studied in the future. 


\begin{table}[h]
\begin{tabular}{|c|c|c|c|c|c|c|cl}
\hline
  \rule[-2mm]{0mm}{15pt}     & $a_0$        & $a_k$        & $a_q$          &  $A$       & $R_k$        & $R_q$        \\ \hline
 \rule[-2mm]{0mm}{15pt}$H_1$ & 0.1755(88) & 0.113(30)   & 0.086(24)    &  0.1792(78) & 0.453(88) & 0.40(10)   \\ \hline
 \rule[-2mm]{0mm}{15pt}$H_2$ & 0.199(21) & 0.341(84)  & -0.03(3)       & 0.217(17)  & 0.87(12)    & -0.2(2) \\ \hline
 \rule[-2mm]{0mm}{15pt}$F_A$ & 0.0300(43) & 0.04(4)  & 0.00(1)   & 0.0320(30) & 0.74(50)  & 0.0(3)  \\ \hline
 \rule[-2mm]{0mm}{15pt}$F_V$ & 0.0912(39)  & 0.044(18) & 0.0246(59)  & 0.0921(38) & 0.38(13)  & 0.233(49)  \\ \hline
\end{tabular}
\caption{\it Values of the fit parameters for all the form factors, as obtained from the polynomial and pole-like fits of Eqs.\,(\ref{poly}) and (\ref{pole}).}
	\label{fitpar}
\end{table}

In Ref.\,\cite{Poblaguev_2002}, a Vector Meson Dominance (VMD) ansatz has been used in order to describe the momentum behavior of the form factors $H_{1},\ F_{A},\ F_{V}$, and has been then used in order to reproduce the experimental data\footnote{The authors of Ref.\,\cite{Poblaguev_2002} assumed that $H_2$ only contributes through the kaon's electromagnetic form factor and neglected the other unknown SD contributions. These however, are suppressed in ChPT \cite{Bijnens:1994me}.}. Within the VMD framework, the momentum dependence of the form factors is assumed to be determined by the masses of the low-lying resonances created by the electromagnetic and  weak currents. In Ref.\,\cite{Poblaguev_2002}, for each of the three form factors, the fitting function has been taken to be of the form
\begin{align}
\label{eq:VMD}
F_{\mathrm{VMD}}\left(x_{k},x_{q}\right)=\frac{F(0,0)}{\left(1-x_{k}^{2}\, m_{K}^{2}/m_{\rho}^{2}\right)\left(1-x_{q}^2\, m^2_{K} /m_{K^{*}}^{2}\right)}\,,
\end{align}
where $F(0,0)$ is the only free fitting parameter. In Eq.\,(\ref{eq:VMD}), $m_{\rho}$ is the mass of the $\rho$ meson, while $m_{K^*}$ is the mass of the $K^{*}(1270)$ in the axial channel, and that of the $K^{*}(892)$ in the vector one. Thus, the VMD model corresponds to fixing, in the pole-like fit of Eq.~(\ref{pole}), $R_{k} = (m_{K}/m_{\rho})^{2}\simeq 0.4116$, and $R_{q} = (m_{K}/m_{K^{*}}(1270))^{2} \simeq 0.1513$ for the axial channel and $R_{q}= (m_{K}/m_{K^{*}}(892))^{2} \simeq 0.3064$ for the vector one. 

\begin{figure}[t]
\hspace{-0.5cm}
	\subfloat{%
		\includegraphics[scale=0.34]{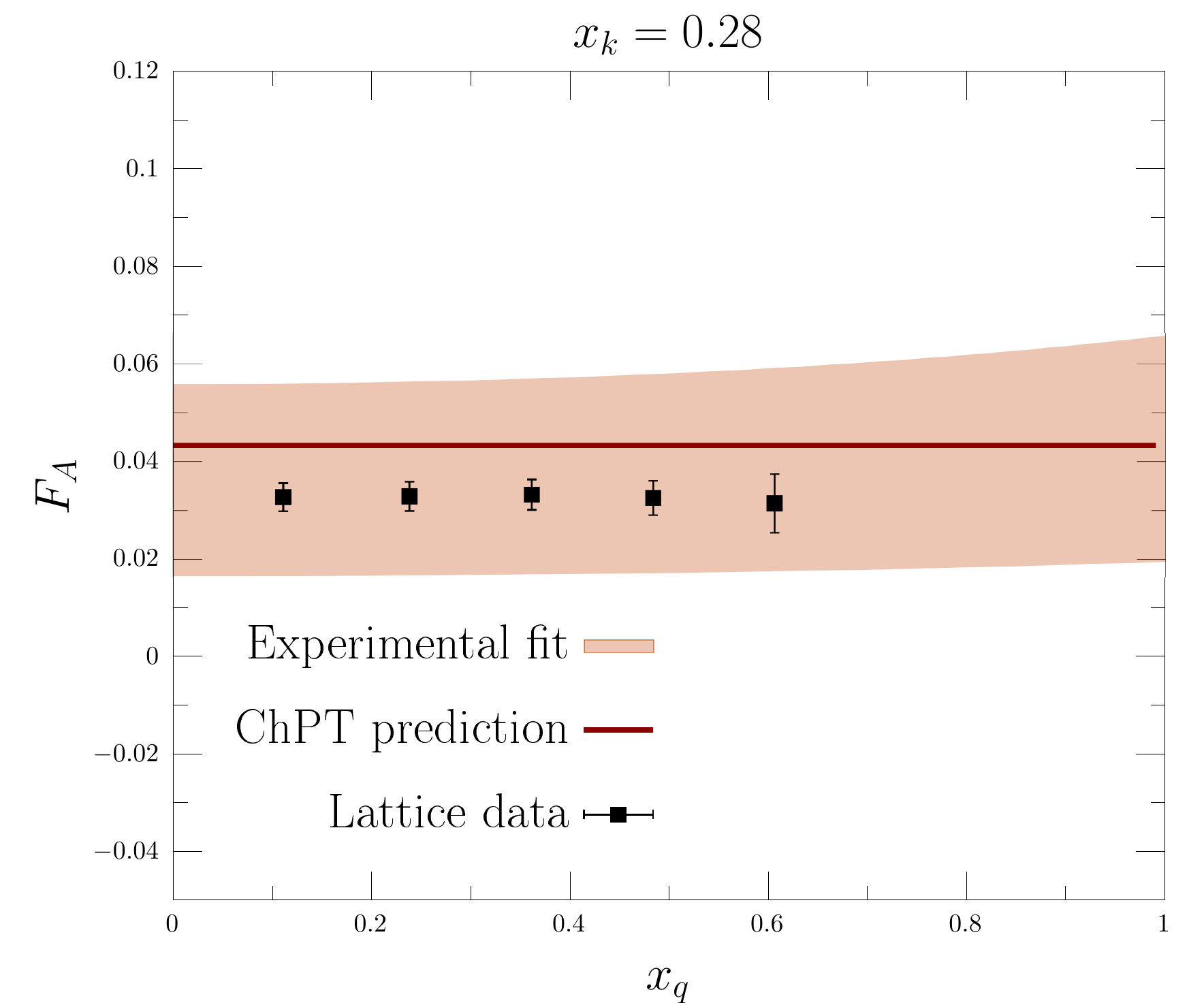}}
	\subfloat{%
		\includegraphics[scale=0.34]{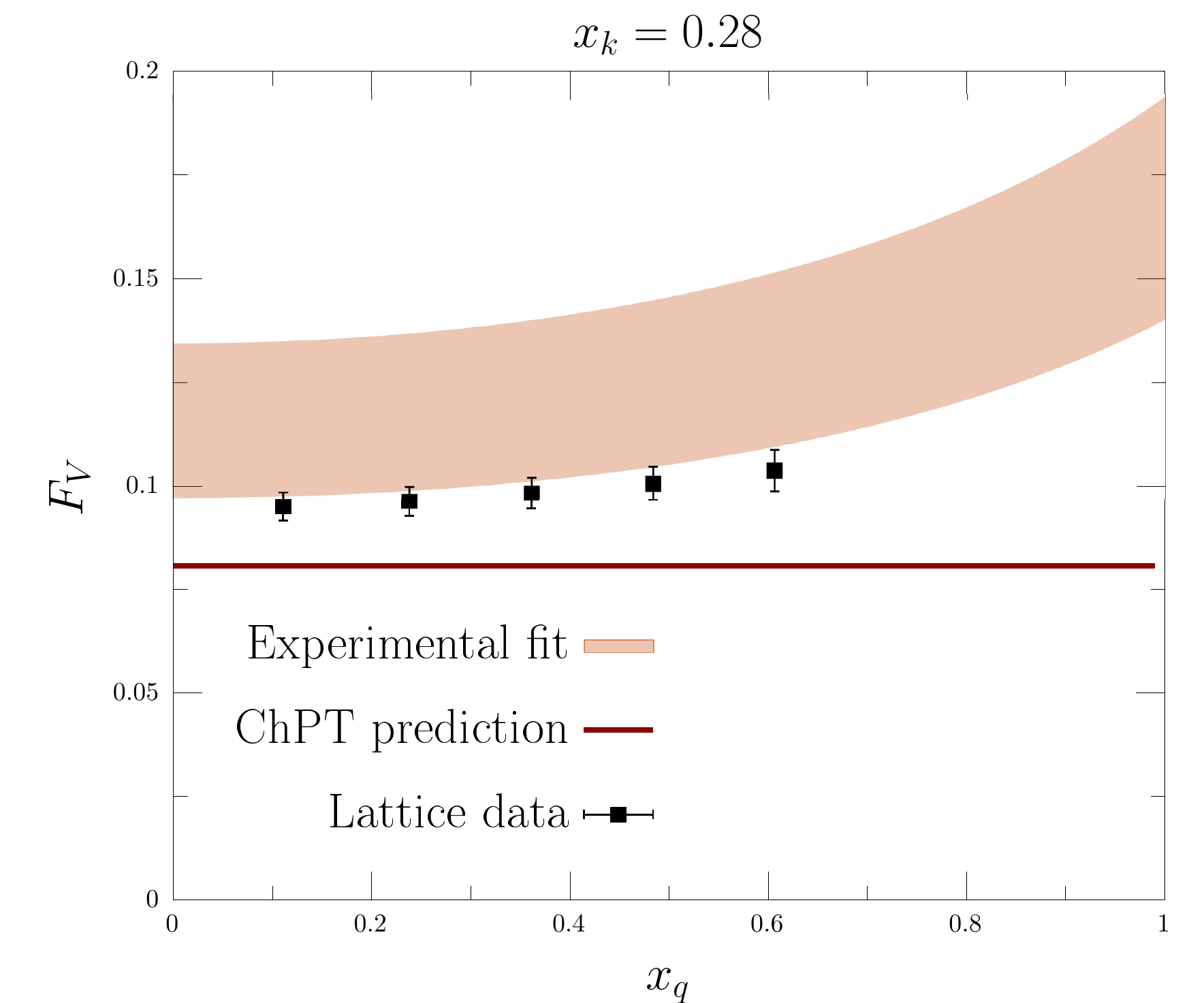}}
	\subfloat{%
		\includegraphics[scale=0.34]{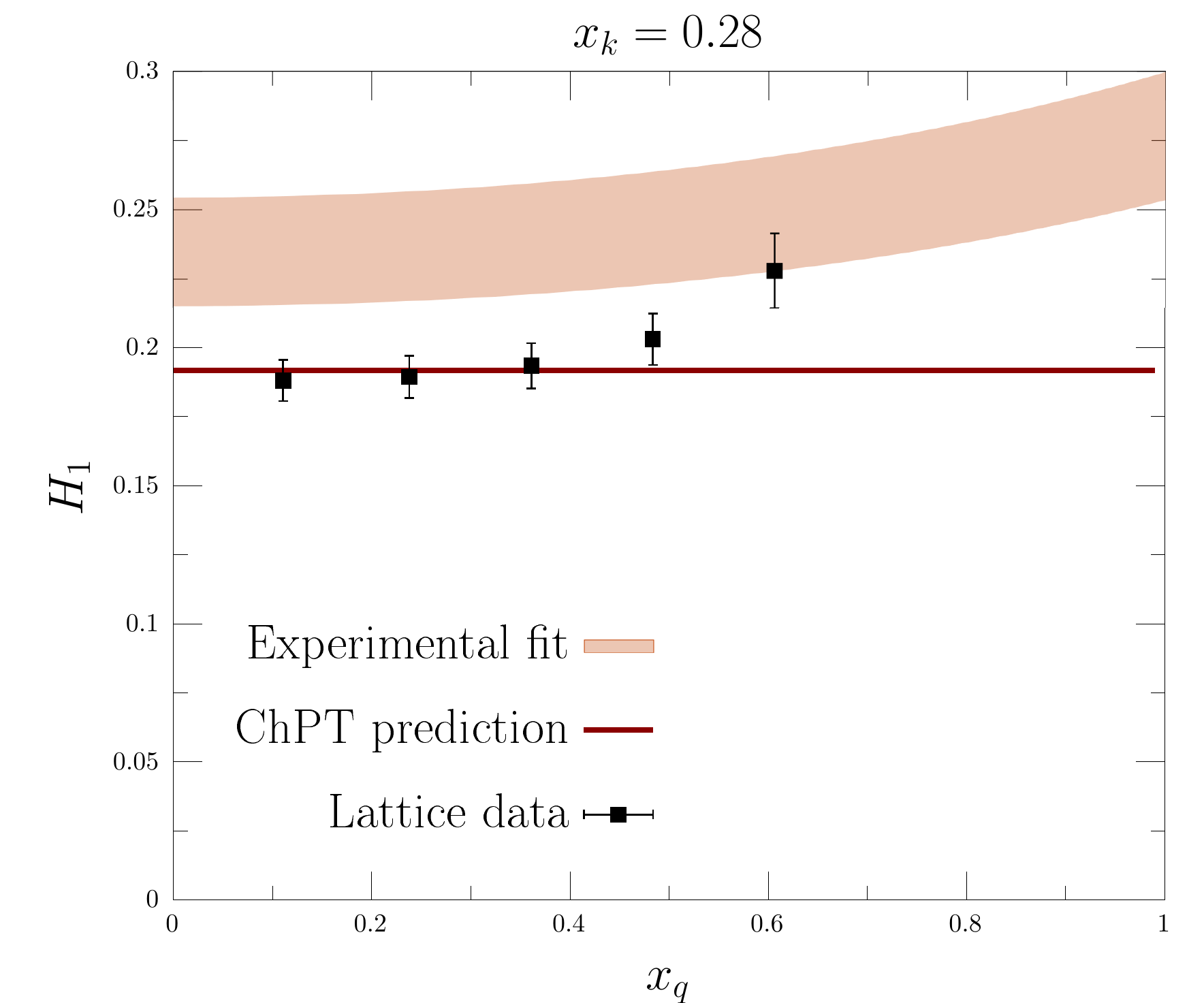}} \\
\hspace{-0.5cm}
	\subfloat{%
		\includegraphics[scale=0.34]{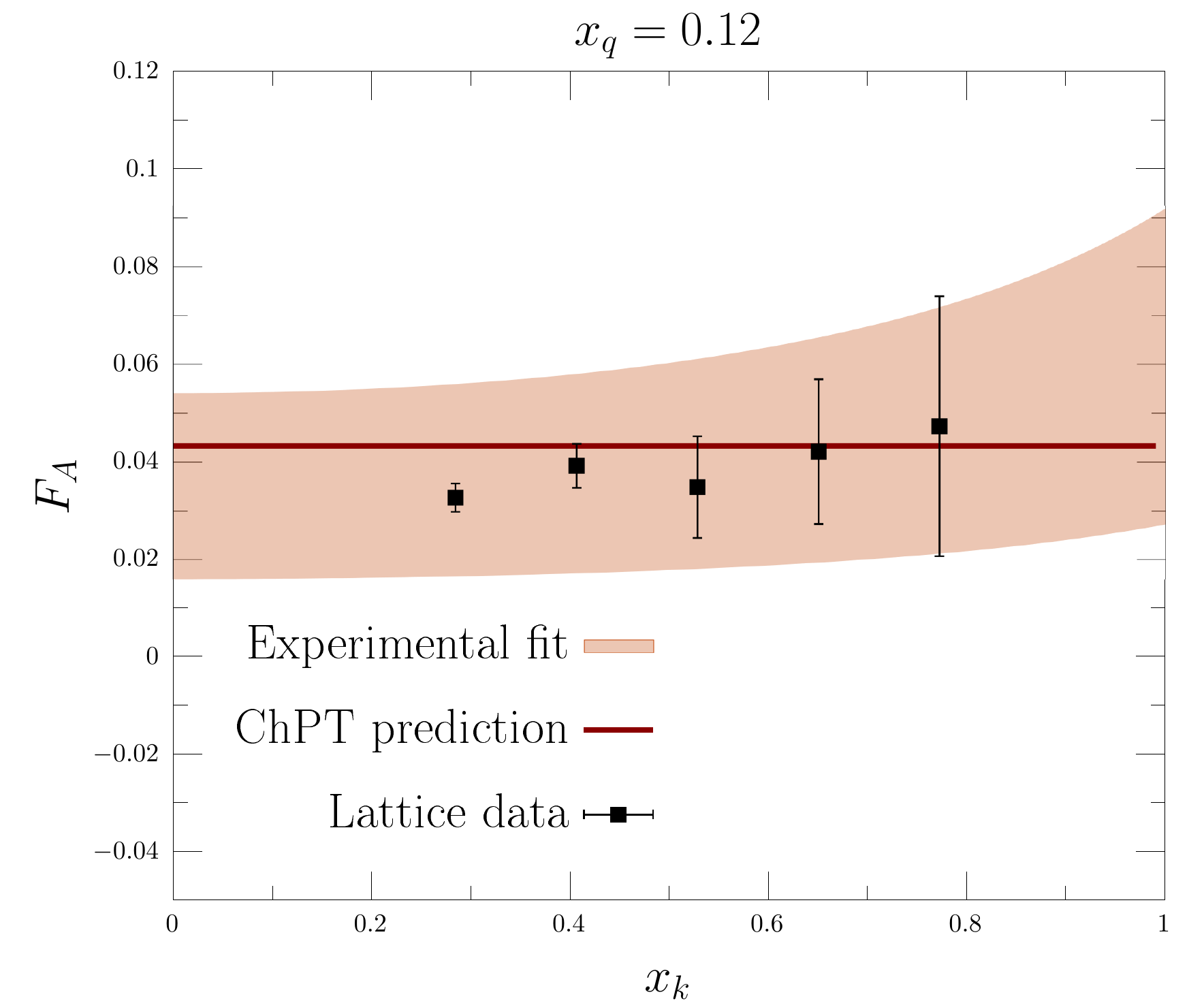}}
	\subfloat{%
		\includegraphics[scale=0.34]{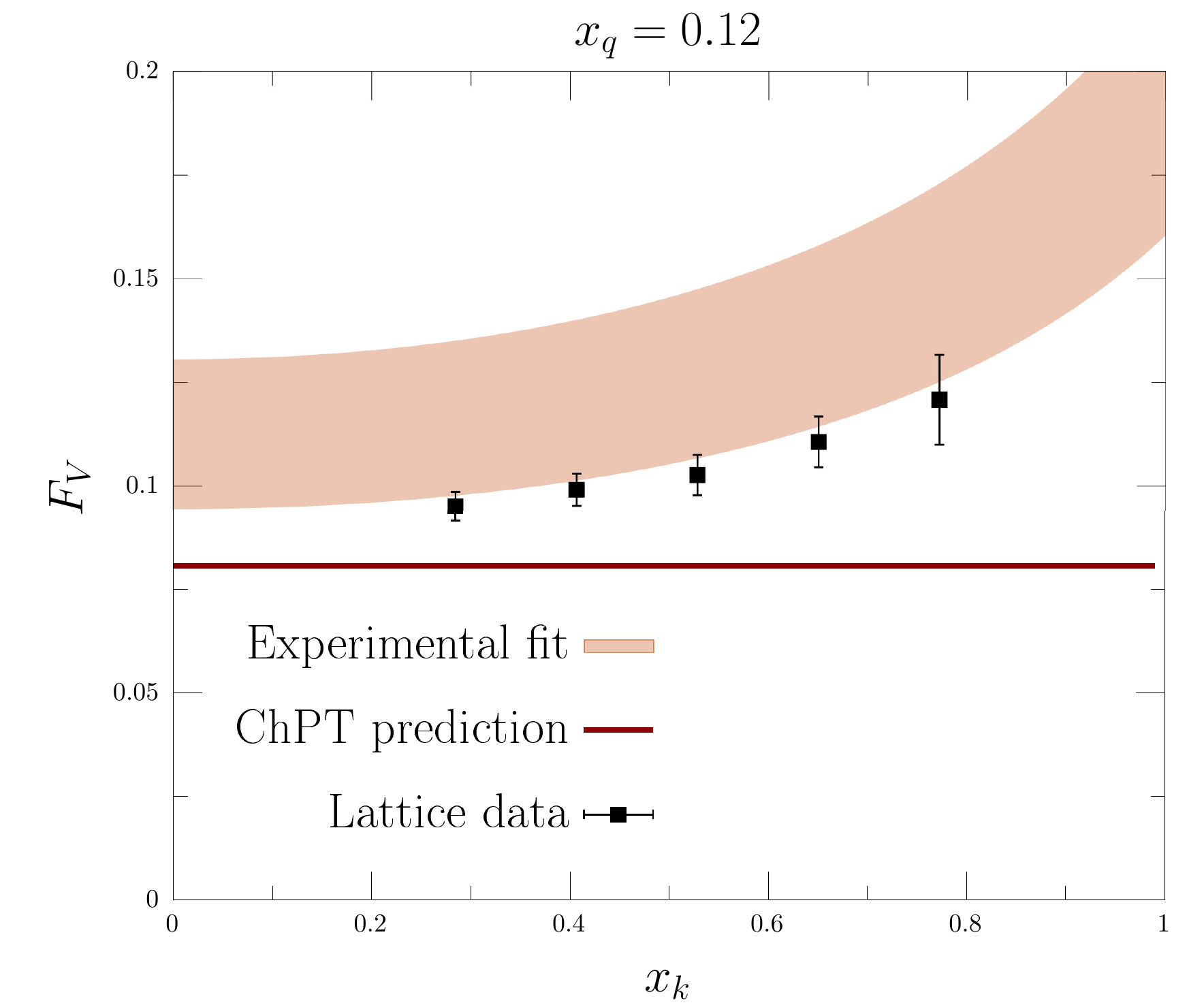}}
	\subfloat{%
		\includegraphics[scale=0.34]{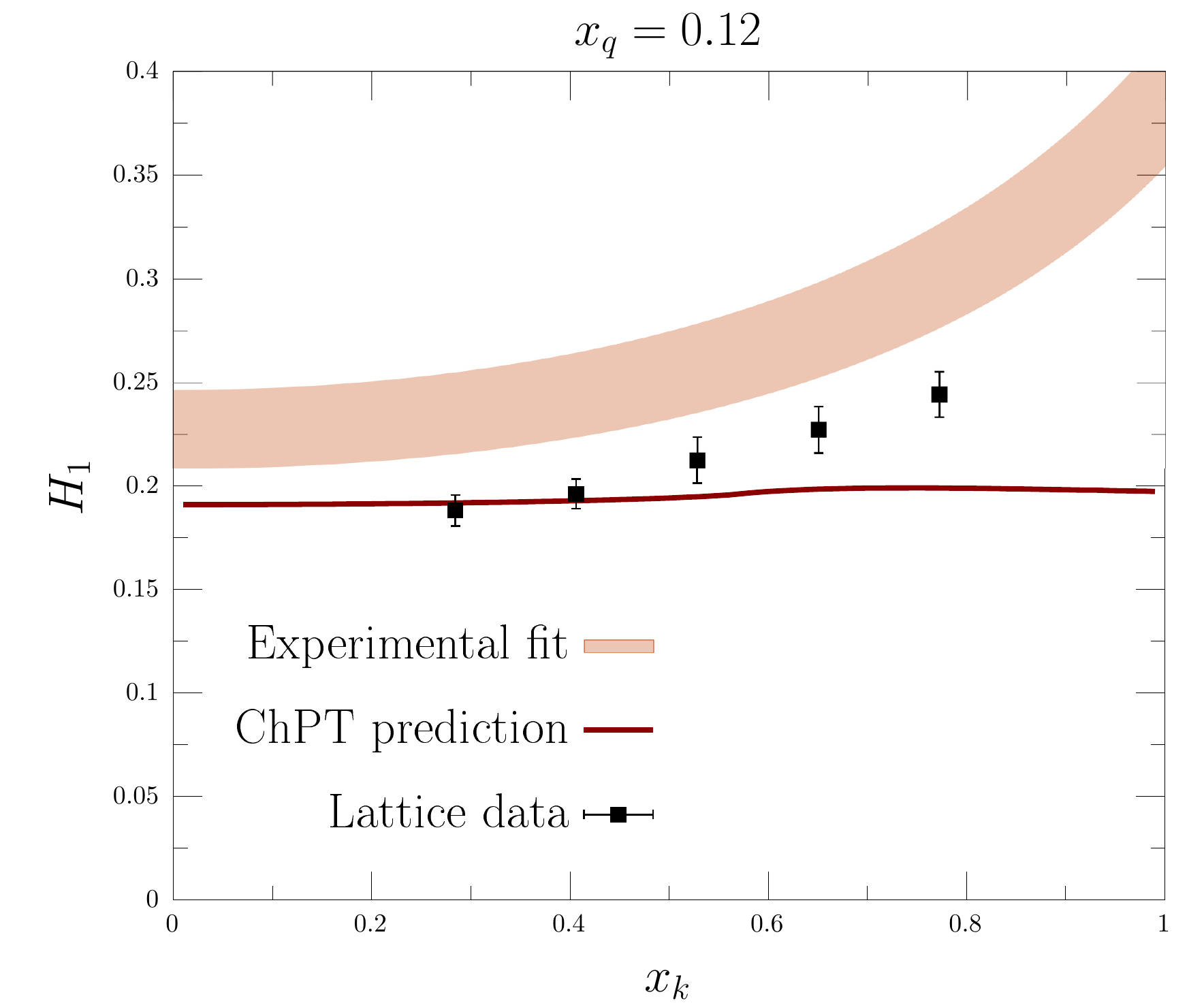}}
	\caption{\it The experimental VMD fits for the form factors $H_{1}, F_{A}$ and $F_{V}$ performed in \cite{Poblaguev_2002} are plotted, together with our lattice data, as a function of $x_{q}$ at a fixed value of $x_k=0.28$ (top) and as function of $x_{k}$ at a fixed value of $x_q=0.12$ (bottom). The red line corresponds to the 1-loop ChPT prediction with $F=f_K/\sqrt{2}$.}
	\label{vmd_xk_xq}
\end{figure}

In Tab.\,\ref{vmd_par} we compare the values of $F(0,0)$ values obtained from experiment assuming VMD and presented in Ref.\,\cite{Poblaguev_2002}, with the corresponding values of the form factors at zero $x_k$ and $x_q$ that we obtained from our lattice data using the pole fit of Eq.\,(\ref{pole}), i.e. the values for the parameter $A$ reported in Tab.\,\ref{fitpar}. In Fig.\,\ref{vmd_xk_xq} we compare our lattice data for the form factors $H_{1},\ F_{A},\ F_{V}$, with the result of the experimental VMD fits performed in \cite{Poblaguev_2002}.
Despite the systematic uncertainties affecting our lattice computation, the results are in reasonably good agreement. The largest discrepancy that we observe, which is of $\mathcal{O}(20\%)$, is for the form factor $H_{1}$.

We also notice  that the fitted values of $R_{k}$ and $R_{q}$ reported in Tab.\,\ref{fitpar}, are in qualitative agreement with the values predicted by VMD, albeit with large errors. In particular we are not able to see a clear $x_{k}$ and $x_{q}$ dependence in the lattice data for $F_{A}$, while the value of the fit parameter $R_{q}$,  for the form factor $H_{1}$, seems to be larger than the expectation based on VMD. Concerning the form factor $H_{2}$, the dominant contribution from the low-lying intermediate states comes from the virtual $K^{+}$ state created by the weak current ($|n_{f}\rangle= |K^{+}\rangle$ in Eq.~(\ref{eq:HmunuM1})). The resulting contribution $H_{2; K^{+}}$ to $H_{2}$ is then proportional to the electromagnetic kaon form factor $F_{V}^{K}(x_{k})$\,\cite{Bijnens:1994me}
\begin{align}
\label{H2_pred_Kplus}
H_{2; K^{+}} = \frac{ 2f_{K}}{m_{K}}~\frac{ F_{V}^{K}(x_{k}) - 1}{x_{k}^{2}}\,,
\end{align}
in agreement with the prediction from ChPT presented in Eq.\,(\ref{eq:chpt_pred}).
As shown in Tab.\,\ref{fitpar} and in Fig.\,\ref{xk_0.2845_0.4064}, we do not see any clear $x_{q}$ dependence in our lattice data for $H_{2}$ in agreement with the prediction of Eq.\,(\ref{H2_pred_Kplus})\footnote{This depends on the choice we had made in Eq.\,(\ref{Hmunu_sd}) for the kinematic prefactor in the definition of $H_2$ in the decomposition of the hadronic tensor in terms of form factors. If instead we had employed the same parametrization as in Refs.\,\cite{Bijnens:1994me} or \cite{Carrasco_2015}, we would have had a pole $1/(1-x_q^2)$ in the expression in Eq.\,(\ref{H2_pred_Kplus}).}. Moreover, making use of Eq.\,(\ref{eq:FVlinear})
one has that the contribution from the intermediate kaon to $H_{2}$ at $x_{k}=0$ is given by
\begin{align}
H_{2;K^{+}}(x_{k}=0) = f_{K}m_{K}\frac{ \langle r_{K}^{2}\rangle   }{3}~.
\end{align}
Using the value from the PDG, $\langle r_{K}^{2}\rangle = (0.560 \pm 0.031~{\rm fm})^{2}$\,\cite{pdg} and the physical values of the kaon mass and decay constant, one obtains $H_{2;K^{+}}(x_{k}=0) = 0.206(23)$, which nicely agrees with the value we obtained for the parameter $A$ in the pole-like fit of $H_{2}$ presented in Tab.\,\ref{fitpar}. Assuming the dominance of the rho-meson pole in the electromagnetic form factor $F_{V}^{K}$, one has that $H_{2} \propto 1/(1-R_{k}x_{k}^{2})$ with $R_{k} = (m_{K}/m_{\rho})^{2} \simeq 0.4116$. In this case, our fitted value of $R_{k}$ turns out to be larger than the value predicted by VMD. 

\begin{table}[]
\begin{tabular}{c|c|c|c}
& \rule[-2mm]{0mm}{15pt}$H_1(0,0)$ &  $F_A(0,0)$ & $F_V(0,0)$  \\ \hline
\rule[-2mm]{0mm}{15pt} This work  &~ $0.1792(78) $ ~& ~$0.0320(30)$~&~ $0.0921(38)$~ \\ \hline
\rule[-2mm]{0mm}{15pt}Experiment \cite{Poblaguev_2002} & $0.227(19)$ &  $0.035(19)$  & $0.112(18)$   \\ \hline
\end{tabular}
\caption{\it Comparison of the values of the VMD fit parameters $F(0,0)$ for the form factors $H_{1}, F_{A}$ and $F_{V}$ as obtained in Ref.\,\cite{Poblaguev_2002} with the lattice results from this work (using the pole fit in Eq.\,(\ref{pole})).\label{vmd_par}}
\end{table}

\begin{figure} 
    \centering
    \includegraphics[scale=0.5]{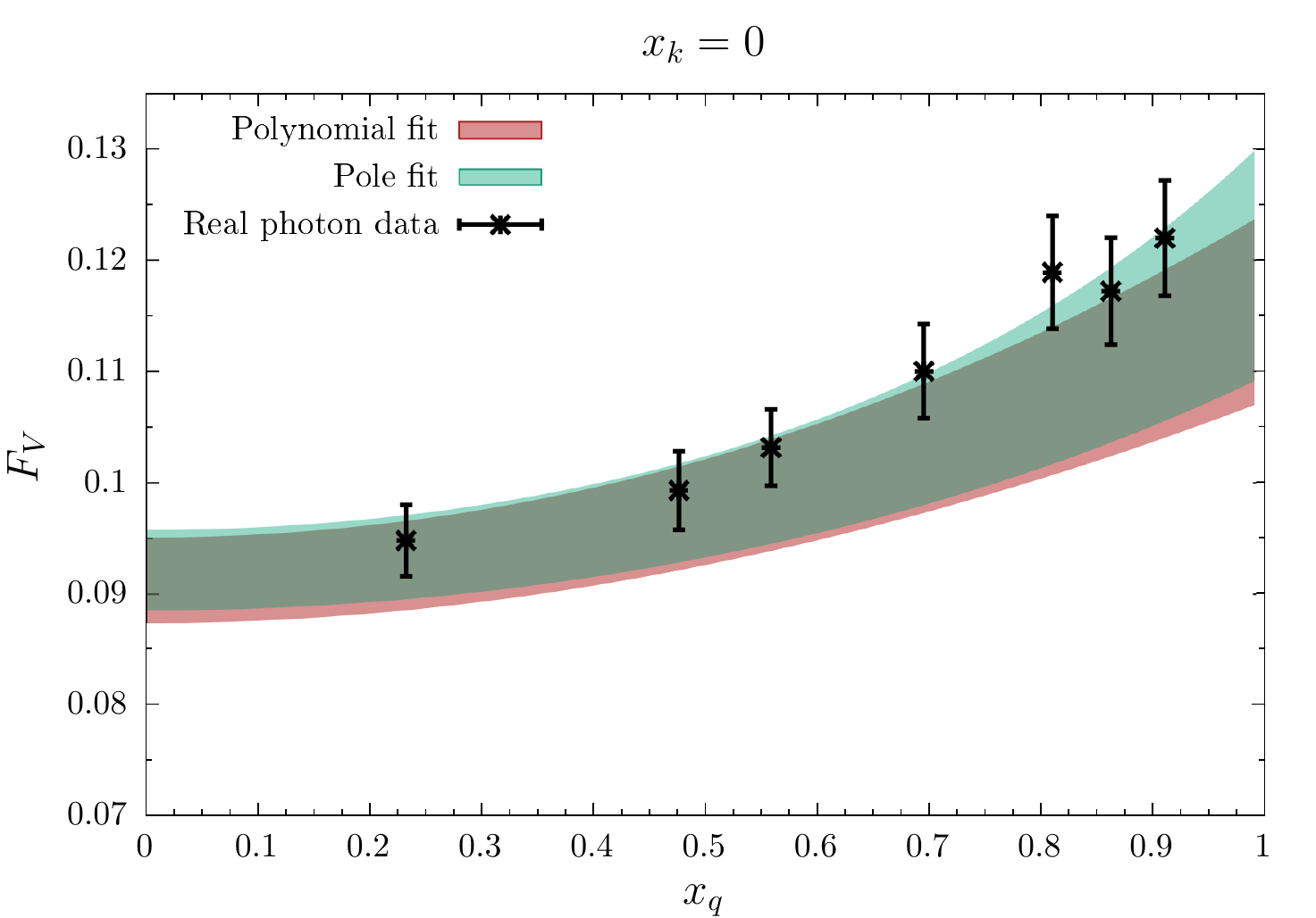}
\hspace{0.5cm}
    \includegraphics[scale=0.5]{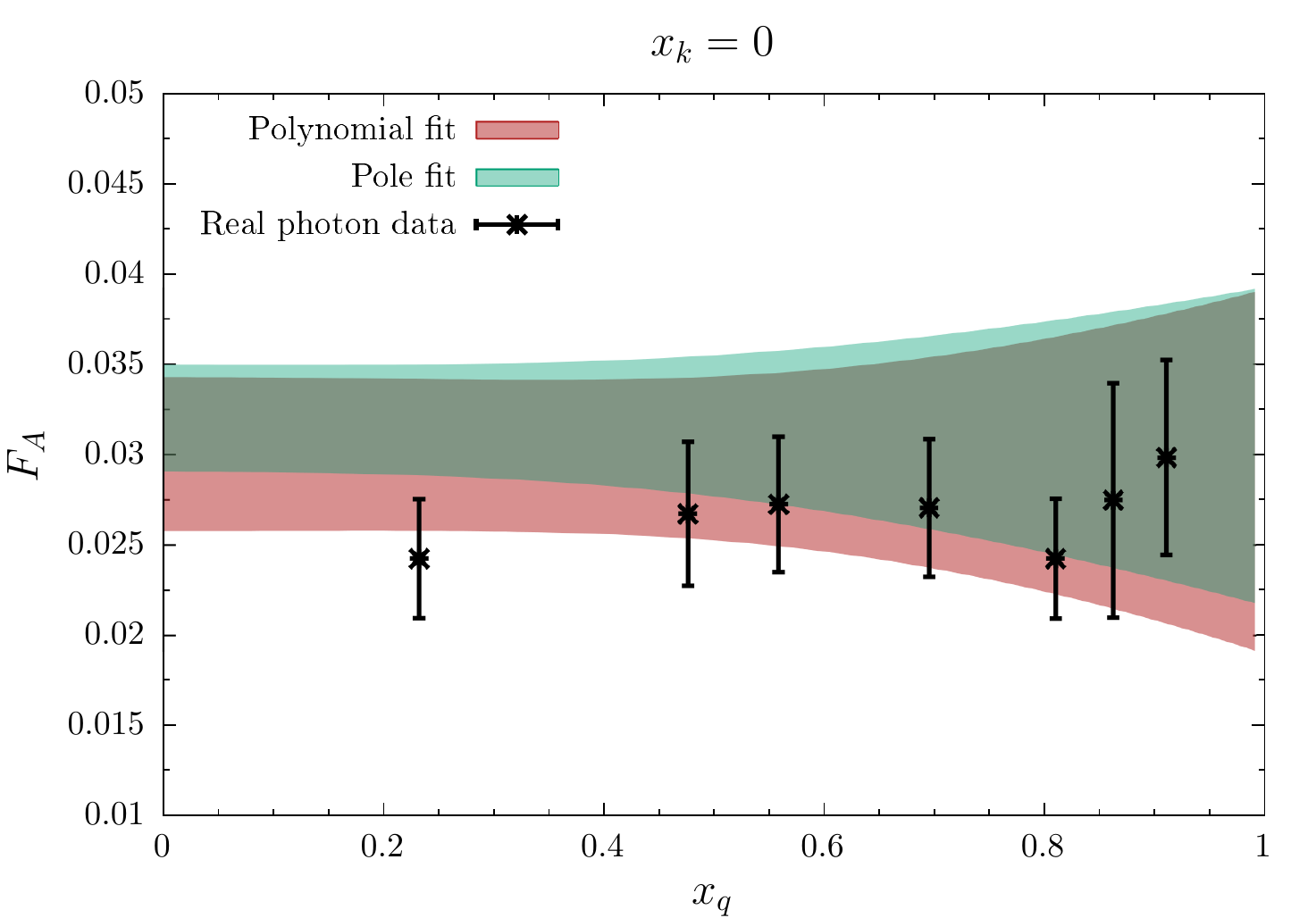}
    \caption{\it Extrapolation of our lattice results for $F_{V}$ (left) and $F_{A}$ (right) to $x_{k}=0$ using the polynomial and pole fit ansatzes defined in Eqs.\,(\ref{poly}) and~(\ref{pole}) (coloured bands). The black points correspond to the lattice results for $F_{V}$ and $F_{A}$ obtained directly at $x_{k}=0$ in our study of $K\to\ell\nu_\ell\gamma$ decays\,\cite{Desiderio2020}.}
    \label{fig:FVA_real_comp}
\end{figure}

We end this section by comparing the results for the form factors $F_V$ and $F_A$ obtained in this paper and extrapolated to $x_k=0$ using Eqs.\,(\ref{poly}) and (\ref{pole}) with those reported on the same configurations in our earlier paper with a real photon in the final state, i.e. for the decays $K\to\ell\nu_\ell\gamma$\,\footnote{The form factors $H_1$ and $H_2$ do not contribute to the amplitude for $K\to\ell\nu_\ell\gamma$ decays.}. The comparison is shown in Fig.\,\ref{fig:FVA_real_comp} and shows good agreement, in spite of the fact that the ansatzes and parameters in Eqs.\,(\ref{poly}) and (\ref{pole}) were obtained from fits to data with $x_k\ge 0.28$.

\section{\boldmath{$K^{+} \to \ell^{+} \nu_{\ell}\, \ell^{\prime\,+}\ell^{\prime\,-}$} decay rates}\label{brnum}
From the knowledge of the hadronic tensor $H^{\mu\nu}$, the $K^{+} \to \ell^{+} \nu_{\ell}\, \ell^{\prime+}\ell^{\prime-}$ decay rate is obtained by integrating the square of the unpolarised amplitude, $\sum_{\mathrm{spins}}|\mathcal{M}|^{2}$, over the phase space of the final-state charged leptons and neutrino. When the two positively-charged leptons are different, i.e. when $\ell\neq\ell^\prime$, the amplitude $\mathcal{M}$ is given by 
\begin{equation}
\label{l_not_lprime_amplitude}
\mathcal{M}(p_{\ell^{\prime+}},p_{\ell^{\prime-}},p_{\ell^+}, p_{\nu_{\ell}})  =-\frac{G_F}{\sqrt{2}}\,V^*_{us}\,\frac{e^2}{k^2}~
\bar{u}(p_{\ell^{\prime-}}\!)\gamma_\mu v(p_{\ell^{\prime+}}\!)~\Big[f_{K}\,\!L^\mu(p_{\ell^{\prime+}},p_{\ell^{\prime-}},p_{\ell^{+}}, p_{\nu_{\ell}})-H_{\mathrm{SD}}^{\mu\nu}(p,q)\,l_{\nu}(p_{\ell^{+}}, p_{\nu_{\ell}})\Big]\,,    
\end{equation}
where the leptonic vectors are given by
\begin{eqnarray}
L^\mu(p_{\ell^{\prime+}},p_{\ell^{\prime-}},p_{\ell^{+}}, p_{\nu_{\ell}})
  &=& m_\ell\,\bar{u}(p_{\nu_{\ell}})(1+\gamma_5)\left\{\frac{2p^\mu-k^\mu}{2p\cdot k-k^2}-\frac{2p_{\ell^{+}}^\mu+\slashed{k}\gamma^\mu}{2p_{\ell^{+}}\!\cdot k+k^2}\right\}v(p_{\ell^{+}})\,,\\[8pt]
l^{\mu}(p_{\ell^{+}}, p_{\nu_{\ell}}) &= &\bar{u}(p_{\nu_{\ell}})\gamma^\mu(1-\gamma_5)v(p_{\ell^{+}})\,.\label{eq:lmudef}
\end{eqnarray}
In Eqs.\,(\ref{l_not_lprime_amplitude})\,-\,(\ref{eq:lmudef}),
$p$ is the four-momentum of the kaon, $k=p_{\ell^{\prime+}}+p_{\ell^{\prime-}}$, and $q= p_{\ell^{+}}+ p_{\nu_{\ell}}$. In Eq.\,(\ref{l_not_lprime_amplitude}), the first term in the square parentheses gives the decay rate in the approximation in which the decaying kaon is treated as a point-like particle and includes the radiation from both the meson and charged lepton\,\footnote{This term is frequently referred to as the \textit{inner-brehmstrahlung} contribution.}.
Except for the 
kaon decay constant $f_{K}$, the non-perturbative contribution to the rate is entirely contained in the second term of Eq.\,(\ref{l_not_lprime_amplitude}).
The SD part of the hadronic tensor $H_{\mathrm{SD}}^{\mu\nu}$ is defined in Eq.\,(\ref{Hmunu_sd}). 

When $\ell=\ell^\prime$, since the final-state positively-charged leptons are indistinguishable, the exchange contribution, in which the momenta $p_{\ell^{\prime +}}$ and $p_{\ell^{+}}$ are interchanged, must be added to the amplitude $\mathcal{M}$ resulting in the replacement
\begin{equation}
\label{l_ugual_lprime_amplitude}
\mathcal{M}(p_{\ell^{\prime+}},p_{\ell^{\prime-}},p_{\ell^+}, p_{\nu_{\ell}})
\to \mathcal{M}(p_{\ell^{\prime+}},p_{\ell^{\prime-}},p_{\ell^+}, p_{\nu_{\ell}})-
\mathcal{M}(p_{\ell^{+}},p_{\ell^{\prime-}},p_{\ell^{\prime+}}, p_{\nu_{\ell}})\,.
\end{equation}

The branching ratio for $K^{+}\to l^{+}\,\nu_l\,l'^+\,l'^-$ decays is given by 
\begin{equation}
\textrm{BR}\left[K^+\to l^+\,\nu_l\,l'^+\,l'^-\right] 
=\frac{\mathcal{S}}{2m_K\Gamma_K (2\pi)^{8}}\int \sum_{\mathrm{spins}}|\mathcal{M}|^2~\delta\left(p-p_{l^{+}}-p_{\nu_{l}}-p_{l'^+}-p_{l'^-}\right)\frac{d^3p_{l^{+}}}{2E_{l^{+}}}\frac{d^3p_{\nu_{l}}}{2E_{\nu_{l}}}\frac{d^3p_{l'^+}}{2E_{l'^+}}\frac{d^3p_{l'^-}}{2E_{l'^-}}~,
\end{equation}
where $\Gamma_{K}= 5.3167(86)\times 10^{-17}\,{\rm GeV}$ is the total decay rate of the $K^{+}$ meson\,\cite{pdg} and $\mathcal{S}$ is a symmetry factor that takes the value $\mathcal{S}=1$ for $\ell \ne \ell^\prime$ and $\mathcal{S}= 1/2$ for $\ell = \ell^\prime$. Since the phase-space integration is considerably easier for the case $\ell \ne \ell^\prime$, in which a significant part of the integration can be performed analytically, we will discuss the two cases separately. 
\subsection{Decays with \boldmath{$\ell\neq \ell^\prime$}}\label{subsec:lnotlprime}
When the final state leptons have different flavours, the integral over the spatial momenta of the final-state particles can be partially performed analytically using invariance arguments and the fact that in $\sum_{\mathrm{spins}}|\mathcal{M}|^{2}$ the form factors only depend on $k^{2}= (p_{\ell^{\prime+}}+p_{l^{\prime-}})^{2}$ and $q^{2}= (p_{\ell^{+}}+ p_{\nu_{\ell}})^{2}$\,\cite{krishna}. This leads to the following simplified expression for the differential decay rate\,\cite{Bijnens:1994me}
\begin{equation}
\label{S16}
d\Gamma \left[K^+ \to \ell^+ \nu_{\ell} \ell^{\prime+}\ell^{\prime-}\right] =
\alpha^2 G_F^2 |V_{us}|^2 m_K^5\, 2x_{k}x_{q}G(x_k,r_{\ell^\prime})\, \bigg\{\!\!
   -\!\sum_{\mathrm{spins}} T_\mu^* T^\mu \bigg\}
    dx_k dx_q dy\,,
\end{equation}
where
\begin{eqnarray}
r_\ell = \frac{m_\ell^2}{m_K^2}\,,\qquad
r_{\ell^\prime}&=&\frac{m_{\ell^\prime}^2}{m_K^2}\,,\qquad
y = \frac{2p_\ell\cdot p}{m_K^2},\nonumber\\[10pt]
&& \hspace{-1.22in}G(x_k,r_{\ell^\prime})= \frac{1}{192 \pi^3 x_k^{2}} \left\{
     1 + \frac{2r_{\ell^\prime}}{x_k^{2}}\right\}
  \sqrt{1 - \frac{4r_{\ell^\prime}}{x_k^{2}}}\,,\label{eq:defs}
\\[10pt]
&&\hspace{-0.75in}T^\mu = \frac{\sqrt{2}}{m_K^2} \left\{
f_{K} L^\mu -
H_{SD}^{\mu\nu}l_\nu\right\}\,.\nonumber
\end{eqnarray}
The dimensionless integration variables $x_{k}$ and $x_{q}$ have been defined in Eq.\,(\ref{eq:xkqdef}). The integration domain is given by
\begin{align}
A - B ~\le~ ~ y ~ \le A+B~,
\end{align}
where
\begin{equation}
A=\frac{(2-x_{\gamma})(1+x_k^{2}+r_\ell -x_{\gamma})}{2 ( 1+x_k^{2}-x_{\gamma})},\qquad
B=\frac{(1+x_k^{2}-x_{\gamma}-r_\ell)\sqrt{x_{\gamma}^2 - 4 x_k^{2}}}{2(1+x_k^{2}-x_{\gamma})}
\end{equation}
\begin{equation}
x_{\gamma} \equiv \frac{2p\cdot k}{m_{K}^{2}} = 1+x_k^{2}-x_q^{2}\,,
\end{equation}
and the limits of integration for $x_{k}$ and $x_{q}$ are given in  Eq.\,(\ref{xkq_range}).
Since the form factors only depend on the invariant mass of the lepton-antilepton pair ($x_{k}m_{K}$) and on the invariant mass of the lepton-neutrino pair ($x_{q}m_{K}$), the integral over the variable $y$ can also be performed analytically, leaving the following expression for the double differential decay rate:
\begin{align}\label{Gamma}
 \frac{\partial^{2} }{\partial x_k\partial x_q}\Gamma\left[K^+ \to \ell^+ \nu_{\ell} \ell^{\prime+}\ell^{\prime-}\right]= \alpha^{2}G_{F}^{2}|V_{us}|^{2}m_{K}^{5}\left[ \Gamma^{\prime\prime}_{\mathrm{pt}}(x_k, x_q) + \Gamma^{\prime\prime}_{\mathrm{int}}(x_k, x_q) + \Gamma^{\prime\prime}_{\mathrm{SD}}(x_k, x_q)\right]\,. 
\end{align}
The differential rate is written as a sum of three different contributions. The first term, $\Gamma^{\prime\prime}_{\mathrm{pt}}(x_k, x_q)$, is the point-like contribution proportional to $f_{K}^{2}$ and gives the total differential decay rate in absence of any SD terms (i.e. if $H_{\mathrm{SD}}^{\mu\nu} = 0$).  The third term, $\Gamma^{\prime\prime}_{\mathrm{SD}}(x_k, x_q)$, is the contribution to the decay rate coming entirely from $H_{\mathrm{SD}}^{\mu\nu}$, and corresponds to a quadratic expression of the form factors $H_{1}, H_{2}, F_{A}, F_{V}$.
Finally, $\Gamma^{\prime\prime}_{\mathrm{int}}(x_k, x_q)$ is the interference term between the point-like and SD components of the amplitude. It arises from contributions of the form $H_{\mathrm{SD}}^{\mu\nu}L_{\mu}l_{\nu}$ in $T_{\mu}^{*}T^{\mu}$ and is  proportional to
$f_{K}$ and depends linearly on the form factors. 
Clearly, all the information from the internal structure of the kaon (i.e. from $H_{\mathrm{SD}}^{\mu\nu}$) is contained in
$\Gamma^{\prime\prime}_{\mathrm{int}}(x_k, x_q)$ and $\Gamma^{\prime\prime}_{\mathrm{SD}}(x_k, x_q)$.
The $\Gamma^{\prime\prime}$ functions are all dimensionless quantities which can be evaluated directly from the knowledge of the form factors and of the dimensionless ratio $f_{K}/m_{K}$, for which we use our lattice value $f_{K}/m_{K} =0.3057(11)$. Their explicit expressions in terms of $H_{1}, H_{2}, F_{A}, F_{V}$, and $f_{K}/m_{K}$ are presented in Appendix\,\ref{kernels}. Using these formulae and the form factors obtained from the polynomial and pole-like fits described in the previous section, we are able to evaluate each of the terms on the right-hand side of Eq.\,(\ref{Gamma}). In order to obtain the total decay rates, we rely on numerical integration using Gaussian quadrature rules.   \\

\begin{figure}
    \centering
    \includegraphics[scale=0.50]{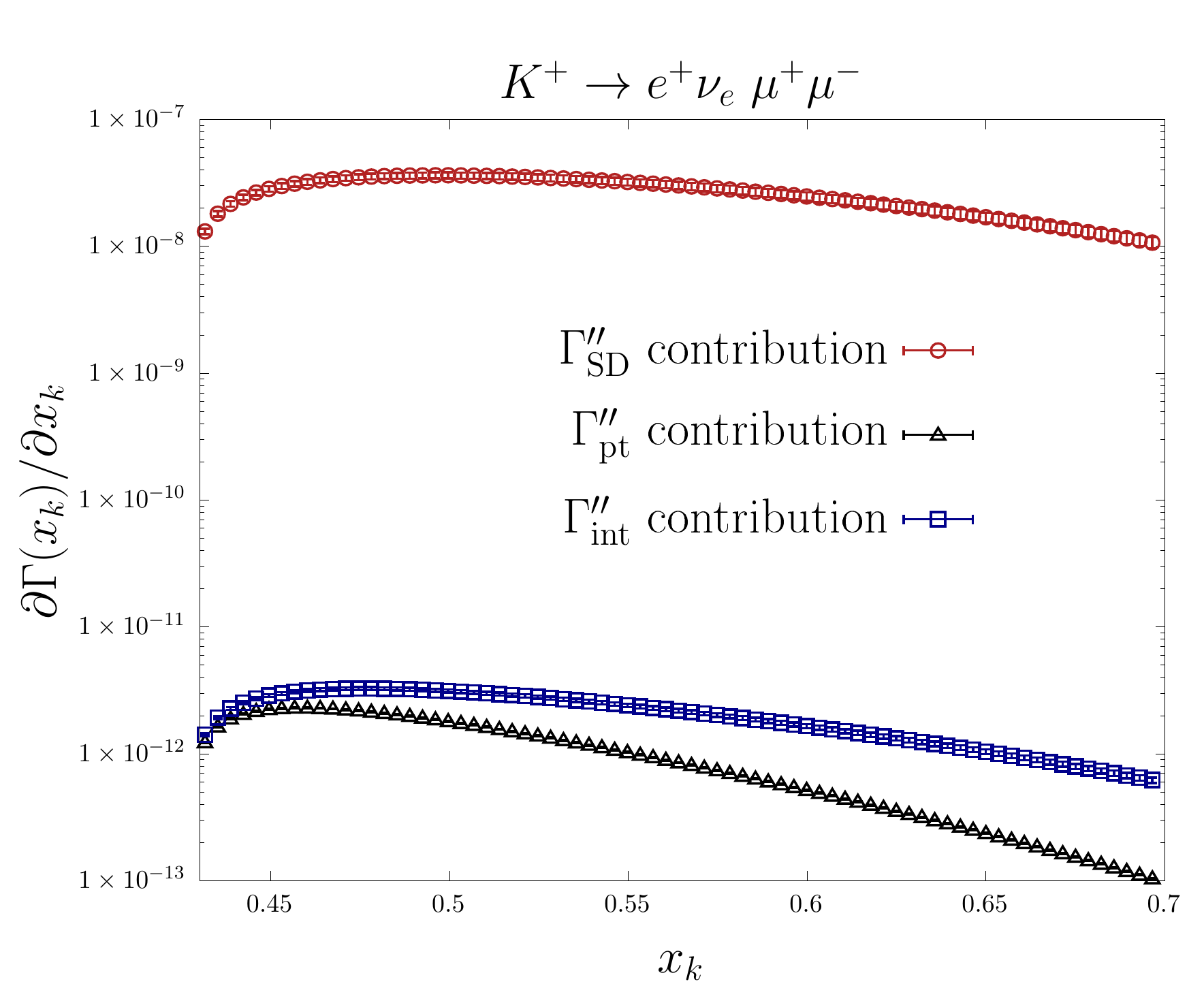}
    \caption{\it{The contributions from $\Gamma^{\prime\prime}_{\mathrm{pt}}$, $\Gamma^{\prime\prime}_{\mathrm{int}}$ and  $\Gamma^{\prime\prime}_{\mathrm{SD}}$ to the differential rate $\partial \Gamma (x_{k})/\partial x_{k}$, are shown for the decay channel $K^{+}\to e^{+}\nu_{e}\mu^{+}\mu^{-}$. Although not shown in the figure, all contributions to $\partial \Gamma (x_{k})/\partial x_{k}$ are zero at $x_{k}=2\sqrt{r_{\ell}} \simeq 0.4280$ but grow rapidly as $x_k$ is increased.}}
    \label{fig:diff_rate_mumu}
\end{figure}

For the decay $K^{+} \to e^{+}\nu_{e}\,\mu^{+}\mu^{-}$ the differential decay rate is completely dominated by the SD terms since the point-like contribution is helicity suppressed ($L_{\mu} \propto m_{e}$). This is shown in Fig.\,\ref{fig:diff_rate_mumu}, where we plot, as functions of $x_{k}$, the contributions from $\Gamma^{\prime\prime}_{\mathrm{pt}}$,  $\Gamma^{\prime\prime}_{\mathrm{int}}$  and from $\Gamma^{\prime\prime}_{\mathrm{SD}}$ to the 
partially-integrated differential decay rate $\partial \Gamma (x_{k})/\partial x_{k} = \int dx_{q}\, \partial^{2}\Gamma/\partial x_{k}\partial x_{q}$. 
Furthermore, we find that the dominant term in the integral of $\Gamma^{\prime\prime}_{\mathrm{SD}}$ is that proportional to $H_{1}^{2}$, while the contribution to the rate from the form factor $H_{2}$ turns out to be negligible. 
The remaining linear and quadratic terms in the form factors give subdominant contributions to the branching ratio of about 5\% in total.
Integrating the double differential decay rate of Eq.~(\ref{Gamma}), we obtain the following value for the branching ratio
\begin{align}\label{eq:epnueepem}
 \textrm{BR}\left[K^+\to e^+\,\nu_{e}\,\mu^+\,\mu^-\right]= 0.762~(49) \times 10^{-8}\,.
 \end{align}
In Tab.\,\ref{tab_Br1} we compare this result with the recent lattice value from Ref.\,\cite{xu}, with the predictions from ChPT  and with the measurement from the E865 experiment at the Brookhaven AGS\,\cite{Ma_2006}. As the table shows, our value of the branching ratio is in tension with the experimental measurement at the level of about $2\sigma$ and is a little smaller than the determination of Ref.\,\cite{xu}. However, it should be noted that both our computation and that of Ref.\,\cite{xu} are limited to a single value of the lattice spacing, a single volume and to unphysically large light-quark masses. Given that the branching ratio is dominated by the quadratic term proportional to $H_{1}^{2}$ an increase of about 25\% in the value of $H_{1}$, due to the missing continuum, chiral and infinite-volume extrapolations, would reduce the tension between our result and the experimental measurement to about $1\sigma$. 
It will be very interesting in the future, once these extrapolations have been performed, to learn whether $H_1$ does indeed increase. 
We also not however, that there is a $1.6\sigma$ difference between the values of $H_1(0,0)$ obtained in Refs.\,\cite{Poblaguev_2002} and \cite{Ma_2006}.
The value of $H_1(0,0)$ deduced by the E865 collaboration from the experimental study of the decay $K^+\to e^+\nu_e\,\mu^+\mu^-$ is $H_1(0,0)=0.303\pm 0.043$\,\cite{Ma_2006}\,\footnote{We have combined the different errors quoted in Eq.\,(7) of Ref.\,\cite{Ma_2006} in quadrature.}. This value is somewhat higher than that obtained, from studies of the decays $K^+\to e^+\nu_e e^+e^-$ and $K^+\to\mu^+\nu_\mu e^+e^-$ also in  the E865 experiment,
$H_1(0,0)=0.227\pm 0.019$\,\cite{Poblaguev_2002}, which is quoted in Tab.\,\ref{vmd_par}.


\begin{table}[]
    \centering
    \begin{tabular}{c|c| c| c| c| c}
    \hline
     \multicolumn{6}{c}{\rule[-2mm]{0mm}{15pt}$\textrm{BR}\left[K^+\to e^+\,\nu_{e}\,\mu^+\,\mu^-\right]$}  \\
     \hline 
     \rule[-2mm]{0mm}{15pt}This work & point-like approximation & Tuo et al.~\cite{xu} & ChPT($f_\pi$) & ChPT($f_{K}$) & experiment~\cite{Ma_2006} \\
     \hline
      \rule[-2mm]{0mm}{15pt} $0.762(49)\times 10^{-8}$ & $3.0\times 10^{-13}$ &  $0.94(8) \times 10^{-8}$ & $1.19\times 10^{-8}$ & $ 0.62\times 10^{-8}$  & $1.72(45)\times 10^{-8}$ \\
    \hline 
    \end{tabular}
    \caption{\it{Comparison of our result for the branching ratio $\textrm{BR}\left[K^+\to e^+\,\nu_{e}\,\mu^+\,\mu^-\right]$ with the one coming from the point-like approximation, the result from Ref.\,\cite{xu} and with the results for the branching ratio obtained using the NLO ChPT predictions for the SD form factors (Eq.~(\ref{eq:chpt_pred})) setting either $F=f_{\pi}/\sqrt{2}$ or 
    $F=f_{K}/\sqrt{2}$ (denoted by ChPT($f_\pi$) and ChPT($f_K$) respectively). In the last column we show the experimental result from the E865 experiment~\cite{Ma_2006}. We stress that both our lattice result and that from Ref.\,\cite{xu} are affected by systematic uncertainties due to the missing chiral, continuum and infinite-volume extrapolations.}}
    \label{tab_Br1}
\end{table}

\begin{figure}
    \centering
    \includegraphics[scale=0.50]{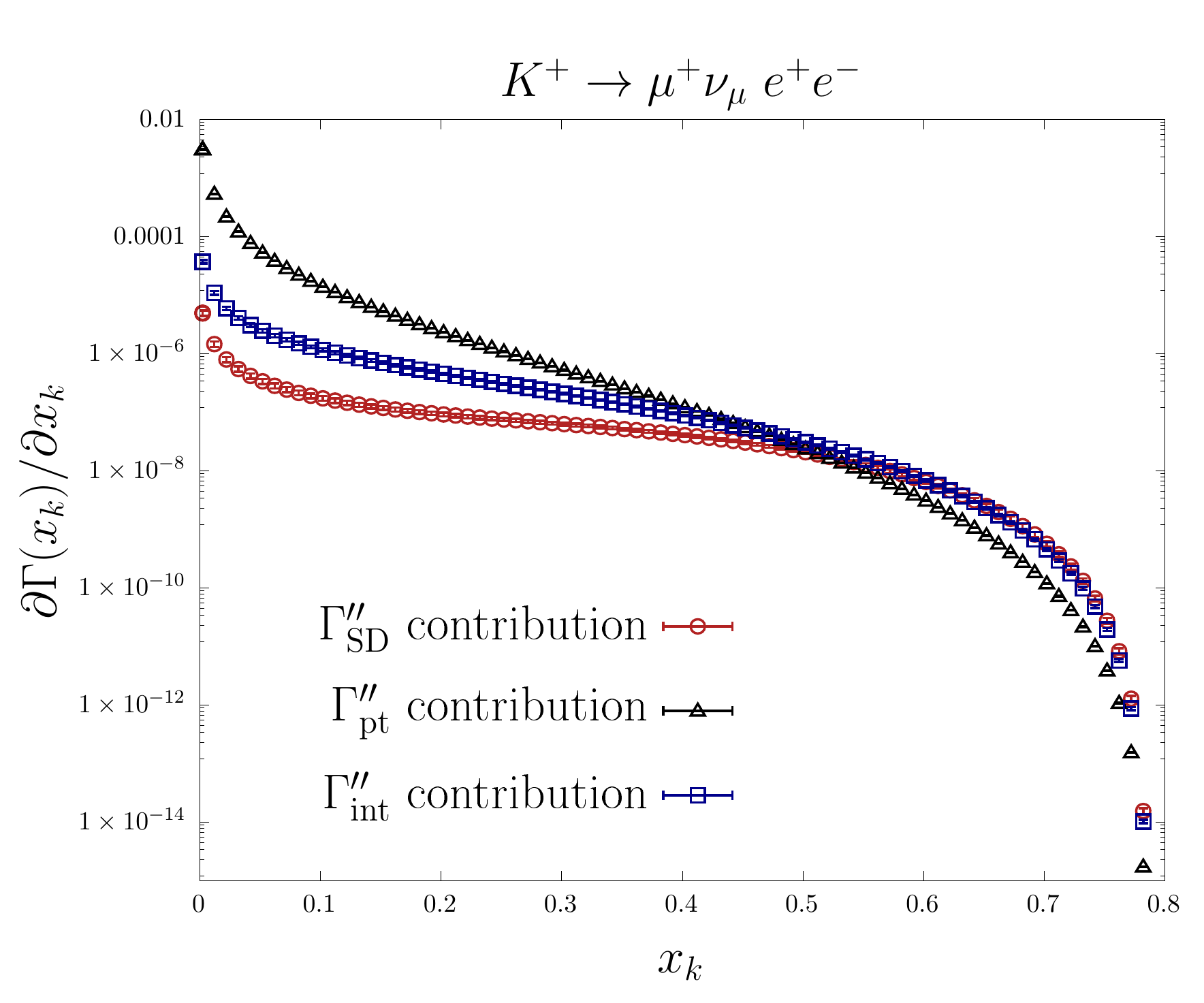}
    \caption{\it{The contributions from $\Gamma^{\prime\prime}_{\mathrm{pt}}$, $\Gamma^{\prime\prime}_{\mathrm{int}}$ and  $\Gamma^{\prime\prime}_{\mathrm{SD}}$ to the differential rate $\partial \Gamma (x_{k})/\partial x_{k}$, are shown for the decay channel $K^{+}\to \mu^{+}\nu_{\mu}e^{+}e^{-}$. Even if not shown in the figure, all contributions to $\partial \Gamma (x_{k})/\partial x_{k}$ are zero at $x_{k}=2\sqrt{r_{\ell}} \simeq 0.00207$ but grow rapidly as $x_k$ is increased.}}
    \label{fig:diff_rate_ee}
\end{figure}

For the decay channel $K^{+} \to \mu^{+}\nu_{\mu}e^{+}e^{-}$, the point-like contribution is not helicity suppressed ($L_{\mu} \propto m_{\mu}$), and gives the dominant contribution to the differential decay rate at small values of the $e^{+}e^{-}$ invariant mass. This is illustrated in Fig.\,\ref{fig:diff_rate_ee}, where we plot the contributions of $\Gamma^{\prime\prime}_{\mathrm{pt}}$, $\Gamma^{\prime\prime}_{\mathrm{int}}$ and $\Gamma^{\prime\prime}_{\mathrm{SD}}$ to the partially integrated differential decay rate $\partial \Gamma(x_{k})/\partial x_{k}$. 
The contributions from $\Gamma^{\prime\prime}_{\mathrm{pt}}$ and $\Gamma^{\prime\prime}_{\mathrm{int}}+ \Gamma^{\prime\prime}_{\mathrm{SD}}$
become of similar size at values of $x_{k} \simeq 0.3-0.4$, which corresponds approximately to the cut on the $e^{+}e^{-}$ invariant mass $\sqrt{k^{2}} > 145,150\,{\rm MeV}$ ($x_{k} > 0.294, 0.304$) adopted in the E865 experiment~\cite{Poblaguev_2002}. For such values of the cut on $x_{k}$, we find that the contribution to the decay rate from $\Gamma^{\prime\prime}_{\mathrm{int}}$ is greater than that of $\Gamma^{\prime\prime}_{\mathrm{SD}}$ and that the contribution from the form factor $H_{2}$ is again negligible. 
Imposing a cut on the $e^{+}e^{-}$ invariant mass of $x_{k} > 0.284$, we obtain the following value for the branching ratio
\begin{align}
 \textrm{BR}\left[K^+\to \mu^+\,\nu_{\mu}\,e^+\,e^-\right]= 8.26~(13) \times 10^{-8}\,.
 \end{align}
 In Tab.\,\ref{tab_Br2} we compare our result for the branching ratio with the lattice determination of Ref.\,\cite{xu}, with the ChPT prediction and with the experimental result of Ref.\,\cite{Poblaguev_2002}. In this case we find a remarkable agreement with both the experimental result and the ChPT predictions, while the lattice result of Ref.\,\cite{xu} is a little larger than ours. In this case, since the inference term dominates over $\Gamma^{\prime\prime}_{\mathrm{SD}}$, systematic effects in our determination of the form factors, due to lattice artefacts and to the unphysical quark masses, will only reflect linearly in the result for the branching ratio; for example an increase in $H_1$ of 20\% would increase the branching ratio by about 7\%.

\begin{table}[]
    \centering
    \begin{tabular}{c|c| c| c| c| c}
    \hline
     \multicolumn{6}{c}{ \rule[-2mm]{0mm}{15pt}  $\textrm{BR}\left[K^+\to \mu^+\,\nu_{\mu}\,e^+\,e^-\right]$ for $x_{k}> 0.284$}  \\
     \hline 
 \rule[-2mm]{0mm}{15pt}    This work & point-like approximation & Tuo et al.~\cite{xu} & ChPT($f_\pi$)& ChPT($f_{K}$) & experiment\,\cite{Poblaguev_2002} \\
  \hline
 \rule[-2mm]{0mm}{15pt}         $8.26(13)\times 10^{-8}$ & $4.8\times 10^{-8}$ &  $11.08(39) \times 10^{-8}$ & $9.82\times 10^{-8}$ & $8.25\times 10^{-8}$ & $7.93(33)\times 10^{-8}$ \\
    \hline 
    \end{tabular}
    \caption{\it{Comparison of our result for the branching ratio $\textrm{BR}\left[K^+\to \mu^+\,\nu_{\mu}\,e^+\,e^-\right]$ with the one coming from the point-like approximation, the result from Ref.\,\cite{xu} and with the results for the branching ratio obtained using the NLO ChPT predictions for the SD form factors (Eq.~(\ref{eq:chpt_pred})) setting either $F=f_{\pi}/\sqrt{2}$ or 
    $F=f_{K}/\sqrt{2}$ (denoted by ChPT($f_\pi$) and ChPT($f_K$) respectively).
  In the last column we show the experimental result from the E865 experiment\,\cite{Poblaguev_2002}, which has been extrapolated from $x_{k} > 0.294$ to $x_{k} > 0.284$ using the formula presented in Ref\,\cite{Poblaguev_2002}. We stress that both our lattice result and that from Ref.\,\cite{xu} are affected by systematic uncertainties due to the missing chiral, continuum and infinite-volume extrapolations.}}
    \label{tab_Br2}
\end{table}

\subsection{Decays with \boldmath{$\ell= \ell^\prime$}}
When the final state leptons have same flavor, the exchange contribution must be added as shown in Eq.\,(\ref{l_ugual_lprime_amplitude}). In this case $\sum_{\textrm{spins}}|\mathcal{M}|^{2}$, depends on products of form factors evaluated at $k^{2}= (p_{\ell^{\prime+}}+p_{\ell^{\prime-}})^2$ and $q^{2}= (p_{\ell^{+}}+p_{\nu_{\ell}})^{2}$ as before, but also at the exchanged invariant masses $k^{\prime\,2}= (p_{\ell^{+}}+p_{\ell^{\prime-}})^{2}$ and $q^{\prime\,2}= (p_{\ell^{\prime+}}+p_{\nu_{\ell}})^{2}$.
It is therefore not possible to integrate analytically as many variables as before. For the decay $K^{+}\to \ell^{+}\nu_{\ell}\ell^{\prime+}\ell^{\prime-}$, the four-body phase space $d\Phi_{4}$ can be written in terms of five Lorentz invariant quantities $x_{k}, x_{q}, y_{12}, y_{34}, \phi$ as~\cite{xu}
\begin{align}
d\Phi_{4} = \frac{\mathcal{S}\lambda \omega m_{K}^{4}}{2^{14}\pi^{6}}~ dx_{k}dx_{q}dy_{12}dy_{34}d\phi~,   
\end{align}
where $\omega= 2x_{k}x_{q}$, the symmetry factor $\mathcal{S}=\frac12$ for the case $\ell=\ell^\prime$, $\lambda = \sqrt{   (1- x_{k}^{2} - x_{q}^{2})^{2} - 4x_{k}^{2}x_{q}^{2}}$
and the three additional integration variables $y_{12}, y_{34}$ and $\phi$, are defined as
\begin{align}
y_{12} &\equiv \frac{2}{m_{K}^{2}\lambda}~(p_{\ell^{\prime-}} - p_{\ell^{\prime+}})\!\cdot\! ( p_{\ell^{+}} + p_{\nu_{\ell}})\,, \nonumber \\[10pt]
y_{34} &\equiv \frac{2}{m_{K}^{2}\lambda}~\Big( (1+ \frac{r_{\ell}}{x_{q}^{2}})\,p_{\nu_{\ell}} -(1-\frac{r_{\ell}}{x_{q}^{2}})\,p_{\ell^{+}}\Big)\!\cdot\! \Big( p_{l^{\prime+}} + p_{\ell^{\prime-}}\Big)\,, \\[10pt] 
\sin{\phi} &\equiv -\frac{16}{\lambda\,\omega\, m_{K}^{4}}~\frac{1}{\sqrt{ (\lambda_{12}^{2}-y_{12}^{2})(\lambda_{34}^{2}-y_{34}^{2}) }}~\epsilon_{\mu\nu\rho\sigma}~p_{\ell^{\prime-}}^{\mu}p_{\ell^{\prime+}}^{\nu}p_{\nu_{\ell}}^{\rho}p_{\ell^{+}}^{\sigma}\nonumber
\end{align}
where 

\begin{align}
\lambda_{12} = \sqrt{ 1- 4\frac{r_{\ell}^{\prime}}{x_{k}^{2}}}~,\qquad   \lambda_{34} = 1 - \frac{r_{\ell}}{x_{q}^{2}}\,. 
\end{align}
\begin{table}[]
    \centering
    \begin{tabular}{c|c| c| c| c| c}
    \hline
     \multicolumn{6}{c}{\rule[-2mm]{0mm}{15pt}$\textrm{BR}\left[K^+\to \mu^+\,\nu_{\mu}\,\mu^+\,\mu^-\right]$ }  \\
     \hline 
   \rule[-2mm]{0mm}{15pt}  This work & point-like approximation & Tuo et al.\,\cite{xu} & ChPT($f_\pi)$ & ChPT($f_K$) & experiment \\
     \hline
 \rule[-2mm]{0mm}{15pt}      $1.178(35)\times 10^{-8}$ & $3.7\times 10^{-9}$ &  $1.52(7) \times 10^{-8}$ & $1.51\times 10^{-8}$ & $ 1.10 \times 10^{-8} $  & -- \\
    \hline 
    \end{tabular}
    \caption{\it{Comparison of our result for the branching ratio $\textrm{BR}\left[K^+\to \mu^+\,\nu_{\mu}\,\mu^+\,\mu^-\right]$ with the one coming from the point-like approximation, the result from Ref.\,\cite{xu} and with the results for the branching ratio obtained using the NLO ChPT predictions for the SD form factors (Eq.~(\ref{eq:chpt_pred})) setting either $F=f_{\pi}/\sqrt{2}$ or 
    $F=f_{K}/\sqrt{2}$ (denoted by ChPT($f_\pi$) and ChPT($f_K$) respectively). We stress that both our lattice result and that from Ref.\,\cite{xu} are affected by systematic uncertainties due to the missing chiral, continuum and infinite-volume extrapolations.}}
\label{tab_Br3}
\end{table}


\begin{figure}
    \centering
    \includegraphics[scale=0.50]{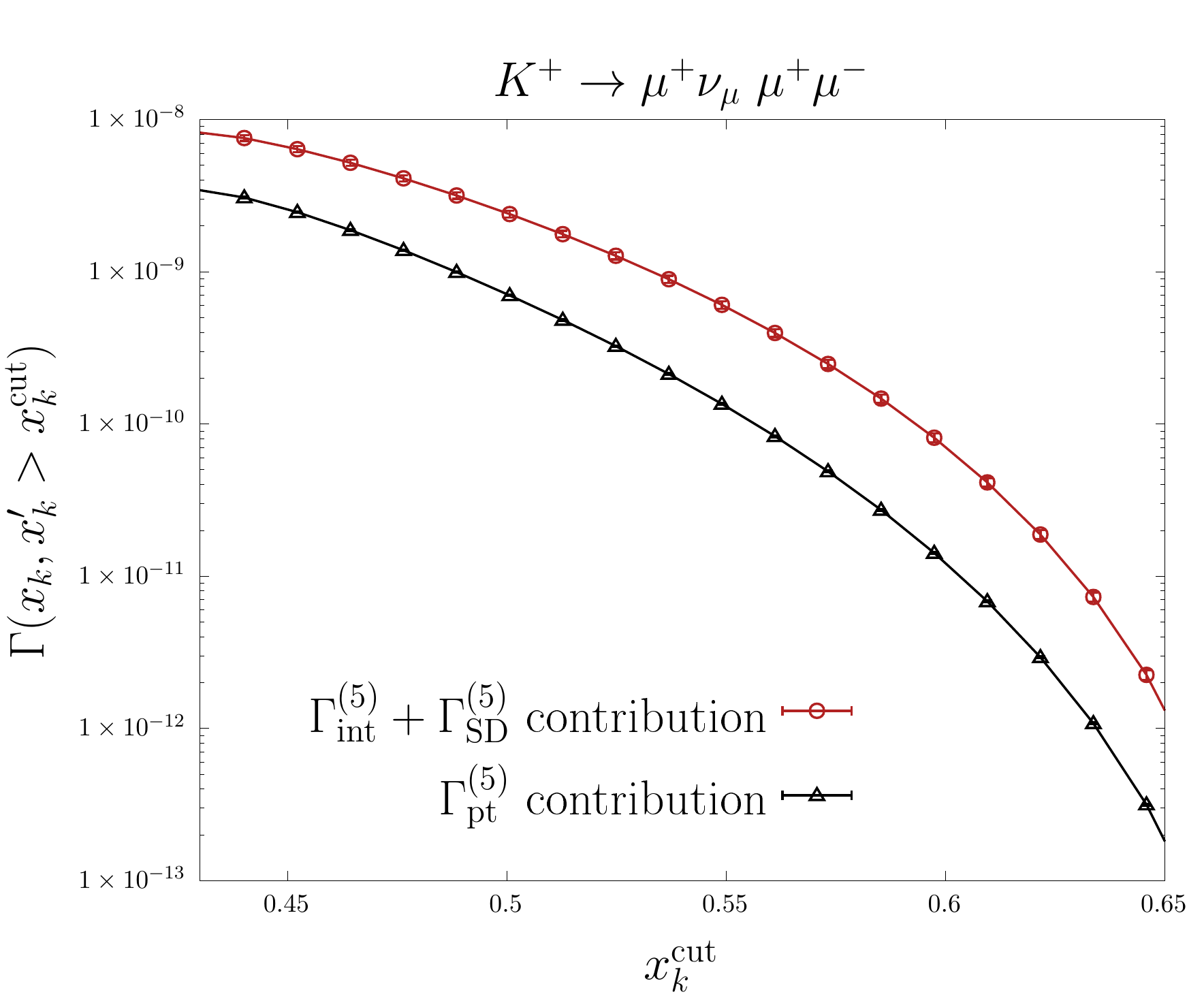}
    \caption{\it{The contributions from $\Gamma^{(5)}_{\mathrm{pt}}$,  and  $\Gamma^{(5)}_{\mathrm{int}}+\Gamma^{(5)}_{\mathrm{SD}}$ to the integrated decay rate $\Gamma (x_{k},\,x_{k'}>x_k^{\mathrm{cut}})$ are shown for the decay channel $K^{+}\to \mu^{+}\nu_{\mu}\mu^{+}\mu^{-}$, as a function of the common lower cut, $x_{k}^{\mathrm{cut}}$, on the values of $x_{k}$ and $x_{k}^{\prime}$. }}
    \label{fig:rate_cut_mumumu}
\end{figure}

The integration domain is given by
\begin{align}
-\lambda_{12}\leq y_{12}\leq \lambda_{12},\qquad -\lambda_{34}\leq y_{34}\leq \lambda_{34},\qquad \phi \in [0,2\pi]~,     
\end{align}
while for $x_{k}$ and $x_{q}$ the limits of integration are as defined in Eq.\,(\ref{xkq_range}). In order to determine the decay rate, we have evaluated the square of the unpolarised amplitude, $\sum_{\mathrm{spins}}|\mathcal{M}|^{2}$, in terms of the five integration variables using FeynCalc\,\cite{2020FenyCalc}. As for the case when $\ell \ne \ell^\prime$, we decompose the differential rate as a sum of a point-like, an interference and a quadratic term ($\mathrm {SD}$) in the form factors, i.e. as
\begin{align}
\frac{\partial^{5}\Gamma}{\partial x_{k}\partial x_{q} \partial y_{12} \partial y_{34} \partial \phi } =  \alpha^{2}G_{F}^{2}|V_{us}|^{2}m_{K}^{5}\left[ \Gamma^{(5)}_{\mathrm{pt}}(x_k, x_q, y_{12},y_{34}, \phi) + \Gamma^{(5)}_{\mathrm{int}}(x_k, x_q, y_{12}, y_{34}, \phi) + \Gamma^{(5)}_{\mathrm{SD}}(x_k, x_q, y_{12}, y_{34}, \phi)\right]\,. 
\end{align}
The explicit, very lengthy, expressions for the three contributions $\Gamma_{\mathrm{pt}}^{(5)}, \Gamma_{\mathrm{int}}^{(5)}$ and $\Gamma_{\mathrm{SD}}^{(5)}$, written in terms of the five integration variables and the form factors, are not presented here but are available on request from the authors. The total rate can be obtained through standard Monte Carlo integration of these expressions over the five-dimensional phase space.  This has been done employing the GSL implementation of the VEGAS algorithm of G.P.\,Lepage\,\cite{1978JCoPh..27..192L}. \\

For the decay channel $K^{+}\to \mu^{+}\nu_{\mu}\mu^{+}\mu^{-}$, we find that the point-like contribution corresponds to about $30\%$ of the total rate. This is shown in Fig.\,\ref{fig:rate_cut_mumumu}, where we plot the contributions to the decay rate as a function of the lower cutoff on the invariant mass of the $\mu^{+}\mu^{-}$ pair,  from $\Gamma^{(5)}_{\mathrm{pt}}$ alone and from $\Gamma^{(5)}_{\mathrm{int}} + \Gamma^{(5)}_{\mathrm{SD}}$. For decays into identical leptons, the same cuts are always applied to both invariant masses 
\begin{equation}
\sqrt{k^{2}} = m_{K}x_{k} 
= \sqrt{( p_{\ell^{\prime+}} + p_{\ell^{\prime-}})^{2}}\,,\qquad
\sqrt{k^{\prime\,2}} \equiv m_{K}x_{k}^{\prime} =  \sqrt{( p_{\ell^{+}} + p_{\ell^{\prime-}})^{2}}\,.
\end{equation}
We find that the contribution from the form factor $H_{2}$ is again negligible and that the contribution from the vector form factor $F_{V}$ is also very small. 
For the total branching ratio, we obtain the value
\begin{align}
\textrm{BR}\left[K^+\to \mu^+\,\nu_{\mu}\,\mu^+\,\mu^-\right]= 1.178~(35)\times 10^{-8}\,.
\end{align}
Since for this decay channel there is no experimental measurement available, our result can only be compared with the lattice determination of Ref.\,\cite{xu} and with the ChPT prediction (Tab.\,\ref{tab_Br3}). As the table shows, our result is in reasonably good agreement with the
value predicted by ChPT, while at this stage the observed discrepancy
with the result obtained by Tuo et al.\,\cite{xu}, which is of O(25\%) may perhaps be attributed to the unknown systematics
associated to the missing chiral, continuum and infinite-volume extrapolations.

\begin{figure}
    \centering
    \includegraphics[scale=0.5]{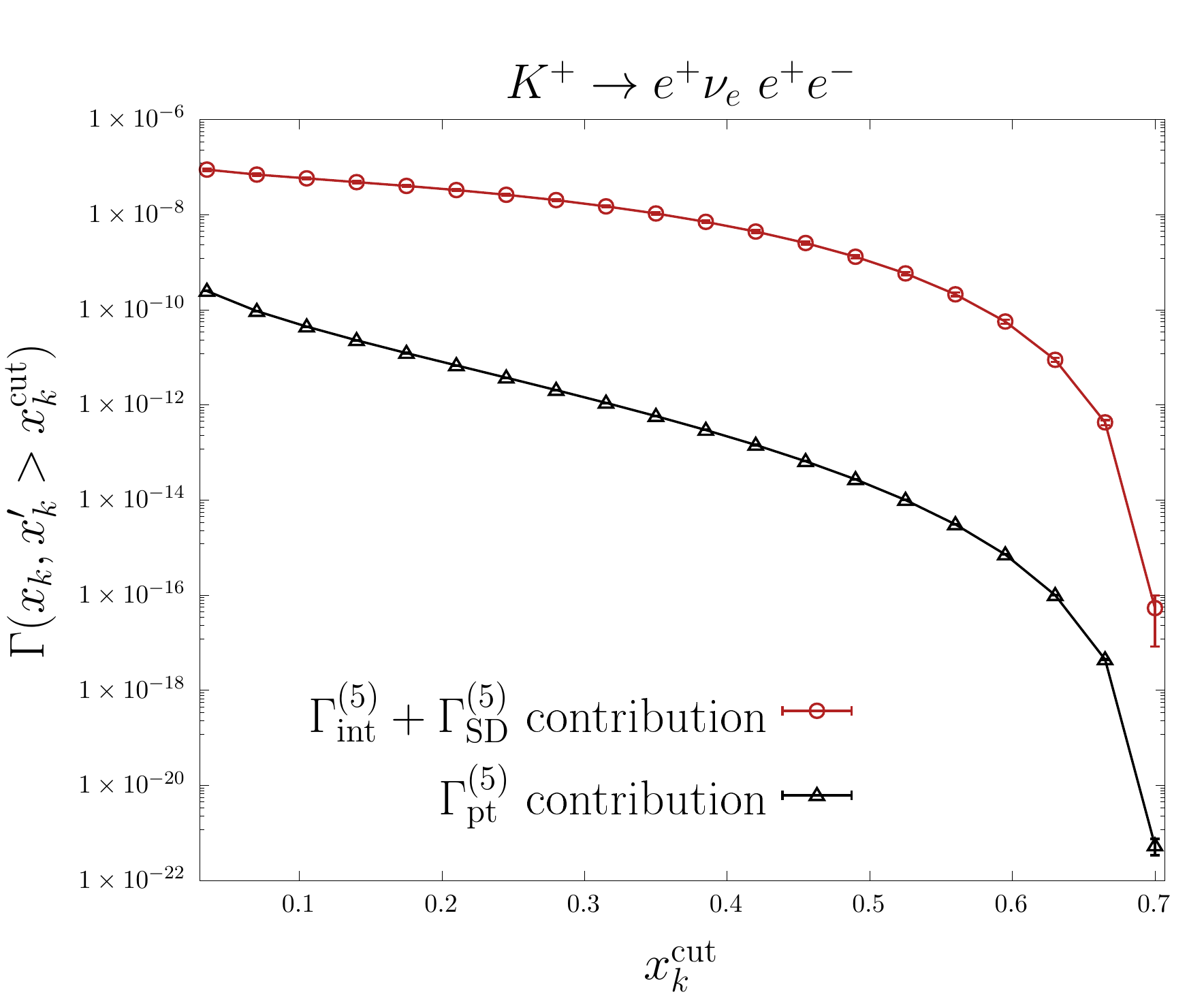}
    \caption{\it{The contributions from $\Gamma^{(5)}_{\mathrm{pt}}$,  and  $\Gamma^{(5)}_{\mathrm{int}}+\Gamma^{(5)}_{\mathrm{SD}}$ to the integrated decay rate $\Gamma (x_{k},\,x_{k'}>x_k^{\mathrm{cut}})$ are shown for the decay channel $K^{+}\to e^{+}\nu_{e}e^{+}e^{-}$, as a function of the common lower cut $x_{k}^{\mathrm{cut}}$ on the values of $x_{k}$ and $x_{k}^{\prime}$.} }
    \label{fig:rate_cut_eee}
\end{figure}

Finally, for the decay channel $K^{+}\to e^{+}\nu_{e}e^{+}e^{-}$, we find again that the point-like contribution is much suppressed compared to that from the SD terms. This is shown in Fig.\,\ref{fig:rate_cut_eee}, where we have plotted, as in the previous case, the contribution from $\Gamma^{(5)}_{\mathrm{pt}}$ and from $\Gamma^{(5)}_{\mathrm{int}}+\Gamma^{(5)}_{\mathrm{SD}}$ to the total decay rate, as a function of the lower cutoff on the $e^{+}e^{-}$ invariant masses. Similarly to the case of the decay $K^{+}\to \mu^{+}\nu_{\mu}e^{+}e^{-}$, we find that the dominant contribution to the rate is given by the term proportional to $H_{1}^{2}$, while the contributions from the form factors $H_{2}$ and $F_{V}$ are very small. Employing the cutoffs $x_{k}, x_{k}^{\prime} > 0.284$, we obtain the following value for the branching ratio
\begin{align}
\textrm{BR}\left[K^+\to e^+\,\nu_{e}\,e^+\,e^-\right]=   1.95~(11) \times 10^{-8}\,.
\end{align}
In Tab.\,\ref{tab_Br4} we compare our determination with the experimental measurement of Ref.~\cite{Poblaguev_2002}, with the lattice result of Ref.~\cite{xu}, and with the ChPT prediction. Our result appears to be slightly smaller than the other determinations as in the case of the $K^{+}\to \mu^{+}\nu_{\mu}e^{+}e^{-}$ decay. Since the term proportional to $H_{1}^{2}$ is also the dominant one in this case, 
this finding is consistent with possible systematic effects of about 20\% on our lattice value.

\begin{table}[]
    \centering
    \begin{tabular}{c|c| c| c| c| c}
    \hline
     \multicolumn{6}{c}{\rule[-2mm]{0mm}{15pt}$\textrm{BR}\left[K^+\to e^+\,\nu_{e}\,e^+\,e^-\right]$ for $x_{k} > 0.284$}  \\
     \hline 
   \rule[-2mm]{0mm}{15pt}  This work & point-like approximation & Tuo et al.\,\cite{xu} & ChPT($f_\pi)$& ChPT($f_K$) & experiment\,\cite{Poblaguev_2002} \\
     \hline
     \rule[-2mm]{0mm}{15pt}  $1.95(11)\times 10^{-8}$ & $2.0\times 10^{-12}$ &  $3.29(35) \times 10^{-8}$ & $3.34\times 10^{-8}$ & $  1.75 \times 10^{-8} $  &  $2.91(23) \times 10^{-8}$  \\
    \hline 
    \end{tabular}
    \caption{\it{Comparison of our result for the branching ratio $\textrm{BR}\left[K^+\to e^+\,\nu_{e}\,e^+\,e^-\right]$ with the one coming from the point-like approximation, the result from Ref.\,\cite{xu} and with the results for the branching ratio obtained using the NLO ChPT predictions for the SD form factors (Eq.~(\ref{eq:chpt_pred})) setting either $F=f_{\pi}/\sqrt{2}$ or 
    $F=f_{K}/\sqrt{2}$ (denoted by ChPT($f_\pi$) and ChPT($f_K$) respectively). 
In the last column we show the experimental result from the E865 experiment\,\cite{Poblaguev_2002}, which has been extrapolated from $x_{k} > 0.304$ to $x_{k} > 0.284$ using the formula presented in Ref.\,\cite{Poblaguev_2002}. We stress that both our lattice result and that from Ref.\,\cite{xu} are affected by systematic uncertainties due to the missing chiral, continuum and infinite-volume extrapolations.}}
    \label{tab_Br4}
\end{table}

\section{Conclusions}\label{sec:concs}
In this paper we have presented a strategy to compute, using Lattice QCD, the amplitudes and branching ratios for the decays $P\to\ell\nu_\ell\,\ell^{\prime\,+}\ell^{\prime\,-}$, where $P$ is a pseudoscalar meson and $\ell$ and $\ell^\prime$ are charged leptons. In particular, we explain how the four structure-dependent (SD) form factors can be determined and separated from the point-like (``inner-bremstrahlung") contribution. Apart from a factor of the leptonic decay constant $f_P$, the point-like contribution to the amplitude can be calculated in perturbation theory, whereas the SD form factors are non-perturbative and describe the interaction of a generic off-shell photon with the internal hadronic structure of the meson.
The present work, studying decays with the emission of a virtual photon, is a natural extension to our recent study of radiative decays $P\to\ell\nu_\ell\gamma$, with a real photon in the final state\,\cite{Desiderio2020} and the subsequent detailed comparison to experimental results\,\cite{Frezzotti:2020bfa}.

We apply the formalism developed in Secs.\,\ref{hadten}\,-\,\ref{sdff} to the four channels of $K\to\ell\nu_\ell\,\ell^{\prime\,+}\ell^{\prime\,-}$ decays, where $\ell$ and $\ell^\prime=\mu$ or $e$, in an exploratory Lattice QCD computation at a single lattice spacing and at unphysical light-quark masses. We demonstrate that all four SD form factors, $F_V,\,F_A,\,H_1$ and $H_2$ can be determined with good precision and used to calculate the corresponding branching ratios. In spite of the unphysical quark masses used in this simulation, our pion and kaon masses are about 320\,MeV and 530\,MeV respectively, it has been interesting and instructive to compare our results with those from experiment (where available) and from NLO ChPT. 
Perhaps surprisingly, as can be seen from Tabs.\,\ref{tab_Br1}\,-\,\ref{tab_Br4},
the results are generally in reasonable semi-quantitative agreement.

The comparison of our results with those from experimental measurements results in an interesting observation to be investigated further in the future. For the decays $K^+\to e^+\nu_e\,\mu^+\mu^-$ and $K^+\to e^+\nu_e\,e^+e^-$ the point-like contribution is negligible as a result of the chiral suppression due to the small electron mass, and the decay rate is dominated by the form-factor $H_1$. In both cases our results are somewhat below the experimental measurement (see Tabs.\,\ref{tab_Br1} and \ref{tab_Br4}) and it would require an increase of order 20\% in the value of $H_1$ to recover consistency\,\footnote{Note also the discrepancy in the values of $H_1(0,0)$ in Refs.\,\cite{Poblaguev_2002} and \cite{Ma_2006} obtained in the E865 experiment. This is discussed in Sec.\,\ref{subsec:lnotlprime}.}. It will be interesting to see whether such an increase will result after the continuum, chiral and infinite-volume extrapolations have been performed in the future. 

A complementary exploratory lattice computation of the branching ratios has been performed by Tuo et al.\,\cite{xu} on a $24^3\times 48$ lattice, with lattice spacing $a=0.093$\,fm and with quark masses similar to our ($m_\pi = 352$\,MeV and $m_K=506$\,MeV). The action with Wilson-Clover Twisted Mass Fermions is different to the one we use which does not include the clover term.
The methodology in Ref.\,\cite{xu} is also different to ours in that the individual form factors are not extracted and the point-like contribution is not separated from structure dependent terms. The aim of our paper on the other hand, is to determine explicitly the (non-perturbative) structure dependent contributions to the decay rate. In Tabs.\,\ref{tab_Br1}\,-\,\ref{tab_Br4} we also compare our results with those of Ref.\,\cite{xu}, but given the different systematics, and in particular the finite-volume effects, we do not speculate on the origin of any differences. 

Having demonstrated the feasibility of the method, our future work will focus on controlling and reducing the systematic uncertainties and in particular those resulting from the current absence of continuum, chiral and infinite-volume extrapolations. 
We will also work to extend the method to heavier pseudoscalar mesons, for which the analytic continuation to Euclidean space gives rise to enhanced finite-volume effects due to the presence of internal lighter states.  
Given the recent results suggesting the violation of lepton-flavour universality and potential new interactions involving leptons (see e.g.\,Ref.\,\cite{Crivellin:2021sff} for a brief introduction), we believe that reliable 
non-perturbative, model-independent theoretical predictions of
decays such as those studied here will be very useful in unravelling the underlying theory Beyond the Standard Model. In particular, experimental measurements of ratios of decay rates of heavy mesons into different final-state leptons, together with the corresponding lattice calculations, would be a significant contribution to the general investigation of lepton-flavour universality.

\section*{Acknowledgements}
We thank Roberto Frezzotti, Marco Garofalo and Cecilia Tarantino for useful discussions, Hong Ma and Andrei Poblaguev for correspondence concerning the experimental results in Refs.\,\cite{Poblaguev_2002,Ma_2006} and Xu Feng for correspondence concerning finite-volume corrections. We acknowledge CINECA for the provision of CPU time under the specific
initiative INFN-LQCD123 and IscrB\_S-EPIC. F.S. G.G and S.S. are supported by the Italian Ministry
of University and Research (MIUR) under grant PRIN20172LNEEZ. F.S. and G.G are supported by
INFN under GRANT73/CALAT. C.T.S. was partially supported by an Emeritus Fellowship from the
Leverhulme Trust and by STFC (UK) grants ST/P000711/1 and ST/T000775/1.

\appendix
\section{Formulae for the decay rates when \boldmath{$\ell\neq\ell^\prime$}}\label{kernels}
In this appendix we present the functions which multiply the form factors in the differential decay rates we computed in section \ref{subsec:lnotlprime}, with different charged leptons in the final state, i.e. for $\ell\neq \ell^\prime$.

We start by defining the following quantity:
\bea
\hspace{1in}\Delta_{\mathrm{log}}(x_k,x_q)=\\[7pt]
&&\hspace{-2in}\log\!\left(\!1+2\frac{x_q^2\sqrt{x_k^4-2x_k^2(x_q^2+1)+(x_q^2-1)^2}-r_\ell\sqrt{-2(x_k^2+1)x_q^2+(x_k^2-1)^2+x_q^4}}{-x_q^2\sqrt{x_k^4-2x_k^2(x_q^2+1)+(x_q^2-1)^2}+r_\ell\sqrt{-2(x_k^2+1)x_q^2+(x_k^2-1)^2+x_q^4}+r_\ell(x_k^2-1)+x_q^2(-x_k^2+x_q^2+r_\ell-1\!)}\right)\!,\nonumber
\eea
\mbox{}\\[0.01in]where $x_k$ and $x_q$ are defined in Eq.\,(\ref{eq:xkqdef}) and $r_\ell$ is defined in Eq.\,(\ref{eq:defs}). 

The point-like contribution to the decay rate is given by 
\bea
\Gamma^{\prime\prime}_{\mathrm{pt}}(x_k,x_q)&=&\frac{f_K^2 r_\ell\,x_q}{24\pi^3m_K^3x_k}\sqrt{1-\frac{4r_{\ell^\prime}}{x_k^2}}\left(\frac{2r_{\ell^\prime}}{x_k^2}+1\right)\Bigg(\frac{2}{x_q^2-1}\left(-x_k^2x_q^2+x_k^2+x_q^4-2x_q^2r_\ell-2(r_\ell-1)r_\ell+1\right)\Delta_{\mathrm{log}}(x_k, x_q)\nonumber\\[7pt]
&&\hspace{-0.7in}+\sqrt{(x_k^2-x_q^2+1)^2-4x_k^2}(x_q^2-r_\ell)\Big(\frac{x_k^2x_q^2-x_k^2r_\ell-2x_q^4+4x_q^2r_\ell-2}{(x_q^2-1)^2x_q^2}+\frac{2(r_\ell-1)(x_k^2+2r_\ell)}{(x_q^2-1)^2r_\ell-x_k^2(r_\ell-1)(x_q^2-r_\ell)}\Big)\Bigg)\,,
\eea
where $r_{\ell^\prime}$ is also defined in Eq.\,(\ref{eq:defs}). 

The interference contribution to the decay rate can be expressed in the form
\begin{align}
\Gamma^{\prime\prime}_{\mathrm{int}}(x_k, x_q)&=&\Big[g_V(x_k, x_q)F_V(x_k, x_q)+g_A(x_k, x_q)F_A(x_k, x_q)+g_1(x_k, x_q)H_1(x_k, x_q)+g_2(x_k, x_q)H_2(x_k, x_q)\Big]\,,
\end{align}
where the interference kernels are:
\bea
g_V(x_k,x_q)&=&\frac{f_Kr_\ell\,x_q}{12\pi^3m_K x_k}\sqrt{1-\frac{4r_{\ell^\prime}}{x_k^2}}\left(\frac{2r_{\ell^\prime}}{x_k^2}+1\right)\Bigg\{\left(x_k^2(x_q^2-2r_\ell+1)-(x_q^2-1)^2\right)\Delta_{\mathrm{log}}(x_k, x_q)\nonumber\\[7pt]
&&\hspace{0.8in}+\,\frac{(x_k^2+x_q^2-1)(x_q^2-r_\ell)\,\sqrt{(x_k^2-x_q^2+1)^2-4x_k^2}}{x_q^2}~\Bigg\}\,,\nonumber\\[10pt]
g_A(x_k,x_q)&=&\frac{f_Kr_\ell\,x_q}{12\pi^3m_K x_k(x_q^2-1)}\sqrt{1-\frac{4r_{\ell^\prime}}{x_k^2}}\left(\frac{2r_{\ell^\prime}}{x_k^2}+1\right)\Bigg\{(x_q^2-1)^2(-x_k^2-x_q^2-2r_\ell+1)\Delta_{\mathrm{log}}(x_k, x_q)\nonumber\\[7pt]
&&\hspace{0.2in}+\,\frac{(x_q^2-1)(x_q^2-r_\ell)(x_k^2+2x_q^2+r_\ell-1)\,\sqrt{-2(x_k^2+1)x_q^2+(x_k^2-1)^2+x_q^4}}{x_q^2}~\Bigg\}\,,\nonumber\\[10pt]
g_1(x_k,x_q)&=&\frac{f_Kr_\ell\,x_k\,x_q}{24\pi^3m_K}\sqrt{1-\frac{4r_{\ell^\prime}}{x_k^2}}\left(\frac{2r_{\ell^\prime}}{x_k^2}+1\right)
\Bigg\{\,4(x_q^2+r_\ell-2)\Delta_{\mathrm{log}}(x_k, x_q)\nonumber\\[7pt]
&&-\frac{(x_q^2-r_\ell)\left(r_\ell(-x_k^2+3x_q^2+1)+x_q^2(x_k^2+5x_q^2-9)\right)\sqrt{-2(x_k^2+1)x_q^2+(x_k^2-1)^2+x_q^4}}{(x_q^2-1)\,x_q^4}\,\Bigg\}\,,\nonumber\\[10pt]
g_2(x_k,x_q)&=&\frac{f_Kr_\ell\,x_k\,x_q}{24\pi^3m_K (x_q^2-1)^2}\sqrt{1-\frac{4r_{\ell^\prime}}{x_k^2}}\left(\frac{2r_{\ell^\prime}}{x_k^2}+1\right)\Bigg\{\,2(x_q^2-1)(x_q^2-r_\ell^2)\Delta_{\mathrm{log}}(x_k, x_q)\nonumber\\[7pt]
&&\hspace{0.1in}-\frac{(x_q^2-r_\ell)\left(r_\ell(x_k^2-3x_q^2-1)+x_q^2(-x_k^2+x_q^2+3)\right)\sqrt{(x_k^2-x_q^2+1)^2-4x_k^2}}{x_q^2}\,\Bigg\}\,.
\eea

The SD contribution to the decay rate can be expressed in the form
\bea
\Gamma^{\prime\prime}_{\mathrm{SD}}(x_k, x_q)&=&g_{VV}(x_k, x_q)F_V^2(x_k, x_q)+g_{AA}(x_k, x_q)F_A^2(x_k, x_q)+g_{11}(x_k, x_q)H_1^2(x_k, x_q)+g_{22}(x_k, x_q)H_2^2(x_k, x_q)\nonumber\\[7pt]
&&+g_{A1}(x_k, x_q)F_A(x_k, x_q)H_1(x_k, x_q)+g_{12}(x_k, x_q)H_1(x_k, x_q)H_2(x_k, x_q)\,,
\eea
where the SD kernels are:
\bea
g_{VV}(x_k, x_q)&=&\frac{1}{24\pi^3x_k\,x_q}\sqrt{1-\frac{4r_{\ell^\prime}}{x_k^2}}\left(\frac{2r_{\ell^\prime}}{x_k^2}+1\right)\left((x_k^2-x_q^2+1)^2-4x_k^2\right)^{3/2}(x_q^2-r_\ell)^2\,,\nonumber\\[10pt]
g_{AA}(x_k, x_q)&=&\frac{1}{144\pi^3x_k\,x_q^3}\sqrt{1-\frac{4r_{\ell^\prime}}{x_k^2}}\left(\frac{2r_{\ell^\prime}}{x_k^2}+1\right) \left(x_k^4+x_k^2(4x_q^2-2)+(x_q^2-1)^2\right)(x_q^2-r_\ell)^2(2x_q^2+r_\ell)\nonumber\\[10pt]
&&\hspace{1.5in}\times\sqrt{(x_k^2-x_q^2+1)^2-4x_k^2}~,\nonumber\\[10pt]
g_{11}(x_k, x_q)&=&\frac{x_k}{144\pi^3x_q^5}\sqrt{1-\frac{4r_{\ell^\prime}}{x_k^2}}\left(\frac{2r_{\ell^\prime}}{x_k^2}+1\right)(x_q^2-r_\ell)^2
\sqrt{(x_k^2-x_q^2+1)^2-4x_k^2}
\nonumber\\[7pt]
&&\times\left(2x_q^4(5x_k^2+r_\ell-1)+x_q^2\left(2(x_k^2-2)r_\ell+(x_k^2-1)^2\right)+2(x_k^2-1)^2r_\ell+x_q^6\right)\,,\nonumber\\[10pt]
g_{22}(x_k, x_q)&=&\frac{r_\ell\,x_k}{96\pi^3(x_q^2-1)^2x_q}\sqrt{1-\frac{4r_{\ell^\prime}}{x_k^2}}\left(\frac{2r_{\ell^\prime}}{x_k^2}+1\right)\left((x_k^2-x_q^2+1)^2-4x_k^2\right)^{3/2}(x_q^2-r_\ell)^2\,,\nonumber\\[10pt]
g_{A1}(x_k, x_q)&=&-\frac{x_k}{24\pi^3x_q^3}\sqrt{1-\frac{4r_{\ell^\prime}}{x_k^2}}\left(\frac{2r_{\ell^\prime}}{x_k^2}+1\right)(x_k^2+x_q^2-1)(x_q^2-r_\ell)^2(2x_q^2+r_\ell)\sqrt{(x_k^2-x_q^2+1)^2-4x_k^2}\,,\nonumber\\[10pt]
g_{12}(x_k, x_q)&=&-\frac{r_\ell\,x_k}{48\pi^3(x_q^2-1)\,x_q^3}\sqrt{1-\frac{4r_{\ell^\prime}}{x_k^2}}\left(\frac{2r_{\ell^\prime}}{x_k^2}+1\right)\left((x_k^2-x_q^2+1)^2-4x_k^2\right)^{3/2}(x_q^2-r_\ell)^2\,.
\eea

\section{Three-point correlation function in the infrared limit \boldmath{$k\to 0$}}\label{kto0}
In this appendix we study the behaviour of the lattice Euclidean correlation function $C^{\mu\nu}(t, E_\gamma, \bs{k}, \bs{p})$ in the limit $k\to 0$ which, as we will see below, is non trivial. From spectral decomposition one obtains
\bea
\label{eq:spectral_C}
C^{\mu\nu}(t, E_\gamma, \bs{k}, \bs{p})=c_1^{\mu\nu}e^{-tE_P(\bs{p}) }+c_2^{\mu\nu}e^{-t\left\{E_P(\bs{p}-\bs{k})+E_\gamma \right\}}+\dots\ \,,
\eea
where the dots represent exponentially suppressed contributions with an energy gap which, in the soft photon limit, is or order $2m_\pi$.
The first exponential corresponds to the on-shell external meson $P(\bs{p})$ with spatial momentum $\bs{p}$, and gives the contribution we aim to isolate, while the second exponential corresponds to the  $P(\bs{p}-\bs{k}) + \gamma$ state, composed of an on-shell meson $P(\bs{p}-\bs{k})$ with spatial momentum $\bs{p}-\bs{k}$, and a virtual photon with spatial momentum $\bs{k}$ and off-shell energy $E_\gamma$. When either $\bs{k}$ or $E_{\gamma}$ are non-zero, it is possible to isolate the matrix element corresponding to the ground state $P(\bs{p})$, since the second exponential in Eq.\,(\ref{eq:spectral_C}) is subleading at large time separations $t$. However, in the exact limit $k\to 0$, the energy-gap between the two states vanishes, and the lattice Euclidean correlator $C^{\mu\nu}(t,0,\bs{0},\bs{p})$ has a non-trivial behaviour which we now discuss, paying special attention to the leading cutoff effects. 
This has been already done for $P\to\ell\nu_\ell\gamma$ decays, with the emission of a real photon, in appendix C of Ref.\,\cite{Desiderio2020}, focusing on the spatial components of $C^{\mu\nu}$, which are the only ones relevant in that case.  In this appendix we generalize the analysis of Ref.\,\cite{Desiderio2020} to the components $C^{0 \nu}$ and $C^{\mu 0}$, with $\mu,\nu=0,1,2,3$. \\

The starting point is the electromagnetic Ward Identity that, for Wilson-like Fermions adopted in this study, reads\,\cite{Desiderio2020}
\bea\label{wi}
\sum_{\mu=0}^3 \frac{2}{a}\sin\left(ak_\mu/2\right)C^{\mu\nu}_A(t,k,\bs{p})=C^\nu_A(t, \bs{p})-C_A^\nu(t, E_\gamma, \bs{p}-\bs{k})\,,
\eea
where we have defined
\bea
C^{\mu\nu}_A(t,k,\bs{p})&=&-i\int d^4y\,d^3\bs{x}\,e^{-ik\cdot(y+\hat{\mu}/2)-i\bs{p}\cdot\bs{x}}\langle 0 | T[J^\nu_A(0)\,J_{\mathrm{em}}^\mu(y)P(-t,-\bs{x})]|0\rangle\,,\label{eq:CmunuAdef}\\[10pt]
C^\nu_A(t, \bs{p})&=&\int d^3\bs{x}\,e^{-i\bs{p}\cdot\bs{x}}\langle 0 | T[J^\nu_A(0)P(-t,-\bs{x})]|0\rangle=p^\nu\frac{\hat{f}_P(\bs{p})\hat{G}_P(\bs{p})}{2\hat{E}_P(\bs{p})}e^{-t\hat{E}_P(\bs{p})}+\dots\,,\label{eq:CnuApdef}\\[10pt]
C_A^\nu(t, \hat{E}_\gamma, \bs{p}-\bs{k})&=&e^{-\hat{E}_\gamma t}\int d^3\bs{x}\,e^{-i(\bs{p}-\bs{k})\cdot\bs{x}}\langle 0 | T[J^\nu_A(0)P(-t,-\bs{x})]|0\rangle\nonumber\\[10pt]
&=&u^\nu\frac{\hat{f}_P(\bs{p}-\bs{k})\hat{G}_P(\bs{p}-\bs{k})}{2\hat{E}_P(\bs{p}-\bs{k})}e^{-t\hat{E}_P(\bs{p}-\bs{k})-t\hat{E}_\gamma}+\dots\,,
\label{eq:CnuApmkdef}
\eea
and where the ellipsis represents sub-leading exponentials with an energy gap that, in the infrared limit, starts at order $2m_\pi$\,\footnote{We take this opportunity to correct a typographical mistake in Eq.\,(C6) of Ref.\,\cite{Desiderio2020}. In that equation, the factor $e^{E_\gamma t}$ should be replaced by $e^{-E_\gamma t}$. This is the corresponding factor to $e^{-\hat{E}_\gamma t}$ in the first line of Eq.\,(\ref{eq:CnuApmkdef}) above.}.
In Eqs.\,(\ref{eq:CmunuAdef})\,-\,(\ref{eq:CnuApmkdef}) the integrals are to be read as lattice sums,  $k=(i\hat{E}_\gamma, \bs{k})$ is 
Euclidean four-momentum of the photon,
the photon's four-momentum, while the on-shell (Euclidean) four-momenta of the mesons $P(\bs{p})$ and $P(\bs{p}-\bs{k})$ are given respectively by
\begin{align}
 p=\left(i\hat{E}_P(\bs{p}),\bs{p}\right),\qquad u=\left(i\hat{E}_P(\bs{p}-\bs{k}), \bs{p}-\bs{k}\right)\,.
\end{align}
In the previous expressions, the hat symbol denote lattice quantities, which are related to their continuum counterpart by\,\footnote{In our Twisted mass formulation, cut-off effects on parity-even observables starts at order $\mathrm{O}(a^2)$.}
\begin{align}
 \hat{f}_P(\bs{p})=f_P+O(a^2)\,,\quad \hat{G}_P(\bs{p})=G_P+O(a^2)\,,\quad \hat{E}_P(\bs{p})=E_P+O(a^2)\,,\quad\hat{E}_\gamma=E_\gamma+O(a^2)\,,
\end{align} 
where $f_P$, $G_P$, $E_P(\bs{p})$ and $E_\gamma$ are respectively the continuum decay constant, the continuum matrix element of the pseudoscalar density used as the interpolating operator ($G_P=\mel{0}{P}{P(\bs{p})}$), the continuum energy of the meson and of the virtual photon. \\

We now differentiate Eq.\,(\ref{wi}) with respect to $k^\mu$ and then set $k=0$. 
Since we are considering a generic off-shell photon, the temporal and spatial component of the photon momentum $k^\mu$ are treated as independent quantities.
When the indices $\mu$ and $\nu$ are spatial, one obtains the result quoted in Eq.\,(C17) of Ref.\,\cite{Desiderio2020},
\begin{align}
C^{ij}_A(t,0,\boldsymbol{p})=\frac{\hat{f}_P(\boldsymbol{p})\hat{G}_P(\boldsymbol{p})}{2\hat{E}_P(\boldsymbol{p})}e^{-t\hat{E}_P(\boldsymbol{p})}\left\{\delta^{ij}+p^{j}\left[\frac{1}{\hat{f}_P(\boldsymbol{p})}\frac{\partial \hat{f}_P(\boldsymbol{p})}{\partial p^i}+\frac{1}{\hat{G}_P(\boldsymbol{p})}\frac{\partial \hat{G}_P(\boldsymbol{p})}{\partial p^i}-\left(t+\frac{1}{\hat{E}_P(\boldsymbol{p})}\right)\frac{\partial \hat{E}_P(\boldsymbol{p})}{\partial p^i}\right]\right\}+\ldots~.
\end{align}
Moreover the $H(3)$ symmetry of the lattice implies
\bea
\frac{\partial \hat{f}_P(\boldsymbol{p})}{\partial p^i}=p^i \times O(a^2),\quad \frac{\partial \hat{G}_P(\boldsymbol{p})}{\partial p^i}=p^i \times O(a^2),\quad \frac{\partial \hat{E}_P(\boldsymbol{p})}{\partial p^i}=\frac{p^i}{E_P(\bs{p})} \times \left( 1 + O(a^2)\right)
\eea
so that
\bea\label{ck=0}
C^{ij}_A(t,0,\boldsymbol{p})=\frac{\hat{f}_P(\boldsymbol{p})\hat{G}_P(\boldsymbol{p})}{2\hat{E}_P(\boldsymbol{p})}e^{-t\hat{E}_P(\boldsymbol{p})}\left\{\delta^{ij}-\frac{p^ip^j}{\hat{E}_P^2(\bs{p})}\left(1+t\hat{E}_P(\bs{p})+O(a^2)\right)\right\}+\dots\,.
\eea
In the rest frame of the meson, $\bs{p}=\bs{0}$, which we use in our study, we therefore obtain for the spatial components of the correlation function:
\bea\label{cji}
C^{ij}_A(t,0,\bs{0})=\delta^{ij}\,\frac{\hat{f}_P(\boldsymbol{0})\hat{G}_P(\boldsymbol{0})}{2\hat{E}_P(\boldsymbol{0})}e^{-t\hat{E}_P(\boldsymbol{0})}+...\,.
\eea

For the component $C^{00}_A$ the same procedure gives
\bea\label{c00}
C^{00}_A(t,0,\boldsymbol{p})=-t\hat{E}_P(\boldsymbol{p})\,\frac{\hat{f}_P(\boldsymbol{p})\hat{G}_P(\boldsymbol{p})}{2\hat{E}_P(\boldsymbol{p})}e^{-t\hat{E}_P(\boldsymbol{p})}\,+\,\dots\,.
\eea
For the components $C_{A}^{i 0}$, for the correlation function with $J_A^0$ and $J^i_{\mathrm{em}}$, we obtain 
\bea
C^{i0}_A(t,0,\boldsymbol{p})&=&\frac{\hat{f}_P(\boldsymbol{p})\hat{G}_P(\boldsymbol{p})}{2\hat{E}_P(\boldsymbol{p})}e^{-t\hat{E}_P(\boldsymbol{p})}\times\hat{E}_P(\boldsymbol{p})\left[\frac{1}{\hat{f}_P(\boldsymbol{p})}\frac{\partial \hat{f}_P(\boldsymbol{p})}{\partial p^i}+\frac{1}{\hat{G}_P(\boldsymbol{p})}\frac{\partial \hat{G}_P(\boldsymbol{p})}{\partial p^i}-t\frac{\partial \hat{E}_P(\boldsymbol{p})}{\partial p^i}\right]+ ...\nonumber\\
&=&-\frac{\hat{f}_P(\boldsymbol{p})\hat{G}_P(\boldsymbol{p})}{2\hat{E}_P(\boldsymbol{p})}e^{-t\hat{E}_P(\boldsymbol{p})}\times\hat{E}_P(\boldsymbol{p})\,p^i\left(t+O(a^2)\right)+\dots\,,
\eea
that in our reference frame becomes
\bea\label{c0i}
C^{i0}_A(t,0,\bs{0})&=&0+\dots\,.
\eea
Similarly, for the components $C^{0 i}_{A}$, differentiating equation Eq.\,(\ref{wi}) results in 
\bea\label{cj0}
C^{0j}_A(t,0,\boldsymbol{p})&=&-p^j t\,\frac{\hat{f}_P(\boldsymbol{p})\hat{G}_P(\boldsymbol{p})}{2\hat{E}_P(\boldsymbol{p})}e^{-t\hat{E}_P(\boldsymbol{p})} +\dots\,,
\eea
which in the rest frame of the meson becomes
\bea
C^{0j}_A(t,0,\bs{0})&=&0+\dots\,.
\eea \\

As explained in Ref.\,\cite{Desiderio2020}, Eq.\,(\ref{cji}) allows one to subtract the point-like contribution from the diagonal spatial components $C_A^{ii}$ components, non-perturbatively to all orders in the lattice spacing $a$. 
From Eqs.\,(\ref{c0i}) and (\ref{cj0}), we see that instead in the limit $k\to 0$ the contribution from the $P(\bs{p}-\bs{k}) + \gamma$ state exactly cancels the signal. Hence, for such components, it is not possible to extract, in the exact limit $k=0$, the physical matrix element from the Euclidean three point function.
\begin{figure}[t]
	\centering
	\includegraphics[scale=0.5]{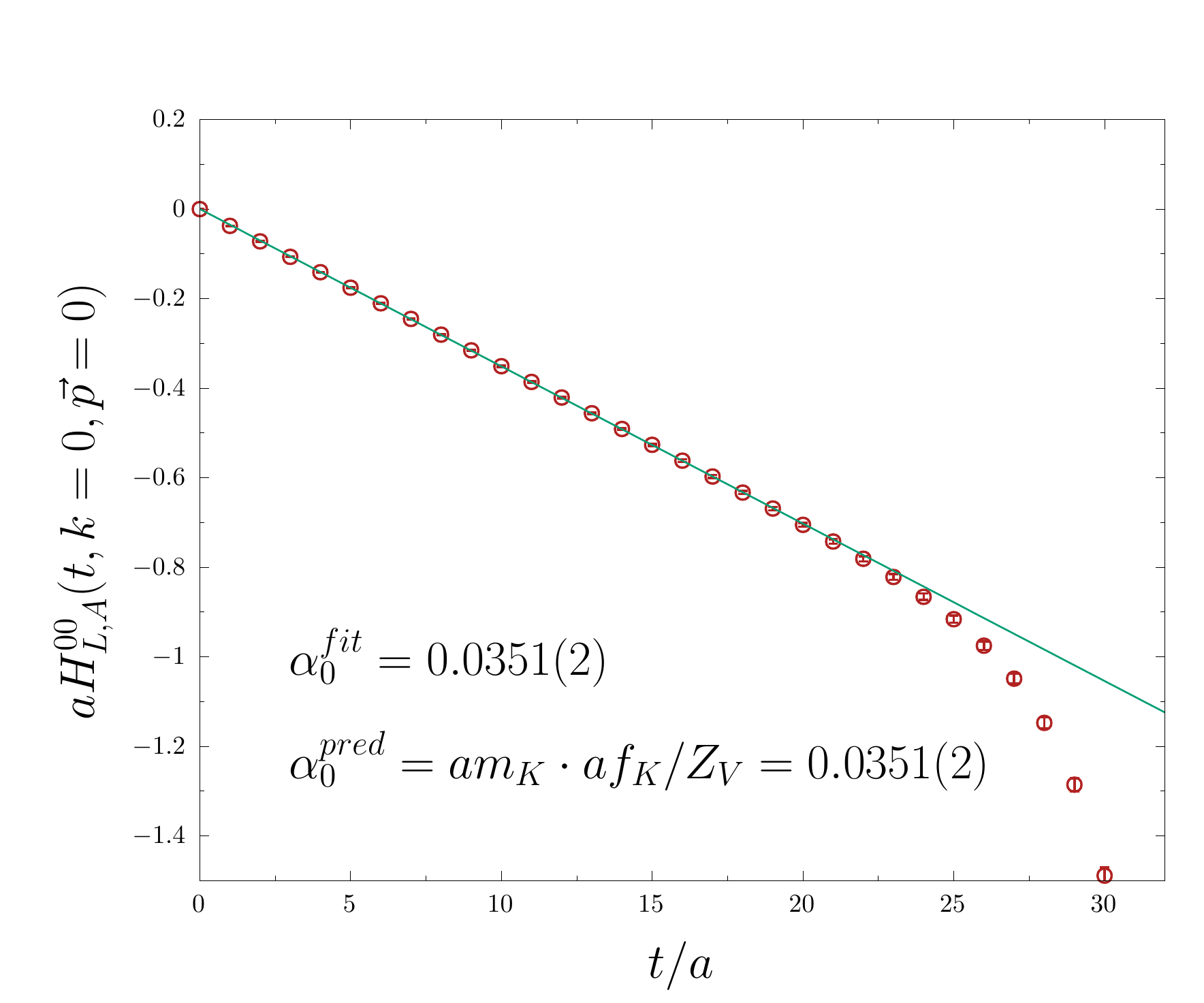}
	\caption{\it Determination of the $00$ component of the hadronic tensor, from the lattice three-point correlation function at $k=0$. The green line is the result of a linear fit in time  $aH_{L,A}^{00}(t,0,\bs{0})=-\alpha_{0}^{fit}t $, where $\alpha_{0}^{fit}$ is a fit parameter, and which is compared with the predicted value, $\alpha_{0}^{pred}$, derived from Eq. (\ref{c00}). The fit was performed in the interval $t=(4,21)$, away from the centre of the lattice where backward propagating contributions to the correlation function become significant.}
	\label{00plot}
\end{figure}/
We remark that in our analysis we have not used the purely temporal component $C_A^{00}$, which would make it difficult to identify the plateaux due to the presence of a large contribution from the excited state $P(\bs{p}-\bs{k}) + \gamma$, at small values of $k$. 

Finally, it is worthwhile noting the peculiar behaviour in $t$ of the purely temporal component of the lattice three point function, $C^{00}_A(t,0,\bs{p})$. From Eq.\,(\ref{c00}) it can be seen that $C^{00}_A(t,0,\bs{p})$ exhibits a time-behaviour of type $te^{-t\hat{E}_{P}(\bs{p})}$, which is a manifestation of the singular behaviour of the correlation function at large distances, and which gives rise to a double pole in momentum space.
In our simulation we found numerical evidence for the presence of such a behaviour.
This is shown in Fig.\,\ref{00plot}, where we compare our numerical data for $H^{00}_{L,A}(t,0,\bs{0})$, defined in Eq.\,(\ref{eq:lat_hadr_tens}), with the prediction of Eq.\,(\ref{c00}).

\bibliographystyle{apsrev4-2}
\bibliography{main}

\begin{thebibliography}{32}%
\makeatletter
\providecommand \@ifxundefined [1]{%
 \@ifx{#1\undefined}
}%
\providecommand \@ifnum [1]{%
 \ifnum #1\expandafter \@firstoftwo
 \else \expandafter \@secondoftwo
 \fi
}%
\providecommand \@ifx [1]{%
 \ifx #1\expandafter \@firstoftwo
 \else \expandafter \@secondoftwo
 \fi
}%
\providecommand \natexlab [1]{#1}%
\providecommand \enquote  [1]{``#1''}%
\providecommand \bibnamefont  [1]{#1}%
\providecommand \bibfnamefont [1]{#1}%
\providecommand \citenamefont [1]{#1}%
\providecommand \href@noop [0]{\@secondoftwo}%
\providecommand \href [0]{\begingroup \@sanitize@url \@href}%
\providecommand \@href[1]{\@@startlink{#1}\@@href}%
\providecommand \@@href[1]{\endgroup#1\@@endlink}%
\providecommand \@sanitize@url [0]{\catcode `\\12\catcode `\$12\catcode
  `\&12\catcode `\#12\catcode `\^12\catcode `\_12\catcode `\%12\relax}%
\providecommand \@@startlink[1]{}%
\providecommand \@@endlink[0]{}%
\providecommand \url  [0]{\begingroup\@sanitize@url \@url }%
\providecommand \@url [1]{\endgroup\@href {#1}{\urlprefix }}%
\providecommand \urlprefix  [0]{URL }%
\providecommand \Eprint [0]{\href }%
\providecommand \doibase [0]{https://doi.org/}%
\providecommand \selectlanguage [0]{\@gobble}%
\providecommand \bibinfo  [0]{\@secondoftwo}%
\providecommand \bibfield  [0]{\@secondoftwo}%
\providecommand \translation [1]{[#1]}%
\providecommand \BibitemOpen [0]{}%
\providecommand \bibitemStop [0]{}%
\providecommand \bibitemNoStop [0]{.\EOS\space}%
\providecommand \EOS [0]{\spacefactor3000\relax}%
\providecommand \BibitemShut  [1]{\csname bibitem#1\endcsname}%
\let\auto@bib@innerbib\@empty
\bibitem [{\citenamefont {Aaij}\ \emph {et~al.}(2021)\citenamefont {Aaij} \emph
  {et~al.}}]{LHCb:2021trn}%
  \BibitemOpen
  \bibfield  {author} {\bibinfo {author} {\bibfnamefont {R.}~\bibnamefont
  {Aaij}} \emph {et~al.} (\bibinfo {collaboration} {LHCb}),\ }\href@noop {} {\
  (\bibinfo {year} {2021})},\ \Eprint {https://arxiv.org/abs/2103.11769}
  {arXiv:2103.11769 [hep-ex]} \BibitemShut {NoStop}%
\bibitem [{\citenamefont {Crivellin}\ and\ \citenamefont
  {Hoferichter}(2021)}]{Crivellin:2021sff}%
  \BibitemOpen
  \bibfield  {author} {\bibinfo {author} {\bibfnamefont {A.}~\bibnamefont
  {Crivellin}}\ and\ \bibinfo {author} {\bibfnamefont {M.}~\bibnamefont
  {Hoferichter}},\ }\href {https://doi.org/10.1126/science.abk2450} {\bibfield
  {journal} {\bibinfo  {journal} {Science}\ }\textbf {\bibinfo {volume}
  {374}},\ \bibinfo {pages} {1051} (\bibinfo {year} {2021})},\ \Eprint
  {https://arxiv.org/abs/2111.12739} {arXiv:2111.12739 [hep-ph]} \BibitemShut
  {NoStop}%
\bibitem [{\citenamefont {Carrasco}\ \emph {et~al.}(2014)\citenamefont
  {Carrasco}, \citenamefont {Deuzeman}, \citenamefont {Dimopoulos},
  \citenamefont {Frezzotti}, \citenamefont {Gim{\'e}nez}, \citenamefont
  {Herdoiza}, \citenamefont {Lami}, \citenamefont {Lubicz}, \citenamefont
  {Palao}, \citenamefont {Picca},\ and\ \citenamefont {et~al.}}]{Carrasco2014}%
  \BibitemOpen
  \bibfield  {author} {\bibinfo {author} {\bibfnamefont {N.}~\bibnamefont
  {Carrasco}}, \bibinfo {author} {\bibfnamefont {A.}~\bibnamefont {Deuzeman}},
  \bibinfo {author} {\bibfnamefont {P.}~\bibnamefont {Dimopoulos}}, \bibinfo
  {author} {\bibfnamefont {R.}~\bibnamefont {Frezzotti}}, \bibinfo {author}
  {\bibfnamefont {V.}~\bibnamefont {Gim{\'e}nez}}, \bibinfo {author}
  {\bibfnamefont {G.}~\bibnamefont {Herdoiza}}, \bibinfo {author}
  {\bibfnamefont {P.}~\bibnamefont {Lami}}, \bibinfo {author} {\bibfnamefont
  {V.}~\bibnamefont {Lubicz}}, \bibinfo {author} {\bibfnamefont
  {D.}~\bibnamefont {Palao}}, \bibinfo {author} {\bibfnamefont
  {E.}~\bibnamefont {Picca}},\ and\ \bibinfo {author} {\bibnamefont {et~al.}},\
  }\href {https://doi.org/10.1016/j.nuclphysb.2014.07.025} {\bibfield
  {journal} {\bibinfo  {journal} {Nuclear Physics B}\ }\textbf {\bibinfo
  {volume} {887}},\ \bibinfo {pages} {19} (\bibinfo {year} {2014})}\BibitemShut
  {NoStop}%
\bibitem [{\citenamefont {Lubicz}\ \emph {et~al.}(2017)\citenamefont {Lubicz},
  \citenamefont {Martinelli}, \citenamefont {Sachrajda}, \citenamefont
  {Sanfilippo}, \citenamefont {Simula},\ and\ \citenamefont
  {Tantalo}}]{Lubicz:2016xro}%
  \BibitemOpen
  \bibfield  {author} {\bibinfo {author} {\bibfnamefont {V.}~\bibnamefont
  {Lubicz}}, \bibinfo {author} {\bibfnamefont {G.}~\bibnamefont {Martinelli}},
  \bibinfo {author} {\bibfnamefont {C.~T.}\ \bibnamefont {Sachrajda}}, \bibinfo
  {author} {\bibfnamefont {F.}~\bibnamefont {Sanfilippo}}, \bibinfo {author}
  {\bibfnamefont {S.}~\bibnamefont {Simula}},\ and\ \bibinfo {author}
  {\bibfnamefont {N.}~\bibnamefont {Tantalo}},\ }\href
  {https://doi.org/10.1103/PhysRevD.95.034504} {\bibfield  {journal} {\bibinfo
  {journal} {Phys. Rev. D}\ }\textbf {\bibinfo {volume} {95}},\ \bibinfo
  {pages} {034504} (\bibinfo {year} {2017})},\ \Eprint
  {https://arxiv.org/abs/1611.08497} {arXiv:1611.08497 [hep-lat]} \BibitemShut
  {NoStop}%
\bibitem [{\citenamefont {Giusti}\ \emph {et~al.}(2018)\citenamefont {Giusti},
  \citenamefont {Lubicz}, \citenamefont {Martinelli}, \citenamefont
  {Sachrajda}, \citenamefont {Sanfilippo}, \citenamefont {Simula},
  \citenamefont {Tantalo},\ and\ \citenamefont {Tarantino}}]{Giusti:2017dwk}%
  \BibitemOpen
  \bibfield  {author} {\bibinfo {author} {\bibfnamefont {D.}~\bibnamefont
  {Giusti}}, \bibinfo {author} {\bibfnamefont {V.}~\bibnamefont {Lubicz}},
  \bibinfo {author} {\bibfnamefont {G.}~\bibnamefont {Martinelli}}, \bibinfo
  {author} {\bibfnamefont {C.~T.}\ \bibnamefont {Sachrajda}}, \bibinfo {author}
  {\bibfnamefont {F.}~\bibnamefont {Sanfilippo}}, \bibinfo {author}
  {\bibfnamefont {S.}~\bibnamefont {Simula}}, \bibinfo {author} {\bibfnamefont
  {N.}~\bibnamefont {Tantalo}},\ and\ \bibinfo {author} {\bibfnamefont
  {C.}~\bibnamefont {Tarantino}},\ }\href
  {https://doi.org/10.1103/PhysRevLett.120.072001} {\bibfield  {journal}
  {\bibinfo  {journal} {Phys. Rev. Lett.}\ }\textbf {\bibinfo {volume} {120}},\
  \bibinfo {pages} {072001} (\bibinfo {year} {2018})},\ \Eprint
  {https://arxiv.org/abs/1711.06537} {arXiv:1711.06537 [hep-lat]} \BibitemShut
  {NoStop}%
\bibitem [{\citenamefont {Di~Carlo}\ \emph {et~al.}(2019)\citenamefont
  {Di~Carlo}, \citenamefont {Giusti}, \citenamefont {Lubicz}, \citenamefont
  {Martinelli}, \citenamefont {Sachrajda}, \citenamefont {Sanfilippo},
  \citenamefont {Simula},\ and\ \citenamefont {Tantalo}}]{DiCarlo:2019thl}%
  \BibitemOpen
  \bibfield  {author} {\bibinfo {author} {\bibfnamefont {M.}~\bibnamefont
  {Di~Carlo}}, \bibinfo {author} {\bibfnamefont {D.}~\bibnamefont {Giusti}},
  \bibinfo {author} {\bibfnamefont {V.}~\bibnamefont {Lubicz}}, \bibinfo
  {author} {\bibfnamefont {G.}~\bibnamefont {Martinelli}}, \bibinfo {author}
  {\bibfnamefont {C.~T.}\ \bibnamefont {Sachrajda}}, \bibinfo {author}
  {\bibfnamefont {F.}~\bibnamefont {Sanfilippo}}, \bibinfo {author}
  {\bibfnamefont {S.}~\bibnamefont {Simula}},\ and\ \bibinfo {author}
  {\bibfnamefont {N.}~\bibnamefont {Tantalo}},\ }\href
  {https://doi.org/10.1103/PhysRevD.100.034514} {\bibfield  {journal} {\bibinfo
   {journal} {Phys. Rev. D}\ }\textbf {\bibinfo {volume} {100}},\ \bibinfo
  {pages} {034514} (\bibinfo {year} {2019})},\ \Eprint
  {https://arxiv.org/abs/1904.08731} {arXiv:1904.08731 [hep-lat]} \BibitemShut
  {NoStop}%
\bibitem [{\citenamefont {Desiderio}\ \emph {et~al.}(2021)\citenamefont
  {Desiderio}, \citenamefont {Frezzotti}, \citenamefont {Garofalo},
  \citenamefont {Giusti}, \citenamefont {Hansen}, \citenamefont {Lubicz},
  \citenamefont {Martinelli}, \citenamefont {Sachrajda}, \citenamefont
  {Sanfilippo}, \citenamefont {Simula},\ and\ \citenamefont
  {et~al.}}]{Desiderio2020}%
  \BibitemOpen
  \bibfield  {author} {\bibinfo {author} {\bibfnamefont {A.}~\bibnamefont
  {Desiderio}}, \bibinfo {author} {\bibfnamefont {R.}~\bibnamefont
  {Frezzotti}}, \bibinfo {author} {\bibfnamefont {M.}~\bibnamefont {Garofalo}},
  \bibinfo {author} {\bibfnamefont {D.}~\bibnamefont {Giusti}}, \bibinfo
  {author} {\bibfnamefont {M.}~\bibnamefont {Hansen}}, \bibinfo {author}
  {\bibfnamefont {V.}~\bibnamefont {Lubicz}}, \bibinfo {author} {\bibfnamefont
  {G.}~\bibnamefont {Martinelli}}, \bibinfo {author} {\bibfnamefont
  {C.}~\bibnamefont {Sachrajda}}, \bibinfo {author} {\bibfnamefont
  {F.}~\bibnamefont {Sanfilippo}}, \bibinfo {author} {\bibfnamefont
  {S.}~\bibnamefont {Simula}},\ and\ \bibinfo {author} {\bibnamefont
  {et~al.}},\ }\bibfield  {journal} {\bibinfo  {journal} {Physical Review D}\
  }\textbf {\bibinfo {volume} {103}},\ \href
  {https://doi.org/10.1103/physrevd.103.014502} {10.1103/physrevd.103.014502}
  (\bibinfo {year} {2021})\BibitemShut {NoStop}%
\bibitem [{\citenamefont {Frezzotti}\ \emph {et~al.}(2021)\citenamefont
  {Frezzotti}, \citenamefont {Garofalo}, \citenamefont {Lubicz}, \citenamefont
  {Martinelli}, \citenamefont {Sachrajda}, \citenamefont {Sanfilippo},
  \citenamefont {Simula},\ and\ \citenamefont {Tantalo}}]{Frezzotti:2020bfa}%
  \BibitemOpen
  \bibfield  {author} {\bibinfo {author} {\bibfnamefont {R.}~\bibnamefont
  {Frezzotti}}, \bibinfo {author} {\bibfnamefont {M.}~\bibnamefont {Garofalo}},
  \bibinfo {author} {\bibfnamefont {V.}~\bibnamefont {Lubicz}}, \bibinfo
  {author} {\bibfnamefont {G.}~\bibnamefont {Martinelli}}, \bibinfo {author}
  {\bibfnamefont {C.~T.}\ \bibnamefont {Sachrajda}}, \bibinfo {author}
  {\bibfnamefont {F.}~\bibnamefont {Sanfilippo}}, \bibinfo {author}
  {\bibfnamefont {S.}~\bibnamefont {Simula}},\ and\ \bibinfo {author}
  {\bibfnamefont {N.}~\bibnamefont {Tantalo}},\ }\href
  {https://doi.org/10.1103/PhysRevD.103.053005} {\bibfield  {journal} {\bibinfo
   {journal} {Phys. Rev. D}\ }\textbf {\bibinfo {volume} {103}},\ \bibinfo
  {pages} {053005} (\bibinfo {year} {2021})},\ \Eprint
  {https://arxiv.org/abs/2012.02120} {arXiv:2012.02120 [hep-ph]} \BibitemShut
  {NoStop}%
\bibitem [{\citenamefont {Zyla}\ \emph {et~al.}(2020)\citenamefont {Zyla} \emph
  {et~al.}}]{pdg}%
  \BibitemOpen
  \bibfield  {author} {\bibinfo {author} {\bibfnamefont {P.}~\bibnamefont
  {Zyla}} \emph {et~al.} (\bibinfo {collaboration} {Particle Data Group}),\
  }\href {https://doi.org/10.1093/ptep/ptaa104} {\bibfield  {journal} {\bibinfo
   {journal} {PTEP}\ }\textbf {\bibinfo {volume} {2020}},\ \bibinfo {pages}
  {083C01} (\bibinfo {year} {2020})}\BibitemShut {NoStop}%
\bibitem [{\citenamefont {Poblaguev}\ \emph {et~al.}(2002)\citenamefont
  {Poblaguev}, \citenamefont {Appel}, \citenamefont {Atoyan}, \citenamefont
  {Bassalleck}, \citenamefont {Bergman}, \citenamefont {Cheung}, \citenamefont
  {Dhawan}, \citenamefont {Do}, \citenamefont {Egger}, \citenamefont
  {Eilerts},\ and\ \citenamefont {et~al.}}]{Poblaguev_2002}%
  \BibitemOpen
  \bibfield  {author} {\bibinfo {author} {\bibfnamefont {A.~A.}\ \bibnamefont
  {Poblaguev}}, \bibinfo {author} {\bibfnamefont {R.}~\bibnamefont {Appel}},
  \bibinfo {author} {\bibfnamefont {G.~S.}\ \bibnamefont {Atoyan}}, \bibinfo
  {author} {\bibfnamefont {B.}~\bibnamefont {Bassalleck}}, \bibinfo {author}
  {\bibfnamefont {D.~R.}\ \bibnamefont {Bergman}}, \bibinfo {author}
  {\bibfnamefont {N.}~\bibnamefont {Cheung}}, \bibinfo {author} {\bibfnamefont
  {S.}~\bibnamefont {Dhawan}}, \bibinfo {author} {\bibfnamefont
  {H.}~\bibnamefont {Do}}, \bibinfo {author} {\bibfnamefont {J.}~\bibnamefont
  {Egger}}, \bibinfo {author} {\bibfnamefont {S.}~\bibnamefont {Eilerts}},\
  and\ \bibinfo {author} {\bibnamefont {et~al.}},\ }\bibfield  {journal}
  {\bibinfo  {journal} {Physical Review Letters}\ }\textbf {\bibinfo {volume}
  {89}},\ \href {https://doi.org/10.1103/physrevlett.89.061803}
  {10.1103/physrevlett.89.061803} (\bibinfo {year} {2002})\BibitemShut
  {NoStop}%
\bibitem [{\citenamefont {Ma}\ \emph {et~al.}(2006)\citenamefont {Ma},
  \citenamefont {Appel}, \citenamefont {Atoyan}, \citenamefont {Bassalleck},
  \citenamefont {Bergman}, \citenamefont {Cheung}, \citenamefont {Dhawan},
  \citenamefont {Do}, \citenamefont {Egger}, \citenamefont {Eilerts},\ and\
  \citenamefont {et~al.}}]{Ma_2006}%
  \BibitemOpen
  \bibfield  {author} {\bibinfo {author} {\bibfnamefont {H.}~\bibnamefont
  {Ma}}, \bibinfo {author} {\bibfnamefont {R.}~\bibnamefont {Appel}}, \bibinfo
  {author} {\bibfnamefont {G.~S.}\ \bibnamefont {Atoyan}}, \bibinfo {author}
  {\bibfnamefont {B.}~\bibnamefont {Bassalleck}}, \bibinfo {author}
  {\bibfnamefont {D.~R.}\ \bibnamefont {Bergman}}, \bibinfo {author}
  {\bibfnamefont {N.}~\bibnamefont {Cheung}}, \bibinfo {author} {\bibfnamefont
  {S.}~\bibnamefont {Dhawan}}, \bibinfo {author} {\bibfnamefont
  {H.}~\bibnamefont {Do}}, \bibinfo {author} {\bibfnamefont {J.}~\bibnamefont
  {Egger}}, \bibinfo {author} {\bibfnamefont {S.}~\bibnamefont {Eilerts}},\
  and\ \bibinfo {author} {\bibnamefont {et~al.}},\ }\bibfield  {journal}
  {\bibinfo  {journal} {Physical Review D}\ }\textbf {\bibinfo {volume} {73}},\
  \href {https://doi.org/10.1103/physrevd.73.037101}
  {10.1103/physrevd.73.037101} (\bibinfo {year} {2006})\BibitemShut {NoStop}%
\bibitem [{\citenamefont {Aaij}\ \emph {et~al.}(2019)\citenamefont {Aaij},
  \citenamefont {Beteta}, \citenamefont {Adeva}, \citenamefont {Adinolfi},
  \citenamefont {Aidala}, \citenamefont {Ajaltouni}, \citenamefont {Akar},
  \citenamefont {Albicocco}, \citenamefont {Albrecht},\ and\ \citenamefont
  {et~al.}}]{Aaij_2019}%
  \BibitemOpen
  \bibfield  {author} {\bibinfo {author} {\bibfnamefont {R.}~\bibnamefont
  {Aaij}}, \bibinfo {author} {\bibfnamefont {C.~A.}\ \bibnamefont {Beteta}},
  \bibinfo {author} {\bibfnamefont {B.}~\bibnamefont {Adeva}}, \bibinfo
  {author} {\bibfnamefont {M.}~\bibnamefont {Adinolfi}}, \bibinfo {author}
  {\bibfnamefont {C.~A.}\ \bibnamefont {Aidala}}, \bibinfo {author}
  {\bibfnamefont {Z.}~\bibnamefont {Ajaltouni}}, \bibinfo {author}
  {\bibfnamefont {S.}~\bibnamefont {Akar}}, \bibinfo {author} {\bibfnamefont
  {P.}~\bibnamefont {Albicocco}}, \bibinfo {author} {\bibfnamefont
  {J.}~\bibnamefont {Albrecht}},\ and\ \bibinfo {author} {\bibnamefont
  {et~al.}},\ }\bibfield  {journal} {\bibinfo  {journal} {The European Physical
  Journal C}\ }\textbf {\bibinfo {volume} {79}},\ \href
  {https://doi.org/10.1140/epjc/s10052-019-7112-x}
  {10.1140/epjc/s10052-019-7112-x} (\bibinfo {year} {2019})\BibitemShut
  {NoStop}%
\bibitem [{\citenamefont {Tuo}\ \emph {et~al.}(2021)\citenamefont {Tuo},
  \citenamefont {Feng}, \citenamefont {Jin},\ and\ \citenamefont {Wang}}]{xu}%
  \BibitemOpen
  \bibfield  {author} {\bibinfo {author} {\bibfnamefont {X.-Y.}\ \bibnamefont
  {Tuo}}, \bibinfo {author} {\bibfnamefont {X.}~\bibnamefont {Feng}}, \bibinfo
  {author} {\bibfnamefont {L.-C.}\ \bibnamefont {Jin}},\ and\ \bibinfo {author}
  {\bibfnamefont {T.}~\bibnamefont {Wang}},\ }\href@noop {} {} (\bibinfo {year}
  {2021}),\ \Eprint {https://arxiv.org/abs/2103.11331} {arXiv:2103.11331
  [hep-lat]} \BibitemShut {NoStop}%
\bibitem [{\citenamefont {Lellouch}\ and\ \citenamefont
  {Luscher}(2001)}]{Lellouch:2000pv}%
  \BibitemOpen
  \bibfield  {author} {\bibinfo {author} {\bibfnamefont {L.}~\bibnamefont
  {Lellouch}}\ and\ \bibinfo {author} {\bibfnamefont {M.}~\bibnamefont
  {Luscher}},\ }\href {https://doi.org/10.1007/s002200100410} {\bibfield
  {journal} {\bibinfo  {journal} {Commun. Math. Phys.}\ }\textbf {\bibinfo
  {volume} {219}},\ \bibinfo {pages} {31} (\bibinfo {year} {2001})},\ \Eprint
  {https://arxiv.org/abs/hep-lat/0003023} {arXiv:hep-lat/0003023} \BibitemShut
  {NoStop}%
\bibitem [{\citenamefont {Kim}\ \emph {et~al.}(2005)\citenamefont {Kim},
  \citenamefont {Sachrajda},\ and\ \citenamefont {Sharpe}}]{Kim:2005gf}%
  \BibitemOpen
  \bibfield  {author} {\bibinfo {author} {\bibfnamefont {C.~h.}\ \bibnamefont
  {Kim}}, \bibinfo {author} {\bibfnamefont {C.~T.}\ \bibnamefont {Sachrajda}},\
  and\ \bibinfo {author} {\bibfnamefont {S.~R.}\ \bibnamefont {Sharpe}},\
  }\href {https://doi.org/10.1016/j.nuclphysb.2005.08.029} {\bibfield
  {journal} {\bibinfo  {journal} {Nucl. Phys. B}\ }\textbf {\bibinfo {volume}
  {727}},\ \bibinfo {pages} {218} (\bibinfo {year} {2005})},\ \Eprint
  {https://arxiv.org/abs/hep-lat/0507006} {arXiv:hep-lat/0507006} \BibitemShut
  {NoStop}%
\bibitem [{\citenamefont {Brice\~no}\ \emph {et~al.}(2020)\citenamefont
  {Brice\~no}, \citenamefont {Davoudi}, \citenamefont {Hansen}, \citenamefont
  {Schindler},\ and\ \citenamefont {Baroni}}]{Briceno:2019opb}%
  \BibitemOpen
  \bibfield  {author} {\bibinfo {author} {\bibfnamefont {R.~A.}\ \bibnamefont
  {Brice\~no}}, \bibinfo {author} {\bibfnamefont {Z.}~\bibnamefont {Davoudi}},
  \bibinfo {author} {\bibfnamefont {M.~T.}\ \bibnamefont {Hansen}}, \bibinfo
  {author} {\bibfnamefont {M.~R.}\ \bibnamefont {Schindler}},\ and\ \bibinfo
  {author} {\bibfnamefont {A.}~\bibnamefont {Baroni}},\ }\href
  {https://doi.org/10.1103/PhysRevD.101.014509} {\bibfield  {journal} {\bibinfo
   {journal} {Phys. Rev. D}\ }\textbf {\bibinfo {volume} {101}},\ \bibinfo
  {pages} {014509} (\bibinfo {year} {2020})},\ \Eprint
  {https://arxiv.org/abs/1911.04036} {arXiv:1911.04036 [hep-lat]} \BibitemShut
  {NoStop}%
\bibitem [{\citenamefont {Bijnens}\ \emph {et~al.}(1994)\citenamefont
  {Bijnens}, \citenamefont {Colangelo}, \citenamefont {Ecker},\ and\
  \citenamefont {Gasser}}]{Bijnens:1994me}%
  \BibitemOpen
  \bibfield  {author} {\bibinfo {author} {\bibfnamefont {J.}~\bibnamefont
  {Bijnens}}, \bibinfo {author} {\bibfnamefont {G.}~\bibnamefont {Colangelo}},
  \bibinfo {author} {\bibfnamefont {G.}~\bibnamefont {Ecker}},\ and\ \bibinfo
  {author} {\bibfnamefont {J.}~\bibnamefont {Gasser}}\ }(\bibinfo {year}
  {1994})\ \Eprint {https://arxiv.org/abs/hep-ph/9411311}
  {arXiv:hep-ph/9411311} \BibitemShut {NoStop}%
\bibitem [{\citenamefont {Danilina}\ \emph {et~al.}(2020)\citenamefont
  {Danilina}, \citenamefont {Nikitin},\ and\ \citenamefont
  {Toms}}]{Danilina_2020}%
  \BibitemOpen
  \bibfield  {author} {\bibinfo {author} {\bibfnamefont {A.}~\bibnamefont
  {Danilina}}, \bibinfo {author} {\bibfnamefont {N.}~\bibnamefont {Nikitin}},\
  and\ \bibinfo {author} {\bibfnamefont {K.}~\bibnamefont {Toms}},\ }\bibfield
  {journal} {\bibinfo  {journal} {Physical Review D}\ }\textbf {\bibinfo
  {volume} {101}},\ \href {https://doi.org/10.1103/physrevd.101.096007}
  {10.1103/physrevd.101.096007} (\bibinfo {year} {2020})\BibitemShut {NoStop}%
\bibitem [{\citenamefont {Carrasco}\ \emph {et~al.}(2015)\citenamefont
  {Carrasco}, \citenamefont {Lubicz}, \citenamefont {Martinelli}, \citenamefont
  {Sachrajda}, \citenamefont {Tantalo}, \citenamefont {Tarantino},\ and\
  \citenamefont {Testa}}]{Carrasco_2015}%
  \BibitemOpen
  \bibfield  {author} {\bibinfo {author} {\bibfnamefont {N.}~\bibnamefont
  {Carrasco}}, \bibinfo {author} {\bibfnamefont {V.}~\bibnamefont {Lubicz}},
  \bibinfo {author} {\bibfnamefont {G.}~\bibnamefont {Martinelli}}, \bibinfo
  {author} {\bibfnamefont {C.}~\bibnamefont {Sachrajda}}, \bibinfo {author}
  {\bibfnamefont {N.}~\bibnamefont {Tantalo}}, \bibinfo {author} {\bibfnamefont
  {C.}~\bibnamefont {Tarantino}},\ and\ \bibinfo {author} {\bibfnamefont
  {M.}~\bibnamefont {Testa}},\ }\bibfield  {journal} {\bibinfo  {journal}
  {Physical Review D}\ }\textbf {\bibinfo {volume} {91}},\ \href
  {https://doi.org/10.1103/physrevd.91.074506} {10.1103/physrevd.91.074506}
  (\bibinfo {year} {2015})\BibitemShut {NoStop}%
\bibitem [{\citenamefont {Maiani}\ and\ \citenamefont {Testa}(1990)}]{maiani}%
  \BibitemOpen
  \bibfield  {author} {\bibinfo {author} {\bibfnamefont {L.}~\bibnamefont
  {Maiani}}\ and\ \bibinfo {author} {\bibfnamefont {M.}~\bibnamefont {Testa}},\
  }\href {https://doi.org/https://doi.org/10.1016/0370-2693(90)90695-3}
  {\bibfield  {journal} {\bibinfo  {journal} {Physics Letters B}\ }\textbf
  {\bibinfo {volume} {245}},\ \bibinfo {pages} {585} (\bibinfo {year}
  {1990})}\BibitemShut {NoStop}%
\bibitem [{\citenamefont {Frezzotti}\ \emph {et~al.}(2001)\citenamefont
  {Frezzotti}, \citenamefont {Grassi}, \citenamefont {Sint},\ and\
  \citenamefont {Weisz}}]{Frezzotti:2000nk}%
  \BibitemOpen
  \bibfield  {author} {\bibinfo {author} {\bibfnamefont {R.}~\bibnamefont
  {Frezzotti}}, \bibinfo {author} {\bibfnamefont {P.~A.}\ \bibnamefont
  {Grassi}}, \bibinfo {author} {\bibfnamefont {S.}~\bibnamefont {Sint}},\ and\
  \bibinfo {author} {\bibfnamefont {P.}~\bibnamefont {Weisz}} (\bibinfo
  {collaboration} {Alpha}),\ }\href
  {https://doi.org/10.1088/1126-6708/2001/08/058} {\bibfield  {journal}
  {\bibinfo  {journal} {JHEP}\ }\textbf {\bibinfo {volume} {08}},\ \bibinfo
  {pages} {058}},\ \Eprint {https://arxiv.org/abs/hep-lat/0101001}
  {arXiv:hep-lat/0101001} \BibitemShut {NoStop}%
\bibitem [{\citenamefont {Frezzotti}\ and\ \citenamefont
  {Rossi}(2004)}]{Frezzotti_2004}%
  \BibitemOpen
  \bibfield  {author} {\bibinfo {author} {\bibfnamefont {R.}~\bibnamefont
  {Frezzotti}}\ and\ \bibinfo {author} {\bibfnamefont {G.}~\bibnamefont
  {Rossi}},\ }\href {https://doi.org/10.1088/1126-6708/2004/08/007} {\bibfield
  {journal} {\bibinfo  {journal} {Journal of High Energy Physics}\ }\textbf
  {\bibinfo {volume} {08}},\ \bibinfo {pages} {007} (\bibinfo {year}
  {2004})}\BibitemShut {NoStop}%
\bibitem [{\citenamefont {de~Divitiis}\ \emph {et~al.}(2013)\citenamefont
  {de~Divitiis}, \citenamefont {Frezzotti}, \citenamefont {Lubicz},
  \citenamefont {Martinelli}, \citenamefont {Petronzio}, \citenamefont {Rossi},
  \citenamefont {Sanfilippo}, \citenamefont {Simula},\ and\ \citenamefont
  {Tantalo}}]{de_Divitiis_2013}%
  \BibitemOpen
  \bibfield  {author} {\bibinfo {author} {\bibfnamefont {G.~M.}\ \bibnamefont
  {de~Divitiis}}, \bibinfo {author} {\bibfnamefont {R.}~\bibnamefont
  {Frezzotti}}, \bibinfo {author} {\bibfnamefont {V.}~\bibnamefont {Lubicz}},
  \bibinfo {author} {\bibfnamefont {G.}~\bibnamefont {Martinelli}}, \bibinfo
  {author} {\bibfnamefont {R.}~\bibnamefont {Petronzio}}, \bibinfo {author}
  {\bibfnamefont {G.~C.}\ \bibnamefont {Rossi}}, \bibinfo {author}
  {\bibfnamefont {F.}~\bibnamefont {Sanfilippo}}, \bibinfo {author}
  {\bibfnamefont {S.}~\bibnamefont {Simula}},\ and\ \bibinfo {author}
  {\bibfnamefont {N.}~\bibnamefont {Tantalo}},\ }\bibfield  {journal} {\bibinfo
   {journal} {Physical Review D}\ }\textbf {\bibinfo {volume} {87}},\ \href
  {https://doi.org/10.1103/physrevd.87.114505} {10.1103/physrevd.87.114505}
  (\bibinfo {year} {2013})\BibitemShut {NoStop}%
\bibitem [{\citenamefont {de~Divitiis}\ \emph {et~al.}(2004)\citenamefont
  {de~Divitiis}, \citenamefont {Petronzio},\ and\ \citenamefont
  {Tantalo}}]{de_Divitiis_2004}%
  \BibitemOpen
  \bibfield  {author} {\bibinfo {author} {\bibfnamefont {G.}~\bibnamefont
  {de~Divitiis}}, \bibinfo {author} {\bibfnamefont {R.}~\bibnamefont
  {Petronzio}},\ and\ \bibinfo {author} {\bibfnamefont {N.}~\bibnamefont
  {Tantalo}},\ }\href {https://doi.org/10.1016/j.physletb.2004.06.035}
  {\bibfield  {journal} {\bibinfo  {journal} {Physics Letters B}\ }\textbf
  {\bibinfo {volume} {595}},\ \bibinfo {pages} {408} (\bibinfo {year}
  {2004})}\BibitemShut {NoStop}%
\bibitem [{\citenamefont {Flynn}\ \emph {et~al.}(2007)\citenamefont {Flynn},
  \citenamefont {J{\"u}ttner}, \citenamefont {Sachrajda}, \citenamefont
  {Boyle}, \citenamefont {Zanotti},\ and\ \citenamefont
  {collaboration}}]{Boyle:2007wg}%
  \BibitemOpen
  \bibfield  {author} {\bibinfo {author} {\bibfnamefont {J.~M.}\ \bibnamefont
  {Flynn}}, \bibinfo {author} {\bibfnamefont {A.}~\bibnamefont {J{\"u}ttner}},
  \bibinfo {author} {\bibfnamefont {C.}~\bibnamefont {Sachrajda}}, \bibinfo
  {author} {\bibfnamefont {P.~A.}\ \bibnamefont {Boyle}}, \bibinfo {author}
  {\bibfnamefont {J.~M.}\ \bibnamefont {Zanotti}},\ and\ \bibinfo {author}
  {\bibfnamefont {U.}~\bibnamefont {collaboration}},\ }\href
  {https://doi.org/10.1088/1126-6708/2007/05/016} {\bibfield  {journal}
  {\bibinfo  {journal} {Journal of High Energy Physics}\ }\textbf {\bibinfo
  {volume} {2007}},\ \bibinfo {pages} {016} (\bibinfo {year}
  {2007})}\BibitemShut {NoStop}%
\bibitem [{\citenamefont {Sachrajda}\ and\ \citenamefont
  {Villadoro}(2005)}]{Sachrajda:2004mi}%
  \BibitemOpen
  \bibfield  {author} {\bibinfo {author} {\bibfnamefont {C.~T.}\ \bibnamefont
  {Sachrajda}}\ and\ \bibinfo {author} {\bibfnamefont {G.}~\bibnamefont
  {Villadoro}},\ }\href {https://doi.org/10.1016/j.physletb.2005.01.033}
  {\bibfield  {journal} {\bibinfo  {journal} {Phys. Lett. B}\ }\textbf
  {\bibinfo {volume} {609}},\ \bibinfo {pages} {73} (\bibinfo {year} {2005})},\
  \Eprint {https://arxiv.org/abs/hep-lat/0411033} {arXiv:hep-lat/0411033}
  \BibitemShut {NoStop}%
\bibitem [{\citenamefont {Jansen}\ \emph {et~al.}(2004)\citenamefont {Jansen},
  \citenamefont {Shindler}, \citenamefont {Urbach},\ and\ \citenamefont
  {Wetzorke}}]{Jansen:2003ir}%
  \BibitemOpen
  \bibfield  {author} {\bibinfo {author} {\bibfnamefont {K.}~\bibnamefont
  {Jansen}}, \bibinfo {author} {\bibfnamefont {A.}~\bibnamefont {Shindler}},
  \bibinfo {author} {\bibfnamefont {C.}~\bibnamefont {Urbach}},\ and\ \bibinfo
  {author} {\bibfnamefont {I.}~\bibnamefont {Wetzorke}} (\bibinfo
  {collaboration} {XLF}),\ }\href
  {https://doi.org/10.1016/j.physletb.2004.01.030} {\bibfield  {journal}
  {\bibinfo  {journal} {Phys. Lett. B}\ }\textbf {\bibinfo {volume} {586}},\
  \bibinfo {pages} {432} (\bibinfo {year} {2004})},\ \Eprint
  {https://arxiv.org/abs/hep-lat/0312013} {arXiv:hep-lat/0312013} \BibitemShut
  {NoStop}%
\bibitem [{\citenamefont {Shindler}(2008)}]{Shindler:2007vp}%
  \BibitemOpen
  \bibfield  {author} {\bibinfo {author} {\bibfnamefont {A.}~\bibnamefont
  {Shindler}},\ }\href {https://doi.org/10.1016/j.physrep.2008.03.001}
  {\bibfield  {journal} {\bibinfo  {journal} {Phys. Rept.}\ }\textbf {\bibinfo
  {volume} {461}},\ \bibinfo {pages} {37} (\bibinfo {year} {2008})},\ \Eprint
  {https://arxiv.org/abs/0707.4093} {arXiv:0707.4093 [hep-lat]} \BibitemShut
  {NoStop}%
\bibitem [{\citenamefont {Bijnens}\ and\ \citenamefont {Ecker}(2014)}]{LEC}%
  \BibitemOpen
  \bibfield  {author} {\bibinfo {author} {\bibfnamefont {J.}~\bibnamefont
  {Bijnens}}\ and\ \bibinfo {author} {\bibfnamefont {G.}~\bibnamefont
  {Ecker}},\ }\href {https://doi.org/10.1146/annurev-nucl-102313-025528}
  {\bibfield  {journal} {\bibinfo  {journal} {Annual Review of Nuclear and
  Particle Science}\ }\textbf {\bibinfo {volume} {64}},\ \bibinfo {pages} {149}
  (\bibinfo {year} {2014})},\ \Eprint
  {https://arxiv.org/abs/https://doi.org/10.1146/annurev-nucl-102313-025528}
  {https://doi.org/10.1146/annurev-nucl-102313-025528} \BibitemShut {NoStop}%
\bibitem [{\citenamefont {Krishna}\ and\ \citenamefont {Mani}(1972)}]{krishna}%
  \BibitemOpen
  \bibfield  {author} {\bibinfo {author} {\bibfnamefont {S.}~\bibnamefont
  {Krishna}}\ and\ \bibinfo {author} {\bibfnamefont {H.~S.}\ \bibnamefont
  {Mani}},\ }\href {https://doi.org/10.1103/PhysRevD.5.678} {\bibfield
  {journal} {\bibinfo  {journal} {Phys. Rev. D}\ }\textbf {\bibinfo {volume}
  {5}},\ \bibinfo {pages} {678} (\bibinfo {year} {1972})}\BibitemShut {NoStop}%
\bibitem [{\citenamefont {Shtabovenko}\ \emph {et~al.}(2020)\citenamefont
  {Shtabovenko}, \citenamefont {Mertig},\ and\ \citenamefont
  {Orellana}}]{2020FenyCalc}%
  \BibitemOpen
  \bibfield  {author} {\bibinfo {author} {\bibfnamefont {V.}~\bibnamefont
  {Shtabovenko}}, \bibinfo {author} {\bibfnamefont {R.}~\bibnamefont
  {Mertig}},\ and\ \bibinfo {author} {\bibfnamefont {F.}~\bibnamefont
  {Orellana}},\ }\href {https://doi.org/10.1016/j.cpc.2020.107478} {\bibfield
  {journal} {\bibinfo  {journal} {Computer Physics Communications}\ }\textbf
  {\bibinfo {volume} {256}},\ \bibinfo {pages} {107478} (\bibinfo {year}
  {2020})}\BibitemShut {NoStop}%
\bibitem [{\citenamefont {{Lepage}}(1978)}]{1978JCoPh..27..192L}%
  \BibitemOpen
  \bibfield  {author} {\bibinfo {author} {\bibfnamefont {G.~P.}\ \bibnamefont
  {{Lepage}}},\ }\href {https://doi.org/10.1016/0021-9991(78)90004-9}
  {\bibfield  {journal} {\bibinfo  {journal} {Journal of Computational
  Physics}\ }\textbf {\bibinfo {volume} {27}},\ \bibinfo {pages} {192}
  (\bibinfo {year} {1978})}\BibitemShut {NoStop}%
\end{thebibliography}%
\end{document}